\documentclass[12pt]{article}
\pdfoutput=1

\usepackage[utf8]{inputenc}

\usepackage{draft}
\usepackage{putex}
 
\usepackage[all]{xy}

\usepackage{stackrel,amssymb}

\usepackage{graphicx}
\usepackage{caption}
\usepackage{amsmath,bm}
\usepackage{array}
\usepackage{subcaption}
\usepackage{epstopdf}
\usepackage{enumerate}
\usepackage{cite}
\usepackage{tensor}
\usepackage{slashed}
\usepackage[utf8]{inputenc}
\usepackage{rotating}
\usepackage[
      colorlinks=true,
      linkcolor=blue,
      urlcolor=blue,
      filecolor=black,
      citecolor=red,
      ]{hyperref}
      
\usepackage{dsfont}

\usepackage{tikz-cd}
\usetikzlibrary{arrows,automata,positioning}

\def\XXint#1#2#3{{\setbox0=\hbox{$#1{#2#3}{\int}$ }
		\vcenter{\hbox{$#2#3$ }}\kern-.6\wd0}}

  \usepackage{adjustbox}
\usepackage{multirow}
\usepackage{tikz-cd}

\numberwithin{equation}{section}

\newcommand{\cI}{{\cal I}}

\definecolor{vert}{rgb}{0.1367 0.543 0.1367}

\newcommand{\half}{{1\over 2}}
\newcommand{\id}{\mathds{1}}
\DeclareMathOperator{\im}{Im}

\newtheorem{theorem}{Theorem}

\def\<{\langle}
\def\>{\rangle}

\def\pa{\partial}

\def\ep{\epsilon}

\newcommand{\leftrarrows}{\mathrel{\raise.75ex\hbox{\oalign{%
				$\scriptstyle\leftarrow$\cr
				\vrule width0pt height.5ex$\hfil\scriptstyle\relbar$\cr}}}}
\newcommand{\lrightarrows}{\mathrel{\raise.75ex\hbox{\oalign{%
				$\scriptstyle\relbar$\hfil\cr
				$\scriptstyle\vrule width0pt height.5ex\smash\rightarrow$\cr}}}}
\newcommand{\Rrelbar}{\mathrel{\raise.75ex\hbox{\oalign{%
				$\scriptstyle\relbar$\cr
				\vrule width0pt height.5ex$\scriptstyle\relbar$}}}}

\newcommand{\lla}{\la\!\la}
\newcommand{\rra}{\ra\!\ra}
\makeatletter
\def\leftrightarrowsfill@{\arrowfill@\leftrarrows\Rrelbar\lrightarrows}
\newcommand{\xleftrightarrows}[2][]{\ext@arrow 3399\leftrightarrowsfill@{#1}{#2}}
\makeatother

\begin{document}

\preprint{}

	\institution{PU}{Princeton Center for Theoretical Science, Princeton University, Princeton, NJ 08544, USA}\institution{IAS}{School of Natural Sciences, Institute for Advanced Study,
Princeton, NJ 08540, USA}
\institution{CMSA}{Center of Mathematical Sciences and Applications, Harvard University, Cambridge, MA 02138, USA}
	\institution{HU}{Jefferson Physical Laboratory, Harvard University,
		Cambridge, MA 02138, USA}
	
\institution{NYU}{Center for Cosmology and Particle Physics, New York University, New York, NY 10003, USA}

\title{
Bootstrapping Boundaries and Branes
}

\authors{ Scott Collier,\worksat{\PU}\,Dalimil Maz\'a\v c,\worksat{\IAS} and 
Yifan Wang\worksat{\CMSA,\HU,\NYU}}

\abstract{
The study of conformal boundary conditions for two-dimensional conformal field theories (CFTs) has a long history, ranging from the description of impurities in one-dimensional quantum chains to the formulation of D-branes in string theory. Nevertheless, the landscape of conformal boundaries is largely unknown, including in rational CFTs, where the local operator data is completely determined.
We initiate a systematic bootstrap study of conformal boundaries in 2d CFTs by investigating the bootstrap equation that arises from the open-closed consistency condition of the annulus partition function with identical boundaries. We find that this deceivingly simple bootstrap equation, when combined with unitarity, leads to surprisingly strong constraints on admissible boundary states. In particular, we derive universal bounds on the tension (boundary entropy) of stable boundary conditions, which provide a rigorous diagnostic for potential D-brane decays. We also find unique solutions to the bootstrap problem of stable branes in a number of rational CFTs. Along the way, we observe a curious connection between the annulus bootstrap and the sphere packing problem, which is a natural extension of previous work on the modular bootstrap. We also derive bounds on the boundary entropy at large central charge. These potentially have implications for end-of-the-world branes in pure gravity on AdS$_3$.
}
\date{}

\maketitle

\begingroup

\baselineskip .25 in
\tableofcontents

\endgroup

\pagebreak

\section{Introduction and Summary}

Boundaries are an integral part of the modern understanding of quantum field theory (QFT). They are pervasive throughout all branches of physics and are responsible for important boundary effects in many-body systems that arise in particle physics, condensed matter and statistical mechanics. Systems with boundaries exhibit universal behavior in the continuum limit and 
are often described in QFT by boundary conditions constrained by the basic principles of locality, unitarity and symmetries. 
The incorporation of boundaries introduces an abundance of new observables in QFT with novel features that pose outstanding challenges for theoretical understanding.
At the same time, studying the boundaries also provides new windows to the structure of bulk dynamics in QFT. 

The boundary effects are the most pronounced when the bulk system is in the critical phase where the correlation length diverges, described by a conformal field theory (CFT). The local boundary conditions of a CFT are organized into universality classes under boundary renormalization group (RG) flows (e.g.\ triggered by relevant deformations on the boundary) which are characterized by RG fixed points that correspond to conformal boundaries. The coupled  bulk-boundary critical system is also known as boundary conformal field theory (BCFT). By virtue of universality, the conformal boundaries are the natural platform to start delineating the landscape of interesting boundary conditions in QFT.\footnote{See \cite{Andrei:2018die} for a recent review on boundaries and more general defects in CFTs.} Thanks to the enhanced conformal symmetry, there is a systematic and efficient way to explore the constraints from locality and unitarity, which falls under the conformal bootstrap program (see \cite{Poland:2018epd} for a review). 

Here we focus on conformal boundaries of 2d CFTs. They are important for describing critical quantum impurities in one-dimensional quantum spin chains, whose intricate dynamics is exemplified by the Kondo effect \cite{10.1143/PTP.32.37,Affleck:1990by,Affleck:1990iv,Affleck:1992ng,Affleck:1995ge}. They also represent D-branes in string theory as boundary conditions for the worldsheet CFT. Via the AdS/CFT correspondence, these conformal boundaries are mapped to end-of-the-world branes for quantum gravity on AdS$_3$ \cite{Karch:2000gx,Takayanagi:2011zk,Fujita:2011fp,Nozaki:2012qd}, which feature prominently in recent studies of black holes and the unitarity of their evaporation (see e.g. \cite{Almheiri:2019hni,Penington:2019kki,Rozali:2019day,Akal:2021foz,Geng:2020fxl,Geng:2021iyq}).

Given the aforementioned broad applications, 
the 2d conformal boundaries are arguably the most well-studied among all BCFTs. This is further facilitated by the fact that they inherit a diagonal Virasoro subalgebra from the 2d CFT in the bulk and consequently 
appear to have a more rigid structure compared to their higher dimensional cousins. In fact it is believed that 2d BCFT is completely specified by the (bulk and boundary) operator spectrum and 
operator-product expansion (OPE) data, which are
solutions to a set of algebraic constraints, known as bootstrap equations, that ensure the consistency of cutting and gluing of the Riemann surface with boundaries on which the BCFT lives. However, explicit solutions to these bootstrap equations are only known in minimal models \cite{Cardy:1989ir,Cardy:1991tv,Runkel:1998he}, for special cases in free CFTs (e.g.\ Dirichlet and Neumann boundary conditions for free compact bosons \cite{Gaberdiel:2001xm,Gaberdiel:2001zq}), in rational CFT (RCFT) for boundaries that preserve a larger chiral algebra that extends the Virasoro algebra \cite{Behrend:1999bn}, and in Liouville theory \cite{Teschner:2000md,Hosomichi:2001xc,Ponsot:2001ng}. Already for free bosons at $c=2$, there is no known classification of the general conformal boundaries (see  \cite{Affleck:2000ws} for some interesting examples\footnote{Here the bulk CFT is taken to be square torus with radius $R=\sqrt{2\A'}$ and no $B$-field which describes the quantum brownian motion \cite{Affleck:2000ws}.}).

Meanwhile, instead of directly solving the bootstrap equations for (B)CFTs, a different strategy, known as the functional bootstrap \cite{Rattazzi:2008pe}, has led to steady and promising progress in constraining 
the operator spectrum and OPE data of general (B)CFT. The bootstrap equations are in general linear in the conformal blocks whose coefficients encode the OPE data of the (B)CFT. The strategy is to rule out certain operator spectrum and OPE data, by studying a small subset of the bootstrap equations (usually those with positive expansions in the conformal blocks in a given OPE channel) and
identifying functionals acting on the conformal blocks with desired properties to eliminate this possibility. 
In 2d CFT, this strategy has been explored extensively both numerically and analytically to deduce universal bounds on the bulk CFT data by studying the torus partition function \cite{Hellerman:2009bu,Friedan:2013cba,Collier:2016cls,Afkhami-Jeddi:2019zci,Hartman:2019pcd,Afkhami-Jeddi:2020hde}, the sphere four-point functions \cite{Collier:2017shs,Lin:2015wcg,Lin:2016gcl,Besken:2021eps} and higher-genus partition functions \cite{Cho:2017fzo}. The numerical approach involves a scan over a vector space of functionals (that are linear in derivatives with respect to the moduli of the CFT observable) using semi-definite programming, while the analytical approach relies on a clever choice of certain analytic functionals \cite{Mazac:2016qev,Mazac:2018mdx,Mazac:2018ycv} which turn out to have curious connections to the sphere packing problem \cite{Hartman:2019pcd,Afkhami-Jeddi:2020hde}. 

For 2d BCFT, there is a distinguished universal quantity among all the bulk-boundary OPE data, given by the (vaccum) disk partition function with the conformal boundary \cite{Affleck:1991tk}. It is commonly denoted by $g$ and also known as the $g$-function or brane tension in string theory applications. Similar to the conformal $c$ anomaly in the bulk, the $g$-function is known to be a measure of the boundary degrees of freedom and the $g$-theorem says $g$ must decrease monotonically under boundary RG flows \cite{Affleck:1991tk,Friedan:2003yc}. A natural target for the bootstrap program is to look for nontrivial bounds on $g$, which are of interest in the context of both general 2d BCFTs and D-branes in string theory. In this paper, we pursue this goal by performing a multi-angle analysis of the bootstrap equation associated to the cylinder (annulus) partition of the BCFT, extending the previous works \cite{Friedan:2012jk,Friedan:2013bha}. We find that this deceivingly simple bootstrap equation, together with unitarity (positivity), produces a wealth of constraints on the BCFT data including $g$ and boundary operator spectrum as well as the bulk $c$ anomaly and operator spectrum.

 \subsection{Summary of the Main Results}
In the rest of the paper, we start by reviewing general boundary states and the $g$-function in Section~\ref{sec:review} along with a description of boundaries in RCFTs and sigma models which provide important reference points for the bootstrap analysis. We also discuss methods to construct new boundaries in general CFTs using (generalized) symmetries. 
In Section~\ref{sec:btssetup}, we explain the functional method to the annulus bootstrap equation and discuss both the linear programming approach and the analytic functional approach. The latter is made possible by an interesting relation between the annulus partition function with two (possibly distinct) boundary conditions and a four-point function of the $\mZ_2$ twist fields at the ends of two corresponding conformal defect lines. In either case, we will see that there exist nontrivial upper and lower bounds in the presence of a sufficiently large gap in the spectrum of boundary and bulk scalar operators, respectively. The bounds we have obtained from both the numerical and analytical approaches are presented in Section~\ref{sec:stableBraneBounds},\ref{sec:analyticFunctionals} and \ref{sec:universalBounds}, where we make comparisons to existing boundary states as well as new boundary states we construct using the methods explained in Section~\ref{sec:review}. Below we provide a brief summary.

\subsubsection*{Upper bound on the tension $g$ of stable elementary branes}
We define stable elementary branes as conformal boundaries that are not a direct sum of other conformal boundaries and furthermore for which the boundary operator spectrum has a gap $h_{\rm gap}\geq 1$ above the vacuum. In string theory, they correspond to elementary D-branes that are free from open string tachyons.  In Section~\ref{subsec:universalUpper}, we derive from the linear programming method a rigorous numerical upper bound on the tension $g$ of a stable elementary brane, as a function of the central charge $c$ that for sufficiently large $c$ grows monotonically. This bound exists for $1\leq c < 25$. We emphasize that our bound is universal and does not depend on details of the BCFT nor of the spectrum of the bulk CFT. It implies in particular that elementary bound states of D-branes in bosonic string theory of sufficiently large tension must decay due to open string tachyons.     

\subsubsection*{Lower bound on the tension $g$ in the presence of a bulk scalar gap}

A nontrivial lower bound on the tension $g$ of general conformal boundaries exists provided that the gap in the spectrum of bulk scalar primaries is sufficiently large \cite{Friedan:2012jk}. This bound is less universal than the previously described upper bound in the sense that it depends on some details of the spectrum of the bulk CFT. In Section~\ref{subsec:universalLower}, as a proof of principle we compute lower bounds on the $g$-function for general conformal boundary conditions as a function of the central charge with a particular choice of the bulk gap that is close to the maximal value allowed by modular invariance. This carves out a narrow window of allowed values of the tension for stable boundaries in CFTs with this large value of the gap. 

\subsubsection*{Two-sided bounds for boundaries of specific CFTs and the uniqueness of stable elementary branes}
Stronger bounds on the tension $g$ are obtained when additional information about the bulk CFT beyond the central charge $c$ is fed into the bootstrap program. In Section~\ref{sec:stableBraneBounds}, we carry out this analysis for the boundaries of a number of RCFTs (and toroidal CFTs) for which little is known beyond the rational branes, namely the special boundary conditions that preserve the maximal chiral algebra. We discover that for $G_1$ WZW CFTs with 
\ie
G=SU(2)\,,~SU(3)\,,~G_2\,,~F_4\,,~Spin(8)\,,~E_6\,,~E_7\,,~E_8\,,
\fe
the stable elementary branes have the same $g$ and boundary spectrum as the identity rational brane, providing strong evidence that all stable elementary branes in these $G_1$ WZW CFTs are related by the ${G_L\times G_R\over  Z(G)} \rtimes {\rm Out}(G)$ global symmetry.\footnote{Here $G_L$ and $G_R$ are the left and right moving $G$-symmetries and their diagonal center $Z(G)$ in the quotient acts trivially. ${\rm Out}(G)$ is the outer-automorphism group of $G$ which acts as $\mZ_2$ charge conjugation for $G=SU(N)$ with $N>2$ and as $S_3$ triality for $G=Spin(8)$. The exceptional groups have trivial ${\rm Out}(G)$ except for $G=E_6$ which has a ${\rm Out}(E_6)=\mZ_2$ symmetry.}

Furthermore, for CFTs with a conformal manifold, we also explore the dependence of the bounds on $g$ on the bulk marginal couplings, by tuning the bulk scalar gap $\Delta_{\rm gap}$. Explicit results are presented for $c=1$, $c=2$, and $c=8$ in Section~\ref{sec:stableBraneBounds}, which are in agreement with known boundary conditions. The latter two cases also provide guidance on finding new conformal boundary conditions in the $c=2$ and $c=8$ compact boson theories (and their discrete orbifolds).

\subsubsection*{Analytic bounds and optimal solutions}
In Section~\ref{subsec:analyticGeneralc}, we use the analytic functional introduced in Sections~\ref{subsec:analyticFunctional} and \ref{subsec:ZfromG} to derive an exact upper bound on $g$ when the boundary gap satisfies $h_{\rm gap}\geq {c+8\over 16}$, and 
an exact lower bound on $g$ when the bulk scalar gap satisfies $\Delta_{\rm gap}\geq {c+8\over 8}$. As we show in Section~\ref{sec:analyticFunctionals}, both bounds are saturated at $c=8$ by the rational branes of the $(E_8)_1$ CFT, and at $c=24$ by the identity brane of the non-holomorphic $|{\rm  Monster}|^2$ CFT. In particular, the corresponding optimal solutions to the bootstrap equation are uniquely determined by the analytic functional and coincide with the annulus partition functions in the $(E_8)_1$ and the $|{\rm  Monster}|^2$ CFT. We thus establish the following two theorems. 

\begin{theorem}
All branes in the $(E_8)_1$ CFT satisfy $g\geq 1$ and $h_{\rm gap}\leq 1$. Moreover the following are equivalent
\begin{enumerate}
    \item The brane tension saturates the lower bound, $g=1$
    \item The boundary gap saturates the upper bound $h_{\rm gap}=1$,
    \item The cylinder partition function with the same boundary conditions is $j(\tau)^{1/3}$.
\end{enumerate}
\label{E8thm}
\end{theorem}

\begin{theorem}
All branes in the $|{\rm Monster}|^2$ CFT satisfy $g\geq 1$ and $h_{\rm gap}\leq 2$. Moreover the following are equivalent
\begin{enumerate}
    \item The brane tension saturates the lower bound $g=1$,
    \item The boundary gap saturates the upper bound $h_{\rm gap}=2$,
    \item The cylinder partition function with the same boundary conditions is $j(\tau)-744$.
\end{enumerate}
\label{monsterthm}
\end{theorem}

Here $j(\tau)$ is the modular $j$-invariant. 
This proves analytically that all branes in the $(E_8)_1$ CFT have tension $g\geq 1$ while the stable elementary branes must have $g=1$, confirming the numerical bounds in Section~\ref{subsec:c8}. It also gives an immediate rigorous proof that all branes in the $|{\rm  Monster}|^2$ CFT must have tension $g\geq 1$, extending the numerical results of \cite{Friedan:2013bha}. Along the way, we also identify new boundary states for the $|{\rm Monster}|^2$ CFT beyond those in \cite{Craps:2002rw}. 
 
\subsubsection*{A certainty relation between bulk and boundary gaps}
So far we have been talking about upper and lower bounds on the tension $g$, by imposing conditions on the gaps $\{\Delta_{\rm gap},h_{\rm gap}\}$ in the bulk scalar and boundary operator spectrum. Conversely, nontrivial constraints on $\{\Delta_{\rm gap},h_{\rm gap}\}$ can be deduced by obvious consistency conditions on the $g$ bounds, namely the upper bound needs to be above the lower bound. For example, we find from both numeric and analytic bootstrap that for $c>24$, either $\Delta_{\rm gap}<{c+8\over 8}$ or  $h_{\rm gap}<{c+8\over 16}$. More generally, as explained in Footnote~\ref{longFootnote}, whenever the assumptions on the value of the boundary gap used to compute the upper bound on $g$ and that of the bulk gap used to compute the lower bound are  \textit{identical}  $h_{\rm gap} = {\Delta_{\rm gap}\over 2}$, then the resulting upper ($g_+$) and lower ($g_-$) bounds on $g$ are inversely related, $g_+g_- = 1$. Mutual compatibility of these assumptions then requires $g_+ \geq 1$.

\subsubsection*{Boundary entropy bounds at large $c$ and the end-of-the-world (ETW) branes in pure AdS$_3$ gravity}
The bootstrap bounds on the boundary entropy for CFTs with large central charge $c$ have implications for the end-of-the-world  (ETW) branes in the holographic dual quantum gravity on AdS$_3$ \cite{Karch:2000gx,Takayanagi:2011zk,Fujita:2011fp,Nozaki:2012qd}. Motivated by a natural extension of the pure gravity on AdS$_3$ that incorporates ETW branes, we illustrate this point by studying the large $c$ limit of the upper and lower bounds on $g$ with a gap assumption $h_{\rm gap} = {\Delta_{\rm gap}\over 2}={c\over 24}$ that is saturated by BTZ black holes (in the presence of an ETW brane). We find that the boundary entropy $\log g$ is bounded from above and below linearly in $c$, which limits the validity of the semi-classical ETW solutions where the entropy can be tuned arbitarily.

\section{Review of Boundary States and D-branes}
\label{sec:review}

\subsection{Conformal Boundaries}

To formulate 2d CFTs on manifolds with boundaries requires specifying the boundary condition on each boundary component. In the context of string theory, this amounts to understanding how the closed string worldsheet ends on D-branes. The simplest setup involves the CFT on the upper half-space $\mH$ with complex coordinate $z$ and a boundary at ${\Im}\,z=0$. By locality properties of 2d CFT with boundaries, it suffices to understand the boundary conditions in this setup in order to determine general CFT observables on Riemann surfaces with boundaries. Here we are particularly interested in conformal boundary conditions for the CFT, which
requires the bulk stress tensor to remain traceless on $\mH$ and satisfy the following ``gluing condition'' on the boundary,
\ie
\left.T(z)\right|_{\Im z=0}=\left.\bar T(\bar z)\right|_{\Im z=0}\,.
\label{Tglue}
\fe
This ensures that the diagonal Virasoro algebra of the 2d CFT is preserved by the boundary. In particular, this implies that the gravitational anomaly $c_L-c_R$ of the CFT must vanish \cite{Watanabethesis,Billo:2016cpy,Jensen:2017eof,Hellerman:2021fla}.\footnote{Gravitational anomalies produce a general obstruction to boundary conditions for QFTs \cite{Jensen:2017eof,Thorngren:2020yht,Hellerman:2021fla}.} In the following, we always have $c_L=c_R=c$ unless otherwise noted.

A useful way to characterize the conformal boundary condition is to map the half-space $\mH$ to a unit disk by the conformal transformation 
\ie
z\to {z-i\over z+i}\,,
\label{HtoD}
\fe
and then the corresponding boundary condition on the unit circle 
defines a state in the CFT Hilbert space $\cH_{S^1}$ accordingly to the usual radial quantization. We refer to such a state as a boundary state $|B\ra$. The condition \eqref{Tglue} after the transformation \eqref{HtoD} becomes
\ie
(L_n-\bar L_{-n})|B\ra=0\,,\quad {\rm for~all~} n\in \mZ\,.
\label{Confc}
\fe
when expanded into Fourier modes on the circle. A basis of solutions to \eqref{Confc} are coherent states in $\cH_{S^1}$ known as Ishibashi states $|\phi_i\rra$ which are in one-to-one correspondence with scalar Virasoro primaries $|\phi_i\ra \in \cH_{S^1}$ of weight $h=\bar h=h_i$ that satisfy $\la \phi_i |\phi_j\ra=\D_{ij} $. We will not need the explicit form of $|\phi_i\rra$ but the following property is useful. 

Let us consider adding a concentric hole of radius $e^{-\pi t}$ to the unit disk, and assigning two Ishibashi states $|\phi_i\rra$ and $|\phi_j\rra$ to the two boundary circles of the annulus. The corresponding CFT partition function on the annulus is equivalent by a conformal transformation  $z \to i \log z$ to the partition function on a cylinder with length $\pi t$ and circumference $2\pi$, with the following simple form
\ie
\lla \phi_i | e^{- \pi t H_{\rm cl}} |\phi_j \rra =\D_{ij} \chi_{\Delta_i\over 2}(t)\,,
\fe
where $\Delta_i$ is the dimension of the scalar Virasoro primary associated with the Ishibashi state $|\phi_i\rra$, $\chi_h$ denotes the Virasoro character and takes the following form for a non-degenerate Virasoro module of weight $h$,
\begin{equation}
    \chi_{h}(t) = {q^{h-{c-1\over 24}}\over \eta(it)},~ q \equiv e^{-2\pi t}
\end{equation} 
where $\eta$ is the Dedekind eta function, and the ``closed string'' Hamiltonian  reads 
\ie
H_{\rm cl}=L_0+\bar L_0-{c\over 12}\,.
\fe

The putative boundary state is in general a linear combination of the Ishibashi states,
\ie
|B\ra=\sum_i B_i  |\phi_i\rra \,,
\label{Bdecomp}
\fe
subject to two important constraints, the Cardy condition \cite{Cardy:1989ir,Cardy:2004hm} and the bulk-boundary bootstrap equation (or Cardy-Lewellen equation) \cite{Cardy:1991tv,Lewellen:1991tb}.

The Cardy condition comes from the two different ways to decompose the cylinder partition function with any two boundary states $|B_\A \ra$ and $|B_\B\ra$ on the two ends. In the ``closed string'' channel, the partition function computes the overlap of the two boundary states,
\ie
Z_{\A\B}(t)\equiv \la B^\A| e^{- \pi t H_{\rm cl}} |B_\B\ra = \sum_i (B^\A_i)^* B^\B_i \chi_{\Delta_i\over 2}(t)\,,
\label{cc}
\fe
while in the ``open string'' channel, it can be seen as the thermal partition function for the Hilbert space $\cH_{\A\B}$ on an interval of length $\pi$ with boundary conditions $B^\A$ and $B^\B$ at the two ends, and a thermal circle of size $2\pi \over t$,
\ie
Z_{\A\B}(t)=\tr_{\cH_{\A\B}} e^{-{2\pi\over t} H_{\rm op}}=\sum_j n^j_{\A\B}\chi_{h_j}(t^{-1})\,.
\label{oc}
\fe
where 
$n^j_{\A\B}\in \mZ_+$ counts the  primaries in $\cH_{\A\B}$ with conformal weight $h_j$. Here the ``open string'' Hamiltonian is
\ie
H_{\rm op}=L_0-{c\over 24}\,.
\fe
The Cardy condition comes from the equality between \eqref{cc} and \eqref{oc}
\ie
\sum_i (B^\A_i)^* B^\B_i \chi_{\Delta_i\over 2}(t)= \sum_j n^j_{\A\B}\chi_{h_j}(t^{-1})\,,
\label{Cardy}
\fe
which is a natural extension of the usual modular crossing equation associated with modular invariance of the torus partition function, and puts strong constraints on the BCFT data.

Clearly a positive integer linear combination of admissible boundary states remains a valid boundary state. The elementary (fundamental) boundaries $|B_\A\ra $ are defined by requiring $n^0_{\A\B}=\D_{\A\B}$ so that there is a unique dimension zero operator (i.e. the identity) on each elementary boundary $|B_\A\ra$ and no dimension zero boundary changing operators for $\A \neq \B$.

\begin{figure}[h]
    \centering
    \includegraphics[width=.9\textwidth]{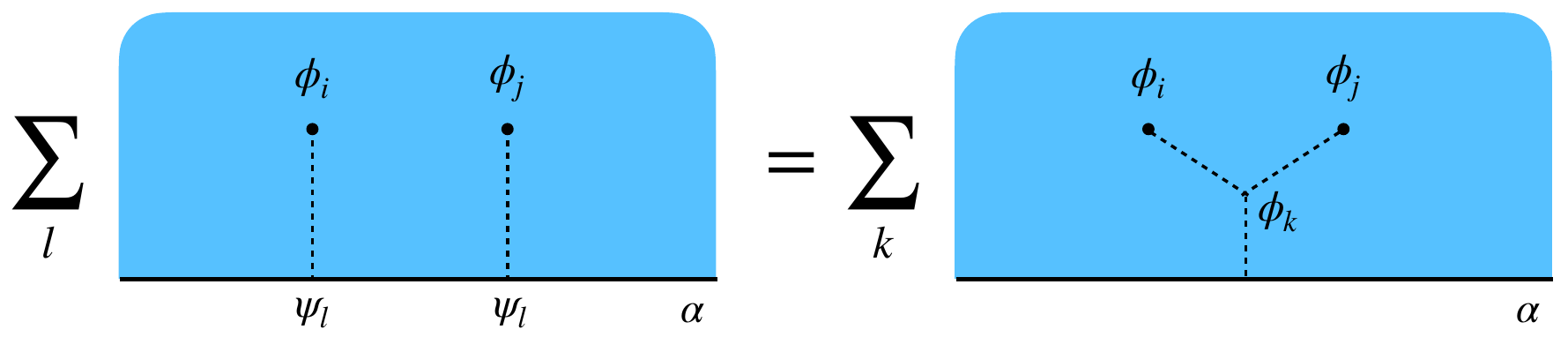}
    \caption{The bootstrap constraint \eqref{eq:CardyLewellen} arising from equating the boundary and bulk OPE of the bulk two-point function $\la \phi_i(z_1)\phi_j(z_2)\ra_\A$.}
    \label{fig:CardyLewellen}
\end{figure}

The constraint from the bulk-boundary bootstrap equation is most easily stated for an elementary boundary $|B_\A\ra$. The correlation function $\la \phi_i(z_1)\phi_j(z_2)\ra_\A$ of two bulk primaries in the presence of the boundary has two OPE channels: the the boundary and bulk OPE channels as in Figure~\ref{fig:CardyLewellen}. The equivalence between the two leads to the following bootstrap equation  \cite{Cardy:1991tv}
\begin{equation}
    \sum\limits_{l}\widetilde{C}^{\alpha}_{il}\widetilde{C}^{\alpha}_{jl}\,\cG^{ii,jj}_{l}(\eta) = \sum\limits_{k}C_{ijk}B^{\alpha}_{k}\,\cG^{ij,ij}_{k}(1-\eta)\,.
\label{eq:CardyLewellen}
\end{equation}
The sum over $l$ on the LHS runs over boundary primary operators $\psi_l$ appearing in the bulk-boundary OPEs of $\phi_i$ and $\phi_j$, and $\widetilde{C}^{\alpha}_{il}$ are the corresponding OPE coefficients. The sum over $k$ on the RHS runs over bulk primaries appearing in the $\phi_i\times\phi_j$ OPE and $C_{ijk}$ are the bulk OPE coefficients. $\cG$ are the conformal blocks.

It is a nontrivial fact that, given OPE data that solve the bulk bootstrap equations (namely sphere four-point crossing and torus one-point modular covariance),
 the Cardy condition \eqref{Cardy} and the bulk-boundary bootstrap equation \eqref{eq:CardyLewellen} are the necessary and sufficient conditions to define correlation functions of local operators on general Riemann surfaces with boundaries that are consistent with cutting and gluing \cite{Lewellen:1991tb,Pradisi:1996yd}.\footnote{For more general correlation functions that include boundary changing operators, the necessary and sufficient constraints are also known and consist of four ``sewing constraints" \cite{Lewellen:1991tb}. In particular, the bulk-boundary bootstrap equation can now involve in addition a boundary changing operator. } 
It is  still to be understood whether the complete set of solutions to the boundary states is unique in general, but so far no counter-example is known.\footnote{See for example \cite{Behrend:1999bn} for a proof of uniqueness of the complete set of conformal boundary states for Virasoro minimal models and rational boundary states for $SU(2)_k$ and $SU(N)_1$ WZW models.} Note that
the equations \eqref{Cardy} and \eqref{eq:CardyLewellen} are invariant under the overall rescaling $B^\A_i\to \lambda B^\A_i$ with $|\lambda|=1$ but  we do not regard this common phase ambiguity as defining physically distinct sets of boundaries. In the following, we fix this ambiguity by requiring that $B^\A_0>0$ which is possible thanks to the reflection-positivity of the disk partition function with boundary state $|B_\A\ra$.

In the later sections of the paper, we will focus on the constraints coming from the Cardy condition on boundary states in general 2d CFTs, and leave the investigation of more general bulk-boundary bootstrap constraints to future.

\subsection{Boundaries in Rational CFTs}
\label{sec:bofRCFT}
It is a very difficult problem to solve the Cardy and bulk-boundary bootstrap constraints for a general 2d CFT. One exception is when the CFT is rational: in this case there is a finite number of conformal blocks with respect to a chiral algebra $\cV$ (and its anti-chiral counterpart $\bar \cV \cong \cV$) that extends the Virasoro algebra. In particular, there is a finite set $\cI$ of $\cV\times \bar\cV$ primaries $\varphi_a$ with $a\in \cI$  and torus characters $\chi^\cV_a(t)$.
In this case, a distinguished set of boundary states are those that preserve a diagonal $\cV$ symmetry,
\ie
(W_n-(-1)^{h_W}\rho(\bar W_{-n}))|B\ra=0\,,\quad {\rm for~all~} n\in \mZ\,,
\fe
for each strong generator $W(z)$ of weight $h_W$ (and $\bar W(\bar z)$) of the chiral algebra $\cV$ (resp. $\bar \cV$), where $\rho$ is an automorphism of $\cV$ that leaves the Virasoro subalgebra invariant. Correspondingly, the general solution is now a combination of the Ishibashi states defined for the larger algebra $\cV\otimes \bar \cV$ which are in one-to-one correspondence with the $\cV\otimes \bar\cV$ primaries,
\ie\label{eq:IshibashiDecomposition}
|B\ra=\sum_{a\in \cI} B_a |\varphi_a \rra \,.
\fe
In this case the bootstrap constraints for boundaries simplify dramatically. In particular, the Cardy-Lewellen condition \eqref{eq:CardyLewellen} can now be projected onto the boundary-channel identity, leading to
\ie
B_a^\A B_b^\A=\sum_k C_{abc}(F^{aab}_b)_{c1} B^\A_0 B^\A_k 
\label{bbbs}
\fe
where $C_{abc}$ are the bulk three-point functions coefficients and $(F^{abc}_d)_{ef}$ is the F-symbol.

When the CFT is defined by a diagonal (charge-conjugation) modular invariant partition function, such as for the $c<1$ Virasoro minimal models, the elementary boundaries $|B_\A\rra$ that solve \eqref{Cardy} and \eqref{bbbs} are simply determined by the modular $S$-matrix, 
\ie
B^{\A}_a={S_{a \A  }\over \sqrt{S_{0 a }}}\,,
\label{diagonalB}
\fe
where $\A$ and $a$ run over the same set $\cI$, and the boundary states are in one-to-one correspondence with the primaries. In particular the bootstrap equation \eqref{bbbs} reduces to the Verlinde algebra. The open string spectrum that solves   the Cardy condition \eqref{Cardy} is determined by the RCFT fusion coefficients $N_{abc}$,
\ie
n^a_{\A\B}= \sum_{b\in \cI}(B^\A_b)^* B^\B_b S_{ba}=N_{\A\B a}\,,
\label{RCFTcardy}
\fe
where the second equality follows from \eqref{diagonalB} and the Verlinde formula
\ie
N_{abc}=\sum_{d\in \cI} {S_{ad} S_{bd} S_{cd}^*\over S_{0d}}\,.
\label{VerF}
\fe

General RCFTs have a two-layer structure consisting of (anti)holomorphic data (e.g. conformal blocks)  from the chiral algebra $\cV$ and topological data (e.g. $S,T$ matrices and $F$-symbols)  from an underlying modular tensor category (MTC) $\cC={\rm Rep}(\cV)$ endowed with a symmetric Frobenius algebra object $A\in\cC$ that specifies the modular invariant (the diagonal one corresponds to the identity object).\footnote{Physically, objects in the category $\cC$ correspond to topological defect lines (TDLs) or Verlinde lines generating a non-invertible global symmetry of the diagonal RCFT. The algebra object $A$ specifies a subset of the TDLs that are non-anomalous and can be gauged (in a generalized sense for the non-invertible TDLs). The RCFT obtained after gauging $A$ is the theory with the new (non-diagonal) modular invariant specified by $A$. \label{foot:algebra}} It turns out that the bootstrap constraints for boundaries lead to interesting algebra structures that only depends on the topological MTC data of the RCFT and  can be solved entirely in terms of such data.

The Cardy condition for the RCFT (first equality of \eqref{RCFTcardy}) demands $n^a_{\A\B}$ is a nonnegative integer
matrix (NIM) of size $|\cI|$, Furthermore, as a consequence of the Verlinde formula \eqref{VerF}, they satisfy the fusion rule (Verlinde algebra) of the RCFT, 
\ie 
n^a \cdot  n^b=\sum_{c\in \cI} N_{abc} n^c\,.
\fe
The matrices $n^a$ are thus called a ``NIM-rep'' of the RCFT fusion ring (for the diagonal modular invariant the NIM-rep coincides with the fusion coefficients). On the other hand, the bulk-boundary bootstrap equation becomes \cite{Behrend:1999bn},
\ie
B^\A_a B^\A_b =\sum_{c\in \cI} M_{abc} B_c^\A B_0^\A\,,
\fe
where $M_{abc}$ is determined by the MTC. The structure constants $M_{abc}$ defines a semi-simple commutative associative algebra that generalizes the Verlinde algebra ($M_{abc}=N_{abc}$ for the diagonal modular invariant) and is known as the classifying algebra for boundary conditions \cite{Fuchs:1997kt,Fuchs:2009zr}.\footnote{In contrast to the Verlinde algebra, the structure constants for a general classifying algebra are neither positive nor integral.} In particular, the elementary rational boundaries correspond to one-dimensional representation   ($B^\A_a/B^\A_0$) of the classifying algebra. 

The solutions to these constraints are in principle known in terms of the topological data from the MTC $\cC$,\footnote{More precisely, the rational boundary states (preserving the chiral algebra $\cV$) correspond to modules over $\cC={\rm Rep}(\cV)$ that are $A$-modules in $\cC$ at the same time. Together with boundary changing operators, they furnish a module category ${\rm Mod}_\cC(A)$ over $\cC$ \cite{modulecat}. The $\cC$-module structure is natural because in the RCFT $\cC$ can be identified with a set of topological defect lines which can fuse with a boundary to produce other boundary conditions preserving the same chiral algebra. The $A$-module structure on the other hand is required since we are gauging $A$ (see Footnote~\ref{foot:algebra}) and thus identifying boundary states related by fusion with $A$.
 } but the latter is not known for general chiral algebras. Thus, in line with the bootstrap philosophy, it maybe useful to first attack the simpler constraints coming from the NIM-rep.  
 However only sporadic results of classification are known beyond the $SU(2)_k$ WZW models.\footnote{For $SU(2)_k$ WZW models, the classification of NIM-reps reproduces all rational boundary states for the CFTs with modular invariants of type A-D-E. However, it also produces a family of spurious solutions for $k=2n-1$ corresponding to the tadpole graphs $T_n=A_{2n}/\mZ_2$
  \cite{DiFrancesco:1989ha,Behrend:1999bn}. }

\subsection{Brane Tension and Boundary Entropy}
Much like the conformal central charge $c$ of 2d CFT, there is a distinguished universal quantity $g_\A$ for any conformal boundary $|B_\A\ra$, given by the coefficient of the identity Ishibashi state in \eqref{Bdecomp},
\ie
|B_\A\ra=g_\A|0\rra + \sum_{i:\,\Delta_i>0} B^\A_i |\phi_i\rra\,.
\fe
The quantity $g_\A$ is often referred to as the $g$-function or brane tension\footnote{More precisely, when the CFT is a part of the worldsheet string theory and the conformal boundary describes a D-brane. In this case the D-brane effective tension and the $g$-function are related by
\ie
T\sim {1\over g_s} g_\A,
\fe
where $g_s$ is the string coupling. See Section~\ref{sigma} for more details.
} for the boundary state $|B_\A\ra$. As explained before, $g_\A>0$ by unitarity and $s_\A\equiv 2\log g_\A$ defines the boundary entropy.\footnote{In what follows we will often refer to $g$ itself as the boundary entropy as a sort of shorthand.} As alluded to, the $g$-function has several nice properties which we elaborate below.

The $g$-function provides a measure of the boundary degrees of freedom. Following \cite{Harvey:1999gq}, we can define the regularized dimension of the open string Hilbert space $\cH_{\A\A}$ from the limit $t\to \infty$ (i.e. $\tilde q = e^{-{2\pi \over t}}\to 1$) of the cylinder partition function,
\ie
\dim_{\rm reg} \cH_{\A \A}\equiv \lim_{t\to \infty} e^{-{c\over 12}\pi t}\sum_i n^i_{\A\A}\chi_i(t^{-1})=g_\A^2
\fe
where the second equality follows from the Cardy condition \eqref{Cardy}. 

The $g$-function also determines the subleading piece in the high-temperature limit of the thermal entropy of a 2d CFT on an interval of length $L$ with inverse temperature $\B$ \cite{Affleck:1991tk}. In general, with the same but not necessarily conformal boundary conditions at the two ends of the interval, the entropy is determined by the cylinder partition function $Z(\B)$ and takes following form
\ie
S(\B)\equiv  (1-\B \pa_\B)\log Z(\B)={\pi\over 3} c {L\over \B} +s(\B)+\dots 
\fe
where we have omitted  terms that vanish in the high temperature limit $L/\B \to \infty$. Here $s(\B)$ is the Affleck-Ludwig boundary entropy which may depend on dimensional parameters of the non-conformal boundary. For conformal boundaries, $s(\B)$ is a constant. Indeed the cylinder partition function with conformal boundaries $|B_\A\ra$ is given by \eqref{oc} with $t={2L\over \B}$,
\ie
\log Z_{\A\A}(\B)={\pi\over 6} c {L\over \B} +2\log g_\A +\dots\,,
\fe
so that $s(\B)=2\log g_\A$ as promised. Furthermore, the boundary entropy $s(\B)$, well-defined away from the boundary conformal fixed points, was proven to be monotonic under boundary RG flows \cite{Affleck:1992ng,Friedan:2003yc},\footnote{The $g$-function (more precisely $\log g$) also contributes to a universal constant term in the ground state entanglement entropy of an interval $x\in [0,r)$ in the large $r$ limit with the conformal boundary at $x=0$  \cite{Calabrese:2009qy}. Relatedly,   an entropic proof of the $g$-theorem was given in \cite{Casini:2016fgb}.} providing a boundary analog of Zamolochikov's $c$-function in the case of bulk RG flows. As a corollary, the $g$-function is independent of boundary marginal couplings.\footnote{The $g$-function can and does depend on bulk marginal couplings in general. This is a common feature of the entropy associated with odd dimensional conformal defects.} 
 
Finally, the $g$-function for conformal boundaries in RCFTs admit simple expressions. For example, if the RCFT is defined by a diagonal modular invariant with respect to a chiral algebra $\cV$, then the elementary Cardy boundaries $|B_\A\ra$ preserving $\cV$ have tension
\ie
\sqrt{S_{00}}\leq g_\A={S_{0\A}\over \sqrt{S_{00}}}\leq {1\over \sqrt{S_{00}}}\,,
\label{gBoundRCFT}
\fe
where the second inequality comes from the unitarity of the S-matrix and is saturated only when the underlying MTC is trivial (e.g. the $(E_8)_1$ CFT or the non-holomorphic $|{\rm Monster}|^2$ CFT) and the first inequality is a consequence of $S_{0\A}\geq S_{00}$ in MTCs and saturation happens when $\A$ labels a simple current (abelian anyon).\footnote{There are many ways to see $S_{0\A}\geq S_{00}$ in  RCFT. For example ${S_{0\A}\over S_{00}}=\la \cL_\A \ra$ is the quantum dimension of the Verlinde line $\cL_\A$ which satisfies $\la \cL_\A \ra \geq 1$ using fusion (see around \eqref{gtopinterface}). Alternatively, the fusion matrix $N_{\A}$ (from $a=\A$ in \eqref{VerF}) is a non-negative integral irreducible square matrix and the inequality follows from the Perron-Frobenius theorem (see for example \cite{Gannon:2001uy}). }

The same inequality \eqref{gBoundRCFT} also holds for general RCFTs.
This is because, for a complete set of elementary Cardy branes, the following matrix
\ie
R_{a\A}=B^\A_a \sqrt{S_{0a}}
\fe
is unitary as a consequence of the Cardy condition \eqref{RCFTcardy} with $n^0_{\A\B}=\D_{\A\B}$.

\subsection{Folding Trick and Conformal Interfaces}
\label{sec:interface}
Conformal boundaries are closely related to conformal interfaces in CFTs by the folding trick. Namely a conformal interface $\cI_{\cT_1|\cT_2}$  between two (possibly identical) CFTs $\cT_1$ and $\cT_2$ corresponds to a conformal boundary in the product CFT $\cT_1 \times \overline{\cT_2}$ where the second factor involves an orientation reversal. Such an interface is admissible only if the neighboring CFTs have identical gravitational anomalies. 
There is a natural notion of interface $g$-function or interface entropy as defined by those of the boundary state obtained after folding, equivalently given by the expectation value in the ground states of the neighboring CFTs,
\ie
g_{\cI_{\cT_1|\cT_2}}=\la \Omega_1 | \cI_{\cT_1|\cT_2} |\Omega_2\ra\,.
\label{interfaceg}
\fe
Correspondingly, the interface $g$-function obeys the $g$-theorem under interface RG flows.\footnote{The $g$-theorem in 2d CFTs have been recently generalized to line defects in higher dimensional CFTs in \cite{Cuomo:2021rkm}. } 
Moreover, the bootstrap results we obtain in the latter sections for the $g$-function of conformal boundaries will automatically apply to conformal interfaces when intepreted accordingly.

Interfaces are categorized by the fraction of energy that is reflected and transmitted at their insertions \cite{Bachas:2001vj,Quella:2006de,Meineri:2019ycm}. At the extremes are interfaces that are totally reflective or transmissive. 
The former are also known as factorized interfaces, and are in one-to-one correspondence with a pair of conformal boundaries $B_\A^1$ and $B_\B^2$ in the $\cT_1$ and $\cT_2$ CFTs respectively,
\ie
\cI_{\cT_1|\cT_2}=|\cB^1_\A\ra \la \cB^2_\B|\,.
\fe
Consequently the factorized interfaces lead to decoupled observables and have a $g$-function given by the product of the corresponding values for the two boundary states.

On the other hand, the purely transmissive interfaces require the stress energy tensor to be continous across their insertions, and thus correspond to topological interfaces. Such interfaces are admissible only if the neighoring CFTs $\cT_1$ and $\cT_2$ have identical central charges $c_L$ and $c_R$. For example, for a conformal manifold $\cM$ of CFTs, they are topological interfaces connecting CFTs at different points on $\cM$ \cite{Bachas:2001vj,Fuchs:2007tx,Bachas:2012bj,Bachas:2013nxa,Becker:2017zai}. Alternatively, if the CFT $\cT$ admits a non-anomalous symmetry $G$, there is natural topological interface between $\cT$ and its orbifold $\cT/G$ (see e.g \cite{Fuchs:2007tx,Thorngren:2021yso}).\footnote{Conventionally $G$ denotes an invertible non-anomalous symmetry associated with a group. In terms of the fusion category symmetry that generalizes invertible symmetries \cite{Bhardwaj:2017xup,Chang:2018iay,Thorngren:2019iar,Thorngren:2021yso}, $G$ is replaced by a symmetric Frobenius algebra object $A$ in the fusion category. The $g$-function of this interface is determined by the quantum dimension of the algebra $\la A\ra $ as $g_\cI=\sqrt{\la A\ra}$. For invertible symmetry $G$, $\la A\ra=|G|$ is simply the order of the group.}
A key feature of the topological interfaces is that they have well-defined fusion products,
\ie
\cI_{\cT_1|\cT_2}\, \cI_{\cT_2|\cT_3}=\sum_a n_a\cI^a_{\cI_1|\cI_3}\,
\label{interfacefusion}
\fe
with $n_a\in \mZ_+$. Evaluating the expectation value of both sides of fusion equation in the CFT ground states and using cluster decomposition, we conclude the interface $g$-functions \eqref{interfaceg} must obey the abelianized fusion rule. 

Topological interfaces between identical (isomorphic) CFTs define topological defect lines (TDL) \cite{Petkova:2000ip,Bachas:2004sy,Frohlich:2006ch,Davydov:2010rm}. They give rise to a vast generalization of symmetries associated with groups in CFTs that is recently axiomatized in \cite{Bhardwaj:2017xup,Chang:2018iay}. In particular they obey fusion rules as in \eqref{interfacefusion} and are mathematically described by fusion categories. The interface (defect) $g$-function of a TDL $\cL$ coincides with its quantum dimension, namely the expectation value of $\cL$ in the CFT ground state,
\ie
g_\cL=\la \cL\ra\,.
\fe
The group-like symmetries correspond to the case $g_\cL=1$ and more general (non-invertible) TDLs have $g_\cL > 1$. In fact this lower bound on $g$ holds for general topological interfaces (and the corresponding boundary states after folding),
\ie
g_{\cI^{\rm top}_{\cT_1|\cT_2}}\geq 1\,.
\label{gtopinterface}
\fe
This follows from inspecting the fusion of a topological interface and its orientation reversal which always contains the identity TDL which has $g=1$.

\subsection{Boundaries from Generalized Symmetries}
\label{sec:bdyfromTDL}

It should come as no surprise that symmetries are extremely powerful in constraining not only the local operator content in a CFT but also its defects.

\subsubsection*{Boundaries from Fusion with Topological Defects}

In particular, boundaries of 2d CFTs form modules of the  generalized fusion category symmetries in the bulk \cite{Graham:2003nc,Kojita:2016jwe,Konechny:2019wff}. This is because when a TDL
$\cL_a$ acts  by fusion on a given boundary  $|B_\A\ra$, it produces another boundary $| \cL_a  B_\A\ra$ (possibly reducible) that obey all the bootstrap axioms thanks to the topological nature and locality properties of the TDLs. In particular, the resulting boundary state $| \cL_a   B_\A\ra$ generally has the following decomposition into elementary branes $|B_\B\ra$,
 \ie
| \cL_a  B_\A\ra \equiv  \cL_a |B_\A\ra =\sum_\B n^a_{\A\B}|B_\B\ra\,,
\label{TDLfusionbdy}
 \fe
with $n^a_{\A\B} \in \mZ_+$.

To see this, let us recall the defect Hilbert space $\cH^{\cL_a}$ associated to a TDL $\cL_a$ intersecting the spatial $S^1$ transversely at one point. Via the operator-state correspondence, states in $\cH^{\cL_a}$ are in one-to-one correspondence with operators that live at the end of $\cL_a$. The defect Hilbert space also generalizes to the open-string channel. Here we have an interval bounded by two conformal boundaries $|B_\A\ra$ and $|B_\B\ra$ and the same TDL $\cL_a$ that passes through a point on the interval. This Hilbert space $\cH^{\cL_a}_{\A\B}$ now contain states that represent operators at the trivalent junction between the boundaries and the TDL. In particular, the weight zero operators correspond to topological junctions whose multiplicity is precisely $n^a_{\A\B}$ in \eqref{TDLfusionbdy}. Such topological junctions endow the set of boundary states $\{|B_\A\ra\}$ the structure of a module category over the fusion category generated by the TDLs $\cL_a$.

By considering consecutive fusions of TDLs with a boundary state, we see $n^a$ must furnish a NIM-rep of the fusion algebra of the TDLs,
\ie 
\cL_a \cL_b=\sum_c\cN_{ab}{}^c \cL_c\,. 
\fe
The well-studied NIM-rep condition for RCFTs reviewed in Section~\ref{sec:bofRCFT} corresponds to the special case when the TDLs are Verlinde lines (i.e. objects in the MTC $\cC$) and the boundary states preserve the chiral algebra $\cV$.

In practice (and as we will see in examples), the TDLs provide a useful way to generate consistent boundary states in a CFT given a subset of them. In particular, the $g$-function of the resulting boundary state from fusion is given by
\ie
g_{\cL_a B_\A}=\la \cL_a \ra g_\A \,.
\label{TDLbg}
\fe
The cylinder partition function is also determined in terms of the interval defect Hilbert spaces for the TDLs as
\ie
\la \cL_a B_\A|e^{-\pi t H_{\rm cl}} |\cL_a B_\A\ra=\sum_{\cL_b\in \cL_a \overline{\cL}_a}
\la   B_\A|e^{-\pi t H_{\rm cl}} |\cL_b B_\A\ra
=\sum_{\cL_b\in \cL_a \overline{\cL}_a}
\tr_{\cH^{\cL_b}_{\A\A}} e^{-{2\pi \over t} H_{\rm op}}\,.
\label{PFfromTDL}
\fe
Note that the interval Hilbert spaces $\cH^{\cL_b}_{\A\A}$ and $\cH_{\A\A'}$ with $|B_{\A'}\ra \equiv |\cL_b B_\A\ra$ are equivalent. 

A simple consequence of the above is that boundary states related by fusion with invertible TDLs all have the same $g$-function and cylinder partition function.

One type of TDLs that appear in many CFTs (and we will see examples later) is the duality defect $\cN$ which explains the self-duality of the theory under gauging an abelian discrete group $G$. The fusion rules are
\ie
\cN^2=\sum_{g\in G}g\,,\quad g\cN=\cN g=\cN\,,
\fe
and the quantum dimension $\la \cN\ra=\sqrt{|G|}$.
The corresponding generalized symmetry is described by a Tambara-Yamagami fusion category \cite{TAMBARA1998692}. In Section~\ref{sec:stableBraneBounds}, we will use these duality defects to illustrate the power of TDLs in generating new boundary states. The simple relations for the $g$-functions \eqref{TDLbg} and the cylinder partition functions \eqref{PFfromTDL} will come in handy.

\subsubsection*{Boundaries from Discrete Gauging}

An anomaly-free symmetry $G$ of the CFT $\cT$ can be gauged to produce another CFT $\cT/G$. Correspondingly, the boundary states of the gauged theory can be understood in terms of those before gauging $G$. This relation is particularly simple when $G$ is a discrete group and the resulting theory is a usual $G$-orbifold.\footnote{Gauging a non-anomalous continous symmetry $G$ corresponds to the coset construction in CFT. In this case, the boundary states in the coset CFT can also be identified from those in the parent theory by standard coset methods    \cite{Stanciu:1997sk,Maldacena:2001ky,Gawedzki:2001ye,Elitzur:2001qd,Fredenhagen:2001kw,Ishikawa:2001zu,Ishikawa:2002wx,Gaberdiel:2004za,Fuchs:2005gp}.} For the cases where $\cT$ is a free theory or an RCFT, the boundary states in the orbifold have been worked out
\cite{Fuchs:1997kt,Fuchs:1999zi,Birke:1999ik,Fuchs:1999xn,Billo:2000yb,Matsubara:2001hz,Cappelli:2002wq,Gaberdiel:2002jr,Yamaguchi:2003yq,Fredenhagen:2004xp,Gaberdiel:2004yn}. Below we present the general construction.

There is an obvious class of boundary states in $\cT/G$ that are given by $G$-invariant combinations of the boundaries $|B_\A\ra$ in $\cT$,
\ie
|(B_\A)_O\ra ={1\over \sqrt{|G|} }\sum_{g\in G} g|B_\A\ra\,,
\label{rbrane}
\fe
which are known as regular branes. The overall normalization is chosen such that the Cardy condition is obeyed in $\cT/G$ and that $|(B_\A)_O\ra$ is elementary for generic boundary moduli. 

The regular branes by themselves do not form a complete basis for boundaries in the orbifold theory $\cT/G$ if $G$ does not freely on the boundary states in $\cT$. In particular the regular brane $|(B_\A)_O\ra$ is no longer elementary when $|B_\A\ra$ is preserved by a subgroup of $G$, and so should decompose into other elementary branes, known as fractional branes. This is because there are twisted sector operators which correspond to extra Ishibashi states in the orbifold theory. They give rise to extra boundary states (fractional branes) of the form
\ie
|(B_\A,r_N)_O\ra= \sum_{h\in N} {\chi_{r_N}^N(h)\over |N|}
\sum_{g\in C_G(h)} {\sqrt{|G|}\over |C_G(h)|}g|B_\A\ra_h\,,
\label{fbrane}
\fe
for each boundary state $|B_\A\ra$ that is invariant under the subgroup $N\subset G$. Here $r_N$ denotes irreducible representations of $N$  and $\chi^N_{r_N}$ is the corresponding group character. The second sum in \eqref{fbrane} is over the centralizer of $h$ in $G$ denoted by $C_G(h)$ which implements the projection to $G$-invariant combinations of the (twisted) boundary states in the CFT $\cT$. The twisted boundary states $|B_\A\ra_h$ is a combination of the $h$-twisted Ishibashi states and obeys the generalized Cardy condition,
\ie
{}_{h} \la B_\A | e^{-\pi t H_{\rm cl}} |B_\B \ra_{h'}=\D_{h,h'} \tr_{\cH_{\A\B}} \left(h e^{-{2\pi \over t} H_{\rm op}} \right)\,,
\fe
where the RHS involves the symmetry defect $h$ acting on the interval Hilbert space $\cH_{\A\B}$.\footnote{Note that the symmetry defect $h$ ends topologically on the $h$-invariant boundaries.} 
In \eqref{fbrane}, we have identified the $h=1$ case with the untwisted boundary state $|B_\A\ra$. One can verify that the regular branes \eqref{rbrane} and the fractional branes \eqref{fbrane} solve the Cardy conditions in the orbifold $\cT/G$. Moreover, the regular branes decompose, when stabilized by $N\subset G$, as
\ie
|(B_\A)_O\ra=\sum_{r_N} \chi_{r_N}^N(1) |(B_\A,r_N)_O\ra\,,
\fe
where the sum is over irreducible representations of $N$ and the equality follows from the identity
\ie
\sum_{r_N} \chi_{r_N}^N(h_1)\chi_{r_N}^N(h_2^{-1})=|N|\D_{h_1,h_2}\,,
\fe
where $\D_{h_1,h_2}=1$ if $h_1$ and $h_2$ are conjugate and otherwise $\D_{h_1,h_2}=0$.

Another perspective on the relation between (twisted) boundary states in the CFT $\cT$ and its orbifold $\cT/G$ is provided by the topological interface $\cI_{\cT|\cT/G}$ introduced in 
Section~\ref{sec:interface}, which also makes manifest the reversibility of the orbifold procedure.\footnote{The orbifold $\cT/G$ has a dual symmetry which can be gauged to recover the original CFT $\cT$. For abelian $G$, this is known as the quantum $G$ symmetry of the orbifold, while for non-abelian $G$, the dual symmetry is a fusion category ${\rm Rep}(G)$ \cite{Bhardwaj:2017xup}.}
Fusion of the topological interface $\cI_{\cT|\cT/G}$ with $G$-twisted boundaries in $\cT$ produce boundary states in the orbifold $\cT/G$. Alternatively, fusing $\cI_{\cT|\cT/G}$ with boundaries in $\cT/G$ twisted by the dual symmetry recovers boundary states in $\cT$. The relations \eqref{rbrane} and \eqref{fbrane} encode the mapping between the (twisted) boundary states in one theory and its orbifold.

Symmetries in 2d CFTs are generally discribed by a fusion category $\cC$ (or a suitable extension thereof)  \cite{Bhardwaj:2017xup,Chang:2018iay}. The general discrete gauging is defined with respect to a symmetric Frobenius algebra object $A\in \cC$ \cite{Frohlich:2009gb,Carqueville:2012dk,Brunner:2013xna}, which can be thought of as a non-anomalous topological defect in the fusion category. The generalized gauging is implemented by decorating the spacetime Riemann surface with a mesh of $A$. For group-like symmetries, $\cC$ is the group algebra $\mC[G]$ and the usual $G$-orbifold corresponds to $A=\oplus_{g\in G}\, g$. It is natural to ponder the mapping between boundary states under the generalized gauging. For rational boundaries in RCFTs, this question is addressed in 
\cite{Fuchs:2000vg,Fuchs:2001qc}. An exposition for the most general case would be interesting but out of the scope of this paper.

\subsection{D-branes in Sigma Models}
\label{sec:sigma}
Let us now consider the case when the CFT has a sigma model description on a $d$-dimensional target space manifold $\cM$, defined by the following action,
\ie
S_{\sigma}={1\over 4\pi \alpha'} \int_\Sigma d^2\sigma  \sqrt{g} (g^{ab}G_{\m\n} +   i\ep^{ab} B_{\m\n})\pa_a X^\m  \pa_b X^\n 
\label{sigma}
\fe
where $X^\m$ denotes target space coordinates with $\m=1,2,\dots,d$, $G_{\m\n}$ and $B_{\m\n}$ are the metric and the $B$-field on $\cM$. The worldsheet $\Sigma$ has coordinates $\sigma^a$ with $a=1,2$ and $g_{ab}$ is the worldsheet metric and $\ep_{ab}$ is the worldsheet two-form. For example, $G$-WZW models can be described by such a sigma model where $\cM$ is the group manifold $G$ with a quantized volume and $B$-field (the torodial CFTs correspond to the case when $G$ is abelian).

Given such a geometric perspective on the CFT, a subset of the boundaries also acquires natural geometric intepretations. Indeed, for $d<26$, one can think of the sigma model \eqref{sigma} as defining the internal part of the bosonic string theory. 
The D$p$-branes in the string theory that extend on a $p$-dimensional submanifold $\cS_p\subset \cM$ and also along the time direction are described by conformal boundaries of the internal sigma models (they specify where the worldsheets can end in the target space $\cM$). In this case, the boundary $g$-function controls the effective tension of the brane, which is in turn determined by the DBI action,\footnote{We expect the formula \eqref{sigbraneg} to hold regardless of the string theory embedding as long as the backgrounds $(G,B)$ in \eqref{sigma} give an exact solution to the vanishing beta function.}
 \ie
g_{\cS_p}=   ( \pi \sqrt{2\A'})^{{d-2p\over 2}} 
 {\int_{\cS_p}  \det^{1\over 2} (G_{ij}+B_{ij}) \over {\rm vol}(\cM)^{1\over 2} }\,.
\label{sigbraneg}
\fe
Here $G_{ij}$ and $B_{ij}$ are the induced metric and $B$-field on $\cS_p\subset \cM$. The normalization is fixed by comparing to explicit answers for the toroidal CFTs \cite{Harvey:1999gq} where the target space coordinates have periodicity $X^\m\sim X^\m+2\pi$ and the D-branes wrap sub-tori,
\ie
g_{\cS_p}=  \left( \A'\over 2 \right)^{{d-2p\over 4}}  {|\det (G_{ij}+B_{ij})|^{1\over 2}\over |\det G_{\m\n}|^{1\over 4} }\,.
\label{sigbranegTd}
\fe
The D0-branes correspond to case when the numerators in \eqref{sigbraneg} and \eqref{sigbranegTd} are 1.

After reducing to the transverse noncompact $(26-d)$-dimensional spacetime, these D-branes give rise to point-like particles and their masses (in the Einstein frame) are determined by the $g$-function of the branes \cite{Harvey:1999gq},
\ie
m_{\cS_p}\sim \sqrt{ \A'^{11-{d\over 2}}\over G_{26-d}}g_{\cS_p}\,,
\fe
where $G_{26-d}$ is the effective Newton constant in the noncompact directions. 

The D-branes for a given sigma model are related to one another by target space dualities such as T-dualities (see \cite{Giveon:1994fu} for a review). Given the discussion in Section~\ref{sec:bdyfromTDL}, one expects these dualites to originate from topological defect lines that act on the branes by fusion. In fact there is a natural geometric construction of such defects as bi-branes for the sigma model \cite{Fuchs:2007fw}, namely D-branes in the doubled target space $\cM\times \cM$, which  is reminiscent of the folding trick in the CFT. The most general bi-branes define conformal defects in the CFT, while a subset of them are topological and explicitly constructed in \cite{Fuchs:2007fw} for WZW models (further examples of duality defects can be found in  \cite{Sarkissian:2008dq,Kapustin:2010zc,Gevorgyan:2013xka,Ji:2019ugf,Thorngren:2021yso}).

The above discussions generalize to supersymmetric sigma models that define supersymmetric CFTs (SCFT) straightforwardly. The preserved supersymmetry opens the door to an immense realm of CFTs defined by highly nontrivial target space $\cM$ such as Calabi-Yau (CY) manifolds. The D-branes provide fine probes of the geometry due to their  non-perturbative nature \cite{Shenker:1995xq}. In particular, $g$-functions of stable D-branes encode the volumes of minimal-volume cycles in the CY target space.\footnote{Here stability means the absence of supersymmetric relevant deformations of the boundary state.} As before, we can embed the CY sigma model into the type II superstring theory. Then an obvious class of stable D-branes is given by the BPS branes wrapping calibrated cycles in the CY. Their stability, as supersymmetric boundary states for the SCFT, is ensured by the spacetime conserved charges due to the nontrivial homology classes of the cycles they wrap. However, not all homology classes of a CY manifold admit calibrated cycles, and consequently the minimal-volume representative would have to be non-BPS (see recent discussions in  \cite{Demirtas:2019lfi,Long:2021lon}). Unlike the calibrated cycles whose volumes, thanks to the BPS condition, are determined by protected data of the CY, the non-BPS minimal-volumes carry much more nontrivial information involving the CY metric about which little is known. It was argued in \cite{Demirtas:2019lfi} that the Weak Gravity Conjecture \cite{Arkani-Hamed:2006emk} leads to interesting upper bounds on the $g$-function of such non-BPS stable D-branes and it will be interesting to compare with bootstrap bounds on boundary states in the SCFT. We hope to report on this in the future.

\section{Bootstrap Methods for Boundaries}
\label{sec:btssetup}

Combined with unitarity, the cylinder crossing equation (\ref{Cardy}) places powerful constraints on the allowed CFT data in the presence of boundaries. In what follows we will consider the Cardy condition associated with the crossing equation for the cylinder with identical boundary conditions $B^\alpha$ on either end for a unitary, compact CFT with central charge $c>1$ and a non-degenerate ground state on the strip Hilbert space $\mathcal{H}_{\alpha\alpha}$. In this setting the Cardy condition can be written as
\begin{equation}\label{eq:cylinderCrossing}
    g_\alpha^2 \widehat \chi_{\rm vac}(t) - \widehat\chi_{\rm vac}(t^{-1}) + \sum_{\substack{i\in\mathcal{I}^{\ell=0}_{\rm bulk}}}|B^\alpha_i|^2\widehat\chi_{\Delta_i\over 2}(t) - \sum_{\substack{j\in\mathcal{I}^{\alpha\alpha}_{\rm bdy}}}n^j_{\alpha\alpha}\widehat \chi_{h_j}(t^{-1}) = 0,
\end{equation}
where $\mathcal{I}_{\rm bulk}^{\ell=0}$ is the spectrum of scalar Virasoro primaries in the Hilbert space of the bulk CFT on the circle, and $\mathcal{I}^{\alpha\alpha}_{\rm bdy}$ is the spectrum of Virasoro primary operators in the strip Hilbert space $\mathcal{H}_{\alpha\alpha}$. Here, the crossing equation has been written in terms of the reduced Virasoro characters
\begin{equation}
    \begin{aligned}
    \widehat\chi_h(t) &= t^{1\over 4}q^{h-{c-1\over 24}}\\
    \widehat\chi_{\rm vac}(t) &= t^{1\over 4}q^{-{c-1\over 24}}(1-q).
    \end{aligned}
\end{equation}
Notice that the vacuum character includes the subtraction of the null descendant of the identity operator. For the case of boundary conditions in unitary CFTs with central charge $c>1$ that are the subject of this paper, this is the only degenerate character needed.
 
We have focused on the case of identical boundaries in order to harness the constraints of unitarity. In this setting the coefficients $|B_i^\alpha|^2$ and $g_\alpha^2$ are non-negative real numbers, and the $n^j_{\alpha\alpha}$ are non-negative integers (the latter constraint is of course not special to the case of identical boundaries). We will see that the Cardy condition (\ref{eq:cylinderCrossing}) together with unitarity and certain simple assumptions about the bulk and/or boundary spectrum lead to highly nontrivial constraints on the allowed boundary CFT data. Indeed we will see that in certain cases these simple inputs are sufficient to uniquely specify the admissable conformal boundary states.
 
\subsection{Linear Programming}\label{subsec:linearProgramming}
In the pioneering work of \cite{Rattazzi:2008pe} it was realized that one can derive powerful constraints on CFT data in the presence of unitarity by acting on crossing equations like the Cardy condition (\ref{eq:cylinderCrossing}) with simple linear functionals. To illustrate the strategy, consider the following setup. Suppose that the boundary entropy is given by some positive number $B^\alpha_0 = g_\alpha$, and that there are nonzero gaps above the identity operator in the bulk scalar primary spectrum and in the boundary primary spectrum, respectively given by $\Delta_{\rm gap}$ and $h_{\rm gap}^{\alpha\alpha}$. If one can construct a linear functional $\omega$ whose action on the reduced characters has the following positivity properties
\begin{equation}
    \begin{aligned}
    \omega\left[g_\alpha^2\widehat\chi_{\rm vac}(t) - \widehat\chi_{\rm vac}(t^{-1})\right] &> 0\\
    \omega\left[\widehat\chi_{\Delta_i\over 2}(t)\right] &\geq 0,\quad \Delta_i \geq \Delta_{\rm gap}\\
    \omega\left[\widehat\chi_{h_j}(t^{-1})\right] & \leq 0, \quad h_j \ge h^{\alpha\alpha}_{\rm gap},
    \end{aligned}
\label{eq:omegaDefinition}
\end{equation}
then by non-negativity of the coefficients $|B^\alpha_i|^2$ and $n^j_{\alpha\alpha}$, the proposed spectrum is incompatible with unitarity. That is, given such a functional one can rigorously conclude that the boundary entropy $g_\alpha$ cannot be realized by any unitary cylinder partition function with the proposed gaps in its bulk and boundary spectra.

A conventional basis for such a functional in numerical approaches to bootstrap consists of derivatives evaluated at the crossing-symmetric point. For example, if we write $t = e^z$, then such a functional could be written as
\begin{equation}
    \omega = \sum_{m=0}^N \omega_m \left.\partial_z^m\right|_{z=0},
\end{equation}
for some real coefficients $\omega_m$, where $N$ is referred to as the ``derivative order'' of the functional. Such functionals can be constructed using semidefinite programming (in this paper we will make use of the SDPB solver \cite{Simmons-Duffin:2015qma}), and the corresponding bounds on the CFT data will improve as the derivative order $N$ is increased and the space of functionals is more fully explored.

In sections \ref{sec:stableBraneBounds} and \ref{sec:universalBounds} we will numerically construct such functionals to derive novel upper and lower bounds on the boundary entropy $g_\alpha$ of possible boundary conditions given some physically motivated assumptions. Let us briefly illustrate how one computes such bounds. The strategy is similar to how one bounds particular squared OPE coefficients in the bootstrap of four-point functions of local operators \cite{Rattazzi:2010gj}. Suppose one can find linear functionals $\omega^{\pm}$ such that
\begin{equation}\label{eq:omegaConditions}
    \begin{aligned}
    \omega^\pm\left[\widehat\chi_{\Delta_i\over 2}(t)\right] & \geq 0, \quad \Delta_i\geq \Delta_{\rm gap} \\
    \omega^\pm \left[\widehat\chi_{h_j}(t^{-1})\right] & \leq 0, \quad h_j \geq h^{\alpha\alpha}_{\rm gap},
    \end{aligned}
\end{equation}
normalized such that
\begin{equation}\label{eq:omegaNormalization}
    \omega^{\pm}\left[\widehat\chi_{\rm vac}(t)\right] = \pm 1.
\end{equation}
Then for a fixed derivative order $N$, the best upper and lower bounds on the possible values of the boundary entropy $g_\alpha$ are given by optimizing the action of $\omega^+$ and $\omega^-$, respectively, on the vacuum character in the open string channel:
\begin{equation}\label{eq:omegaBound}
    \begin{aligned}
    g_\alpha^2 & \leq \min_{\{\omega_m\}}\left(\omega^+\left[\widehat\chi_{\rm vac}(t^{-1})\right]\right)\\
    g_\alpha^2 & \geq \max_{\{\omega_m\}}\left(-\omega^-\left[\widehat\chi_{\rm vac} (t^{-1})\right]\right).
    \end{aligned}
\end{equation}
This optimization procedure is also implemented in SDPB.

It is not at all obvious that functionals satisfying the positivity conditions (\ref{eq:omegaConditions}) should exist in the first place. For one thing, we note that if $h_{\rm gap}^{\alpha\alpha} < {c-1\over 24}$ and $\Delta_{\rm gap} < {c-1\over 12}$, then it is impossible to find such a functional. The reason is that the Virasoro characters in the cross-channel can be expanded in a complete basis of characters in the original channel
\begin{equation}
    \widehat\chi_h(t^{-1}) = \int_{{c-1\over 12}}^{\infty} d\Delta {\cos\left(4\pi \sqrt{(h-{c-1\over 24})({\Delta\over 2}-{c-1\over 24})}\right)\over \sqrt{\Delta-{c-1\over 12}}}\widehat\chi_{\Delta\over 2}(t).
\end{equation}
In particular if $h<{c-1\over 24}$ then the density of states in the integrand on the right hand side is positive everywhere, and so for functionals formed out of a finite number of derivatives evaluated at the crossing-symmetric point, we can commute the action of the functional with that of the integral and arrive at a contradiction. Similar remarks were made in \cite{Friedan:2012jk}, where it was noticed that one needed to assume $\Delta_{\rm gap} > {c-1\over 12}$ in order to compute a lower bound on $g_\alpha$ in the absence of any assumptions about the boundary spectrum.

However, if $h_{\rm gap}^{\alpha\alpha}>{c-1\over 24}$ or $\Delta_{\rm gap} > {c-1\over 12}$, then one may be able to compute bounds on the boundary entropy $g_\alpha$ using the functional method. In particular, we will see that the existence of any upper bound on $g_\alpha$ requries $h_{\rm gap}^{\alpha\alpha} > {c-1\over 24}$, while the existence of a nontrivial lower bound requires $\Delta_{\rm gap} > {c-1\over 12}$. Let us start by explaining why the existence of an upper bound requries a gap in the spectrum of boundary primaries. To derive an upper bound, we require a functional $\omega^+$ that satisfies the positivity conditions (\ref{eq:omegaConditions}) and that moreover acts positively on the vacuum character in the closed-string channel as in (\ref{eq:omegaNormalization}). The closed string vacuum character can be expanded in a complete basis of boundary characters with a positive density of states\footnote{Here we have made use of the Liouville parameterization of the central charge via $c = 1+6(b+b^{-1})^2$. For $c>1$ the density of states on the right-hand side of (\ref{eq:vacuumCrossing}) is positive. To see this, note that for $c>25$ we have $0<b<1$, while for $1<c<25$ $b$ is a pure phase.}
\begin{equation}\label{eq:vacuumCrossing}
    \widehat\chi_{\rm vac}(t) = \int_{c-1\over 24}^\infty {dh\over\sqrt{h-{c-1\over 24}}} \, 2\sqrt{2}\sinh\left(2\pi b\sqrt{h-{c-1\over 24}}\right)\sinh\left(2\pi b^{-1}\sqrt{h-{c-1\over 24}}\right)\widehat\chi_h(t^{-1}).
\end{equation}
So if the gap in the boundary spectrum is sufficiently small, in particular if $h_{\rm gap}^{\alpha\alpha} < {c-1\over 24}$, then it is impossible to satisfy (\ref{eq:omegaConditions}) and (\ref{eq:omegaNormalization}) simultaneously since $\omega^+$ acts non-positively on each boundary character on the right-hand side of (\ref{eq:vacuumCrossing}). 

A nearly identical argument shows that if the gap in the spectrum of bulk Virasoro scalar primaries is sufficiently small, in particular if $\Delta_{\rm gap} < {c-1\over 12}$, then although it may be possible to find a functional $\omega^-$ satisfying (\ref{eq:omegaConditions}) and (\ref{eq:omegaNormalization}), the lower bound on the boundary entropy one infers from (\ref{eq:omegaBound}) will be trivially satisfied in a unitary theory. The reason is that if the bulk gap is $\Delta_{\rm gap} < {c-1\over 12}$, then a similar decomposition to (\ref{eq:vacuumCrossing}) implies that the action of the functional $\omega^-$ satisfying (\ref{eq:omegaConditions}) on the vacuum character in the open string channel is necessarily non-negative
\begin{equation}
    \omega^-\left[\widehat\chi_{\rm vac}(t^{-1})\right] \ge 0\, \text{ if } \Delta_{\rm gap} < {c-1\over 12}.
\end{equation}
Thus (\ref{eq:omegaBound}) implies a trivial lower bound on the boundary entropy $g_\alpha^2$, as the right-hand side is non-positive.

The arguments requiring the existence of sufficiently large gaps in order for bounds on $g$ to exist can be rephrased in terms of the existence of solutions to the cylinder crossing equation that are noncompact (i.e. those with a continuous spectrum and that do not possess a vacuum state) in one channel.\footnote{We can be slightly more explicit as follows. The upper bound on $g$ requires $h_{\rm gap}>{c-1\over 24}$ because otherwise $g^2 \widehat\chi_{\rm vac}(t) = g^2 \int_{c-1\over 24}^\infty dh \rho(h)\widehat\chi_h(t^{-1})$ is a valid solution to crossing for arbitrarily large $g$, where $\rho(h)$ is a positive density of states. Similarly, for the lower bound to exist we need $\Delta_{\rm gap} > {c-1\over 12}$ in order to rule out the solution $\widehat\chi_{\rm vac}(t^{-1}) = \int_{c-1\over 12}^\infty d\Delta \rho(\Delta)\widehat\chi_{\Delta\over 2}(t)$, which has vanishing $g$.} It is possible that bounds on $g$ could exist in the absence of these assumptions provided one was able to insist on the presence of the vacuum state in both channels and compactness of the bulk and boundary spectra. We do not presently know how to accomplish this, and thus avoid spurious noncompact solutions to crossing, in numerical bootstrap.\footnote{Although we note that there is significant motivation to do so from other perspectives. For example, one could address the question of whether ${c-1\over 12}$ is really the optimal upper bound on the twist gap in compact CFT \cite{Collier:2016cls}. More ambitiously, one might hope to bound the total Hodge numbers of Calabi-Yau threefolds by studying modular bootstrap constraints on CY sigma models \cite{Keller:2012mr}.}

\subsection{Analytic Functionals}\label{subsec:analyticFunctional}
In the following section, we will see that numerical bounds on the boundary entropy are often nearly saturated by interesting physical theories. In such cases, it is natural to try to prove analytically that in fact the bound is exactly saturated. Having an analytic proof of exact saturation means explicitly constructing a functional $\omega$ satisfying equations \eqref{eq:omegaDefinition} with the first line replaced by the equality
\begin{equation}
    \omega\left[g_\alpha^2\widehat\chi_{\rm vac}(t) - \widehat\chi_{\rm vac}(t^{-1})\right] = 0\,,
\end{equation}
where $g_\alpha^2$ is the exact value in the saturating theory. It further follows that such $\omega$ must vanish on the entire spectrum exchanged in both channels. We can thus read off the bulk and boundary spectrum of the saturating theory from the zeros of $\omega[\widehat\chi_{\frac{\Delta}{2}}(t)]$ and $\omega[\widehat\chi_{h}(t^{-1})]$ respectively.

Various examples of analytic functionals proving optimal bootstrap bounds have appeared in the literature, starting with the case of the correlator bootstrap in 1D CFTs \cite{Mazac:2016qev,Mazac:2018mdx,Mazac:2018ycv}. Other examples include the correlator bootstrap in general dimension \cite{Mazac:2019shk,Paulos:2019gtx,Caron-Huot:2020adz,Carmi:2020ekr,Caron-Huot:2021enk,Trinh:2021mll}, the correlator bootstrap in the presence of conformal boundaries \cite{Kaviraj:2018tfd,Mazac:2018biw} and the modular bootstrap of the torus partition function \cite{Hartman:2019pcd,Afkhami-Jeddi:2020ezh}. To construct analytic functionals for the annulus partition function, we will closely follow the strategy of \cite{Hartman:2019pcd}. The main idea is to think of the partition function as a suitable four-point function of twist fields. This will allow us to uplift the 1D correlator functionals to the present setting. We will now briefly review the  1D correlator functionals of \cite{Mazac:2016qev,Mazac:2018mdx,Hartman:2019pcd} and explain their uplift to the annulus partition function. Applications to concrete bounds on boundary entropy appear in Section \ref{sec:analyticFunctionals}.

Consider a four-point function of pairwise identical scalar primaries inserted at locations $x_1<x_2<x_3<x_4$ along the real line
\begin{equation}
    \langle \phi_{a}(x_1)\phi_{a}(x_2)\phi_{b}(x_3)\phi_{b}(x_4)\rangle = 
    \langle \phi_{a}(x_1)\phi_{a}(x_2)\rangle\langle\phi_{b}(x_3)\phi_{b}(x_4)\rangle\,
    \mathcal{G}(z)\,,
\end{equation}
where $z=(x_{12}x_{34})/(x_{13}x_{24})$ is the 1D cross-ratio. Let us assume all external operators have the same dimension $\Delta_{\phi}$. The equality between the s- and t-channel OPE takes the form
\begin{equation}
   z^{-2\Delta_{\phi}}\cG(z) =  \sum\limits_{\cO\in\phi_a\times\phi_a}\!\!\!\!f_{a a\cO}f_{bb\cO}z^{-\Delta_{\phi}}G_{\Delta_{\cO}}(z) = 
   \sum\limits_{\cP\in\phi_a\times\phi_b}\!\!\!\!(f_{a b\cP})^2(1-z)^{-\Delta_{\phi}}G_{\Delta_{\cP}}(1-z)\,.
\label{eq:crossing1}
\end{equation}
Here
\begin{equation}
    G_{\Delta}(z) = z^{\Delta}{}_2F_1(\Delta,\Delta;2\Delta;z)
    \label{eq:1Dblock}
\end{equation}
are the 1D conformal blocks and $f_{abc}$ are the structure constants. It is convenient to symmetrize and antisymmetrize the equation under crossing $z\leftrightarrow 1-z$
\ie
&\sum\limits_{\cO\in\phi_a\times\phi_a}\!\!\!\!f_{a a\cO}f_{bb\cO}F^{+}_{\Delta_{\cO}}(z)-\sum\limits_{\cP\in\phi_a\times\phi_b}\!\!\!\!(f_{a b\cP})^2F^{+}_{\Delta_{\cP}}(z) = 0\\
&\sum\limits_{\cO\in\phi_a\times\phi_a}\!\!\!\!f_{a a\cO}f_{bb\cO}F^{-}_{\Delta_{\cO}}(z)+\sum\limits_{\cP\in\phi_a\times\phi_b}\!\!\!\!(f_{a b\cP})^2F^{-}_{\Delta_{\cP}}(z) = 0\,,
\label{eq:crossing2}
\fe
where
\begin{equation}
    F^{\pm}_{\Delta} = z^{-\Delta_{\phi}}G_{\Delta}(z)\pm
    (1-z)^{-\Delta_{\phi}}G_{\Delta}(1-z)\,.
\end{equation}

A general functional $\omega$ for the full crossing equation \eqref{eq:crossing1} is a sum of functionals $\omega^+$, $\omega^-$ acting on the first and second line of \eqref{eq:crossing2}. For each $\Df>0$, there exist distinguished functionals $\beta^{+}_{\Df}$ and $\beta^{-}_{\Df}$, acting respectively on crossing-symmetric and antisymmetric functions and satisfying
\begin{enumerate}
    \item $\beta^{+}_{\Df}[F^{+}_{\Delta}(z)]$ and $\beta^{-}_{\Df}[F^{-}_{\Delta}(z)]$ each have a simple zero at $\Delta = 2\Df+1$ with unit slope.
    \item $\beta^{+}_{\Df}[F^{+}_{\Delta}(z)]$ and $\beta^{-}_{\Df}[F^{-}_{\Delta}(z)]$ each have double zeros at $\Delta = 2\Df+2n+1$ for all $n=1,2,\ldots$.
\end{enumerate}
It turns out that these properties fix $\beta^{+}_{\Df}$ and $\beta^{-}_{\Df}$ uniquely. The functionals take the form of contour integrals in the complex $z$ plane. The construction was reviewed in Section 5 of \cite{Hartman:2019pcd}, to which we refer the reader for details.\footnote{Note that $\beta^{-}_{\Df}$ is denoted $\beta_{\Df}$ there.}

We can combine $\beta^{+}_{\Df}$ and $\beta^{-}_{\Df}$ in a natural way as follows. Consider the functional
\begin{equation}
    \omega_{\Df}[\cF(z)] = \beta^{+}_{\Df}\left[\frac{\cF(z)+\cF(1-z)}{2}\right]+\beta^{-}_{\Df}\left[\frac{\cF(z)-\cF(1-z)}{2}\right]\,.
\end{equation}
It follows from the structure of zeros articulated in points 1 and 2 above that
\begin{enumerate}
    \item $\omega_{\Df}[z^{-\Delta_{\phi}}G_{\Delta}(z)]$ has a simple zero with unit slope at $\Delta = 2\Df+1$ and double zeros at $\Delta=2\Df+2n+1$ for all $n=1,2,\ldots$.
    \item $\omega_{\Df}[(1-z)^{-\Delta_{\phi}}G_{\Delta}(1-z)]$ has double zeros at $\Delta=2\Df+2n+1$ for all $n=0,1,\ldots$.
\end{enumerate}
The key point is that in the first case, the slopes at $\Delta=2\Df+1$ add up, but they cancel out in the second case, yielding a double zero. Besides the structure of double zeros, $\omega_{\Df}$ enjoys the following remarkable positivity properties
\begin{enumerate}
    \item $\omega_{\Df}[z^{-\Delta_{\phi}}G_{\Delta}(z)]\geq 0$ for all $\Delta\geq 2\Df+1$.
    \item $\omega_{\Df}[(1-z)^{-\Delta_{\phi}}G_{\Delta}(1-z)]\leq 0$ for all $\Delta\geq 0$.
\end{enumerate}
These inequalities ensure that after we uplift $\omega_{\Df}$ to a functional for the annulus modular bootstrap, it will satisfy the inequalities listed in \eqref{eq:omegaConditions}.

\subsection{Annulus Partition Function as a Four-point Function of Twist Fields}\label{subsec:ZfromG}

Let us explain how we can represent the annulus partition function $Z_{\alpha\beta}(t)$ as a four-point function of local primary operators inserted on a line. This will allow us to apply $\omega_{\Df}$ from the previous subsection directly to the problem at hand. The construction is closely related to the following familiar interpretation of the torus partition function of a 2D CFT as the four-point function of twist fields in the symmetric product orbifold $(\cT\times\cT)/\mathbb{Z}_2$, where $\cT$ stands for a single copy of the theory. A complex torus of modulus $\tau$ can be represented as the set of solutions of
\begin{equation}
    y^2 = x(x-\lambda(\tau))(x-1)
\end{equation}
in $(x,y)\in\mathbb{C}^2$, where
\begin{equation}
    \lambda(\tau) = \frac{\theta_2(\tau)^4}{\theta_3(\tau)^4}
\end{equation}
is the modular lambda function. This means we have a two-to-one covering of the sphere by the torus $(x,y)\mapsto x$. $\lambda(\tau)$ computes the cross-ratio of the locations of the branch points. Physically, this can be realized by considering the product orbifold $(\cT\times\cT)/\mathbb{Z}_2$ in a plane, with $\mathbb{Z}_2$ twist operators inserted at $x_1=0$, $x_2=\lambda(\tau)$, $x_3=1$ and $x_4=\infty$. Going around any of the twist operators switches the two copies of $\cT$, or equivalently switches the two sheets of the above covering.

\begin{figure}[h]
    \centering
    \includegraphics[width=.9\textwidth]{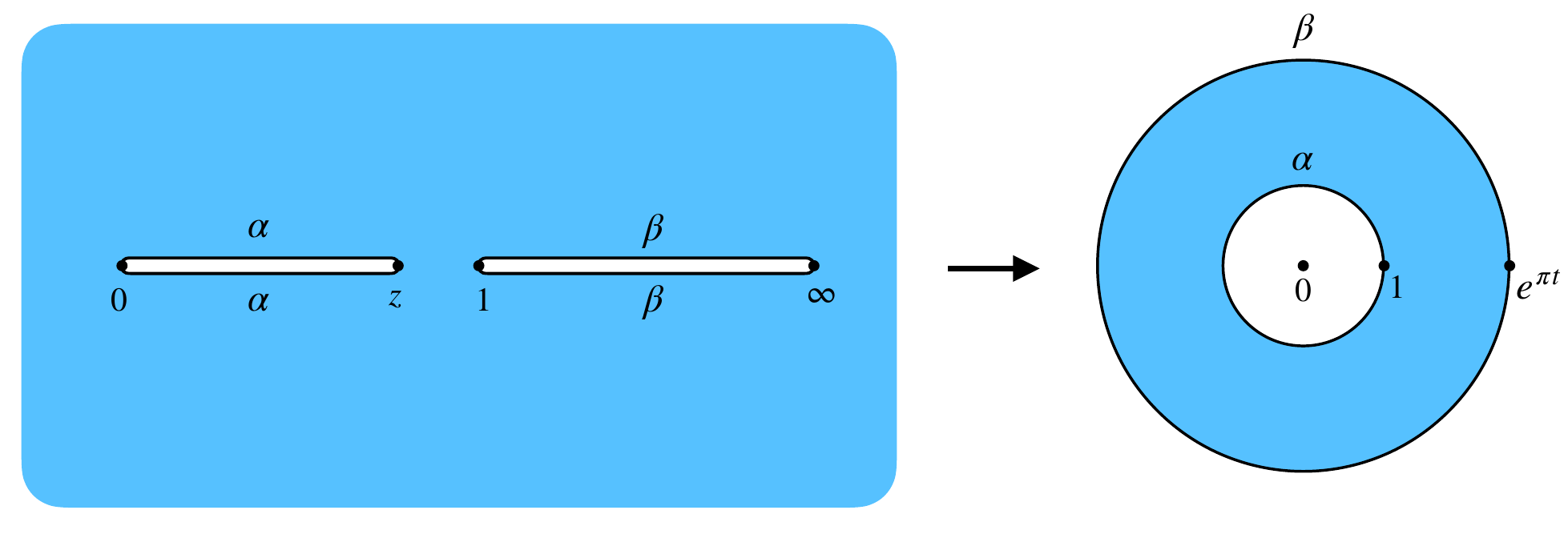}
    \caption{The complex plane with intervals $[0,z]$ and $[1,\infty)$ removed is biholomorphic to the open annulus of modulus $t$, where $z=\lambda(i t)$. This allows us to think of the cylinder partition function as a four-point function of twist fields in the plane.}
    \label{fig:CutPlane}
\end{figure}

Let us now mimic this construction for the annulus. First, note that by sewing two copies of an annulus of modulus $t$ along the boundaries, we get a torus with modulus $\tau = i t$. This implies that the annulus corresponds to a single sheet of the above covering. Indeed, the covering restricts to a biholomorphic map between the open annulus of modulus $t$ and the complex plane with the intervals $[0,\lambda(i t)]$ and $[1,\infty)$ removed. The two boundary circles of the annulus get mapped to the two intervals. If $t\in\mathbb{R}$, then $\lambda(i t)\in (0,1)$, and thus the four points are collinear. The map is illustrated in Figure \ref{fig:CutPlane}.

To proceed, note that for each conformal boundary condition $\alpha$ of theory $\cT$, we can define a conformal line defect $\cD_{\alpha}$ by cutting spacetime along a line and imposing the boundary condition $\alpha$ along both sides of the cut. In other words, $\cD_{\alpha}$ is a factorized conformal interface between two copies of $\cT$ in neighboring half-spaces. This defect can end at a point, say $z=0$. In that case, it can be mapped using $w=\sqrt{z}$ to the situation where we have $\cT$ in a half-plane with boundary condition $\alpha$. Therefore, the local operators that can reside at the endpoint of $\cD_{\alpha}$ are the same as local operators on the boundary. The operator living at the end of $\cD_{\alpha}$ which corresponds to the identity operator on the boundary will be denoted $\phi_{\alpha}$. It is a primary operator with conformal dimension $\Df = c/16$, which can be shown by considering the correlator of the stress tensor in the presence of $\cD_{\alpha}$ stretching along a segment, see \cite{Calabrese:2009qy}.

It follows that the annulus partition function $Z_{\alpha\beta}(t)$ can be computed as the four-point function $\langle \phi_{\alpha}(x_1)\phi_{\alpha}(x_2) \phi_{\beta}(x_3)\phi_{\beta}(x_4) \rangle$ in the plane, where we insert $\cD_{\alpha}$ along the segment $[x_1,x_2]$ and $\cD_{\beta}$ along the segment $[x_3,x_4]$. Without loss of generality, we can take $x_i\in\mathbb{R}$ with $x_1<x_2<x_3<x_4$. Their cross-ratio satisfies $z=\lambda(it)$. The precise relationship is as follows (c.f. \cite{Lunin:2000yv} for the torus example)\footnote{Similar identities have appeared in \cite{Cheng:2020srs} in the study of finite dimensional representations of the mapping class group for a punctured Riemann surface. However the physical picture is different. While  our relation \eqref{eq:GfromZ} here is between quantities in the same CFT,  the identities in \cite{Cheng:2020srs} relate the (normalized) four-point conformal blocks and torus characters of two \textit{distinct} chiral algebras. In particular, there is a curious relation between characters of the $G_1$ WZW CFT in the Deligne-Cvitanovi\'c exceptional series and certain Virasoro four-point blocks in a sequence of minimal models related to the ${G_1\times G_1\over G_2}$ coset \cite{Cheng:2020srs}.
}
\ie
    \langle \phi_{\alpha}(x_1)\phi_{\alpha}(x_2) \phi_{\beta}(x_3)\phi_{\beta}(x_4) \rangle &= 
    \langle \phi_{\alpha}(x_1)\phi_{\alpha}(x_2)\rangle \langle\phi_{\beta}(x_3)\phi_{\beta}(x_4) \rangle\,\cG(z)\\
    &=\langle \phi_{\alpha}(x_1)\phi_{\alpha}(x_2)\rangle \langle\phi_{\beta}(x_3)\phi_{\beta}(x_4) \rangle \left[\tfrac{z^2}{2^8(1-z)}\right]^{\frac{c}{24}}Z_{\alpha\beta}(t)\,.
    \label{eq:GfromZ}
\fe

The expansion of $Z_{\alpha\beta}(t)$ into bulk characters becomes the s-channel OPE of the four-point function $\cG(z)$. Indeed, in that case we are quantizing the theory on a circle surrounding the interval $[x_1,x_2]$, and computing the overlap of states created by $\cD_{\alpha}$ and $\cD_{\beta}$. The conformal blocks for this situation are precisely the bulk characters. The bulk character of a scalar Virasoro primary of dimension $\Delta$ can be expanded in the s-channel 1D conformal blocks \eqref{eq:1Dblock} of dimensions $\Delta+2n$, $n=0,1,\ldots$ with positive coefficients. More concretely,
\begin{equation}
    \left[\tfrac{z^2}{2^8(1-z)}\right]^{\frac{c}{24}}\chi_{\frac{\Delta}{2}}(t(z)) = \sum\limits_{n=0}^{\infty}a_n\,G_{\Delta+2n}(z)
    \label{eq:expansion1}
\end{equation}
with $a_n>0$. Here 
$t(z)$ is the inverse of $z=\lambda(i t)$, namely
\begin{equation}
    t(z) = \frac{K(1-z)}{K(z)}\,,
    \label{eq:zToTau}
\end{equation}
where
\begin{equation}
    K(z) = \frac{\pi}{2}{}_2F_1(1/2,1/2;1;z)
\end{equation}
is the complete elliptic integral of the first kind.

\begin{figure}[h]
    \centering
    \includegraphics[width=.9\textwidth]{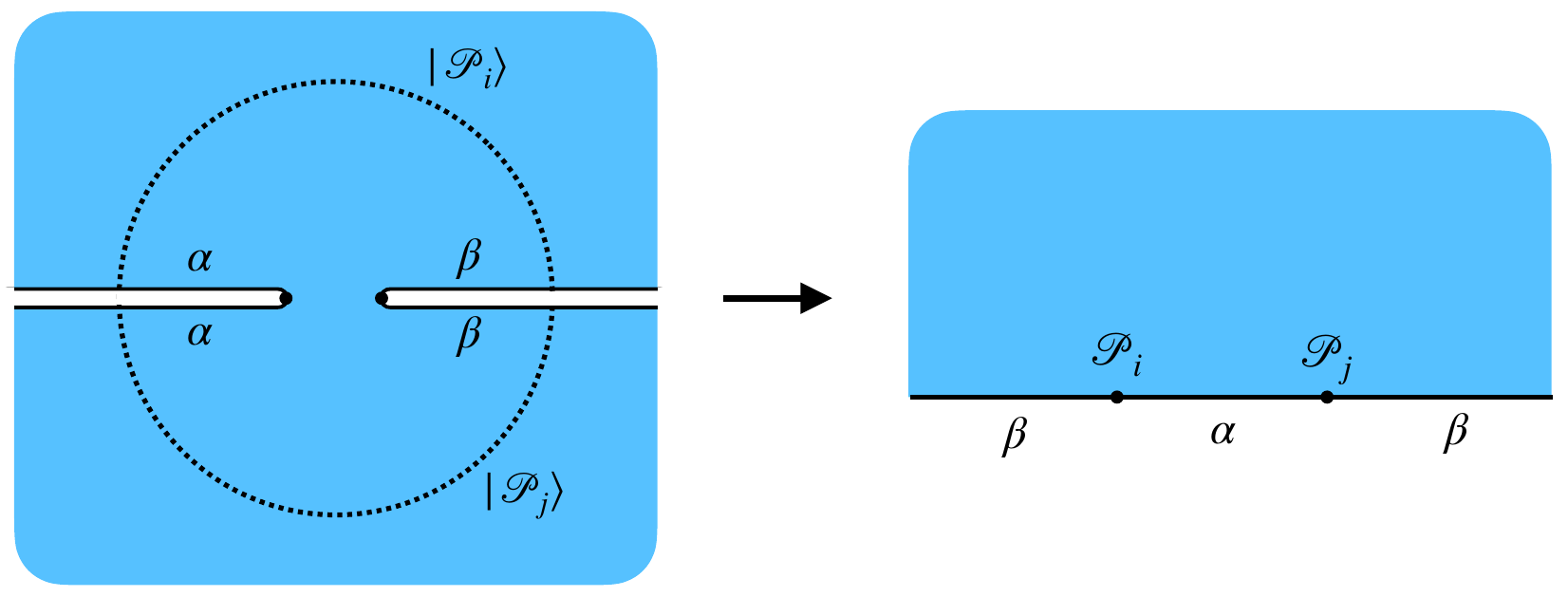}
    \caption{The overlap of the t-channel initial state with the state $|\cP_i\rangle\otimes|\cP_j\rangle\in\cH_{\alpha\beta}\otimes\cH_{\alpha\beta}$ is equal to the boundary two-point function $\langle\beta\,\cP_i\,\alpha\,\cP_j\,\beta\rangle\sim\delta_{ij}$.}
    \label{fig:BoundaryChannel}
\end{figure}

Similarly, the expansion of $Z_{\alpha\beta}(t)$ in the boundary channel is the same as the t-channel OPE of $\cG(z)$. In that case, the Hilbert space in radial quantization is that of $\cT$ on a cylinder with $\cD_{\alpha}$ and $\cD_{\beta}$ inserted along the time direction at opposite points of the $S^1$. This Hilbert space is $\cH_{\alpha\beta}\otimes\cH_{\alpha\beta}$. The primaries in the full Hilbert space are pairs of primaries in $\cH_{\alpha\beta}$, i.e. $(\cP_i,\cP_j)$. However, only the primaries with $\cP_i=\cP_j$ can contribute in the t-channel OPE of \eqref{eq:GfromZ}. This is because the overlap of the in-state with $(\cP_i,\cP_j)$ is equal to the boundary two-point function $\langle\beta\,\cP_i\,\alpha\,\cP_j\,\beta\rangle \sim \delta_{ij}$, where $\alpha,\beta$ denote the boundary conditions, see Figure~\ref{fig:BoundaryChannel}.

The boundary character of a primary of dimension $h$ can be expanded in t-channel 1D conformal blocks of dimensions $2h+2n$ with positive coefficients
\begin{equation}
    \left[\tfrac{z^2}{2^8(1-z)}\right]^{\frac{c}{24}}\chi_{h}(t(z)^{-1})=
    \left[\tfrac{z^2}{2^8(1-z)}\right]^{\frac{c}{24}}\chi_{h}(t(1-z)) = \left(\tfrac{z}{1-z}\right)^{\frac{c}{8}}\sum\limits_{n=0}^{\infty}a_n\,G_{2h+2n}(1-z)\,.
    \label{eq:expansion2}
\end{equation}

The upshot of the present discussion for the functional bootstrap is the following. Consider any functional $\omega$ for the 1D correlator bootstrap, which acts on the crossing equation in the normalization of \eqref{eq:crossing1}, where $\Df=c/16$ is the external dimension. There is a canonical way to apply $\omega$ to the modular crossing equation for $Z_{\alpha\beta}(\tau)$ at central charge $c$, namely
\begin{equation}
    \omega\left[
    \left[2^8 z (1-z)\right]^{-\frac{c}{24}}
    Z_{\alpha\beta}(t(z))\right]
    \label{eq:omegaZ}
\end{equation}
with $Z_{\alpha\beta}$ expanded in bulk and boundary characters. If $\omega$ satisfies positivity properties when acting on 1D conformal blocks, then positivity of the coefficients in \eqref{eq:expansion1} and \eqref{eq:expansion2} guarantees that some positivity is preserved when we apply $\omega$ to the modular crossing equation using \eqref{eq:omegaZ}. We will discuss a concrete incarnation of this idea in Section \ref{sec:analyticFunctionals}.

\section{Bounds on Stable Branes in Specific CFTs}\label{sec:stableBraneBounds}
In this section we will study upper and lower bounds on the boundary entropy of stable boundary conditions in certain specific CFTs from semidefinite programming, following the logic outlined in section \ref{subsec:linearProgramming}. By stable, we mean that the spectrum of boundary primaries includes no relevant operators
\begin{equation}
    h^{\alpha\alpha}_{\rm gap} \geq 1.
\end{equation}
In this sense the corresponding boundary conditions are stable against boundary renormalization group flow. In the case that the bulk CFT is part of a worldsheet string theory, the corresponding D-branes are free from open-string tachyons and thus perturbatively stable against decay. In general, we believe any compact CFT should have at least one stable brane. This is because we expect a lower bound on the boundary entropy $\log g$ and the existence of a stable brane follows by the $g$-theorem. Although we don't have a complete proof, we will provide plenty of evidence for this lower bound here and also in Section~\ref{sec:universalBounds}. In particular, the identity Cardy brane of RCFTs is always stable, and as we will see later in this section, they turn out to saturate various annulus bootstrap bounds. 

In what follows we will pay particular attention  to CFTs where the gap in spectrum of bulk scalar Virasoro primaries saturates the modular bootstrap bound of \cite{Collier:2016cls}. In particular, we will see that in these cases the upper and lower bounds on the boundary entropy for stable branes coincide to a high degree of numerical precision, providing evidence that the boundary entropy for stable branes in these CFTs is uniquely specified. Indeed, in section \ref{sec:analyticFunctionals} we will see that in certain cases we are able to explicitly construct the analytic functional that proves optimality of the upper and lower bounds on the boundary entropy and from which we can reconstruct the spectrum of boundary primary operators. 

For sufficiently small values of the central charge, the lowest-lying local primary operator in the Hilbert space of the bulk CFT on the circle must necessarily be a scalar. Indeed, for $1\leq c \leq 4$, the upper bound on the dimension of the lightest Virasoro primary from modular invariance in unitary CFTs \cite{Collier:2016cls} converged to the following to remarkable numerical precision
\begin{equation}
    \Delta_{\rm gap} \leq {c\over 6} + {1\over 3},\quad 1\leq c \leq 4.
\end{equation}
To the best of our knowledge, the physical origin of this uniform bound has not been understood. For $c>4$, the bound on the dimension of the lightest scalar primary is weaker than the bound on overall gap in the spectrum of Virasoro primaries. In \cite{Collier:2016cls}, several rational CFTs were identified that saturated the upper bound on the scalar gap at the following values of the central charge
\begin{equation}
    c = 1,\, 2,\, {14\over 5},\, 4,\, 8
\end{equation}
corresponding to the $SU(2)_1$, $SU(3)_1$, $(G_2)_1$, $Spin(8)_1$, and $(E_8)_1$ WZW models, respectively. These diagonal theories have the property that there are only two characters with respect to the maximal chiral algebra except for the $(E_8)_1$ case which has only one character.\footnote{One might notice that these groups belong to the Deligne-Cvitanovi\'c exceptional series, and might further wonder whether the WZW models at level one corresponding to the remainder of the Deligne-Cvitanovi\'c exceptional series, namely the $(F_4)_1$, $(E_6)_1$ and the $(E_7)_1$ WZW models at central charges $c = 26/5,\, 6,\, 7$ respectively, also saturate bootstrap bounds. Indeed, in \cite{Bae:2017kcl} it was shown that if one allows for the presence of conserved currents, the level-one WZW models corresponding to the entirety of the Deligne-Cvitanovi\'c exceptional series saturate the bound on the gap in the spectrum of twists $t$ of local primary operators, where $t = \Delta-|\ell|$. }

In this section we will study the upper and lower bounds on the boundary entropy for stable boundary conditions in these CFTs, and will find convincing numerical evidence that in these settings the boundary entropy, and indeed the entire cylinder partition function, is uniquely specified. In some special cases we will find that it is enough to specify the value of the bulk scalar gap in order to pin down $g$. In some other cases we will find that one needs to further input more information about the spectrum of the bulk CFT, for example the dimensions of the first $N$ low-lying scalar operators, in order for the upper and lower bounds to convincingly coincide. We implement this in semidefinite programming as follows. We will seek functionals $\omega^\pm$ that, rather than acting non-negatively on all closed string scalar characters above the gap as in the first line of (\ref{eq:omegaConditions}), instead only act non-negatively on the specific characters corresponding to the first $N$ low-lying scalars, and on all closed-string characters with dimension above that of the $(N+1)^{\rm th}$ operator:
\begin{equation}\label{eq:incorporateLowLying}
    \begin{aligned}
        \omega^{\pm}\left[\widehat\chi_{\Delta_i\over 2}(t)\right] &\geq 0,\quad i=1,\ldots, N,~ \Delta_i\in\mathcal{I}_{\rm bulk}^{\ell=0}\\
        \omega^{\pm}\left[\widehat\chi_{\Delta\over 2}(t)\right] & \geq 0, \quad \Delta\geq \Delta_{N+1}.
    \end{aligned}
\end{equation}

For the particular diagonal rational CFTs that saturate the bound on the scalar gap, we will find evidence that the converging upper and lower bounds on the boundary entropy are saturated by the cylinder partition function associated to the identity Cardy brane
\begin{equation}
    |B^\id\ra=\sum_{j\in \cI} \sqrt{S_{0 j}} |\varphi_j \rra\,,
\end{equation}
as well as those obtained from a global symmetry rotation by ${G\times G\over Z(G)}\rtimes {\rm Out}(G)$  (see Section~\ref{sec:bdyfromTDL}). Here, it is understood that $S$ is the modular $S$ matrix associated with the maximal chiral algebra of the bulk rational CFT and that $\cI$ is the spectrum of local operators that are primary with respect to the maximal chiral algebra (necessarily scalar by the diagonal condition). The corresponding cylinder partition function is then given by the identity character of the maximal chiral algebra in the open-string channel
\begin{equation}
    Z_{\id\id}(t) = \sum_{j\in \cI}S_{0j}\widetilde\chi_{\Delta_j\over 2}(t) = \widetilde \chi_{\rm vac}(t^{-1}),
\end{equation}
where $\widetilde\chi$ are the characters of representations of the maximal chiral algebra. The saturating value of the $g$ function is correspondingly given in terms of the identity-identity element of the modular $S$ matrix as
\begin{equation}
    g_{\id} = \sqrt{S_{00}}.
\end{equation}

\subsection{$c=1$ and the $SU(2)_1$ CFT}\label{subsec:c1}
We start by computing bounds on the boundary entropy for unitary, compact CFTs with $c=1$. To the best of our knowledge, the only known unitary, compact CFTs with $c=1$ are the free compact boson and its orbifolds \cite{Ginsparg:1987eb}. Boundary states for this family of CFTs have been classified \cite{Elitzur:1998va,Recknagel:1998ih,Friedan,Gaberdiel:2001xm,Gaberdiel:2001zq,Janik:2001hb,Cappelli:2002wq,Yamaguchi:2003yq,Gaberdiel:2008zb} which we will review below.

We start with the circle branch of the $c=1$ CFTs, which is characterized by a global symmetry $(U(1)_m\times U(1)_w)\rtimes \mZ_2^C$ where $U(1)_m\times U(1)_w$ are the momentum and winding symmetries respectively while the $\mZ_2^C$ acts by charge conjugation (flips the sign on the boson). The circle branch is a one-dimensional conformal manifold parametrized by the radius $R$ of the compact boson.\footnote{Here we are working in units where $\alpha' = 2$.} At generic $R$, the bulk   spectrum of Virasoro primaries   consists of $\mf{u}(1)$ primaries
\ie
|(m,w)\ra~{\rm with}~h={1\over 2}\left({m\over R}+{wR\over 2} \right)^2,~\bar h={1\over 2}\left({m\over R}-{wR\over 2} \right)^2 
\fe
that carry momentum and winding charges $m,w\in \mZ$, as well as $\mf{u}(1)$ Kac-Moody descendants of the vacuum which are normal ordered symmetric Schur polynomials in the holomorphic and anti-holomorphic $\mf{u}(1)$ currents,
\ie
|n;\bar n\ra~~{\rm with}~h=n^2,~\bar h=\bar n^2
\fe
labelled by $n,\bar n \in \mZ_+$.

The conformal boundaries fall into three disconnected families: the Dirichlet boundaries $|D(\theta)\ra$ with $\theta\in [0,2\pi)$,
\ie
|D(\theta)\ra=\, &
{1\over \sqrt{R}} \left(\sum_{n\in \mZ_+} |n;n\rra+\sum_{m\in \mZ} e^{im \theta} |(m,0) \rra \right)\,,
\fe
the Neumann boundaries $|N(\phi)\ra$ with $\phi\in [0,2\pi)$,
\ie
|N(\phi)\ra=\, &
  \sqrt{R\over 2}\left((-1)^n\sum_{n\in \mZ_+}|n;n\rra+\sum_{w\in \mZ} e^{iw \phi} |(0,w)\rra\right)\,,
  \fe
  and the \textit{exotic} boundaries $|E(x)\ra$ with $x\in [-1,1]$ \cite{Janik:2001hb},
 \ie
  |E(x)\ra =\, & \cN \sum_{n\in \mZ_{\geq 0}} P_n(x) |n;n\rra\,,
\fe
where $P_n(x)$ denotes the Legendre polynomials and $\cN$ is an arbitrary constant. 
Here $|\cdot \rra$ denotes  the Virasoro Ishibashi states. While the Dirichlet and Neumann boundary states preserve the $\mf{u}(1)$ KM algebra and can be written in terms of $\mf{u}(1)$ KM Ishibashi states, the exotic boundaries only preserve the Virasoro algebra. Furthermore the exotic boundaries have a continous open-string spectrum starting at $h=0$ and do not have a normalizable vacuum (similar to the properties of a non-compact boson) and thus do not obey the Cardy condition strictly with integral degeneracies.\footnote{Similar occurence of non-compactness for defects in compact CFTs was recently noticed for topological defect lines on the $c=1$ orbifold branch \cite{Chang:2020imq}.}

To gain a better intuition of these exotic branes, it is helpful to consider the dense rational points on the $c=1$ circle branch with $R={M\over N}\sqrt{2}$ for coprime integers $M,N>0$. This is related to the $SU(2)_1$ WZW CFT at $R=\sqrt{2}$ by first gauging a $\mZ_N$ momentum symmetry and then gauging a $\mZ_M$ winding symmetry. Correspondingly, this gauging picture allows one to classify the boundary states at the rational points from those at $R=\sqrt{2}$. The conformal boundaries at the $SU(2)_1$ point are classified in \cite{Gaberdiel:2001xm} and they are in one-to-one correspondence with group elements of $SU(2)$. Consequently, the boundary states at the $R={M\over N}\sqrt{2}$, include in addition to the Dirichlet and Neumann branes, a family $|h\ra$ parametrized by $h\in {SU(2)/( \mZ_M\times \mZ_N)}$ where $\mZ_M$ acts by $h \to e^{{\pi i \over M}\sigma_3}h e^{-{\pi i \over M}\sigma_3}$ and $\mZ_N$ acts by $h \to e^{{\pi i \over N}\sigma_3}h e^{{\pi i \over N}\sigma_3}$. The $|h\ra$ branes, in the limit $M,N\to \infty$, away from the fixed loci of $ \mZ_M\times \mZ_N$, approach the one parameter family of exotic branes $|E(x)\ra$ at irrational points on the circle branch. Nevertheless, for finite $M,N$, they define ordinary boundary states that obey all of the BCFT axioms \cite{Gaberdiel:2001zq,Gaberdiel:2008zb}. In particular, they have boundary tension $g=2^{-{1\over 4}}\sqrt{MN}\geq1$ and always contain relevant operators in the open string spectrum.

Therefore the potential stable branes on the circle branch are either Dirichlet or Neumann branes.
\begin{table}[!htb]
    \centering		\renewcommand{\arraystretch}{1.8}
    \begin{tabular}{|c|c|c|}
    \hline
       &   $|D(\theta)\ra$ & $|N(\phi)\ra$   \\\hline
    $\Delta_{\rm gap}$  &  \multicolumn{2}{c|}{$ {1\over  R^2}$}    \\\hline
      $h_{\rm gap}$   &  ${R^2\over 2}$    &   ${2\over R^2}$ \\\hline
        $g$     & ${1\over \sqrt{R}}$  & $\sqrt{R\over 2}$ \\\hline
    \end{tabular}
    \caption{Bulk and boundary gaps and brane tensions on the $c=1$ circular branch }
    \label{tab:circleBranes}
\end{table} 

The orbifold branch of the $c=1$ CFTs are described by gauging the $\mZ_2^C$ symmetry of the compact boson. The conformal boundaries in the orbifold theory are built from $\mZ_2^C$ invariant combinations of the ordinary boundary states for the compact boson as well as $\mZ_2^C$ twisted boundary states (see Section~\ref{sec:bdyfromTDL} for a general discussion of branes in orbifold). Restricting to the former leads to so-called regular branes \eqref{rbrane} in the $\mZ_2^C$ orbifold which only have overlaps with untwisted sector bulk operators,
\ie
|D_O(\theta) \ra=&{1\over \sqrt{2}}(|D(\theta)\ra +|D(-\theta)\ra)\,,\quad 
|N_O(\phi)\ra={1\over \sqrt{2}}(|N(\phi)\ra +|N(-\phi)\ra)\,,
\fe
for $0<\phi,\theta<\pi$. Including the $\mZ_2^C$ twisted sector generates another eight branes localized at the $\mZ_2^C$ fixed points (known as fractional branes \eqref{fbrane})
\ie
|D_O(\theta)_\pm\ra=&{1\over \sqrt{2}}|D(\theta)\ra \pm {1\over 2^{1/4}}|D(\theta)\rra_C\,,
\\
|N_O(\phi)_\pm\ra=&{1\over \sqrt{2}}|N(\phi)\ra \pm {1\over 2^{1/4}} |N(\phi) \rra_C\,,
\fe
with $\phi,\theta=0,\pi$ where $|D(\theta)\rra_C$ and $|N(\phi)\rra_C$  are specific combinations of the twised sector Ishibashi states whose explicit forms can be found in \cite{Oshikawa:1996dj}. At rational points on the orbifold branch $R={M\over N}\sqrt{2}$, the $|h\ra$ branes on the circle branch with $h\in SU(2)/(\mZ_M\times \mZ_N)$ leads to additional regular branes,
\ie
|h_O\ra={1\over \sqrt{2}}(|h\ra + C|h\ra  )={1\over \sqrt{2}}(|h\ra + |\sigma_1h\sigma_1\ra  )\,.
\fe
These branes have tension $g=2^{1\over 4}\sqrt{MN}$ and are always unstable. In the limit $M,N\to \infty$, they give rise to analogs of the exotic branes $|E(x)\ra$ at irrational points on the orbifold branch. 

\begin{table}[!htb]
    \centering		\renewcommand{\arraystretch}{1.8}
    \begin{tabular}{|c|c|c|c|c|}
    \hline
       &   $|D_O(\theta)\ra$ & $|N_O(\phi)\ra$ & $|D_O(\theta)_\pm \ra$ & $|N_O(\phi)_\pm\ra$   \\\hline
    $\Delta_{\rm gap}$  &  \multicolumn{4}{c|}{$\min\left({1\over 8},{1\over  R^2}\right)$} \\\hline
      $h_{\rm gap}$   &  ${R^2\over 2\pi^2}\min\left( \theta^2 , (\pi-\theta)^2\right)  $ &   ${2\over R^2\pi^2}\min\left( \phi^2 , (\pi-\phi)^2\right) $
      &
      ${R^2\over 2}$ & ${2\over R^2}$\\\hline
        $g$     & $\sqrt{2\over R}$  & ${\sqrt{R}}$ & ${1\over \sqrt{2R}}$  & ${\sqrt{R}\over 2}$\\\hline
    \end{tabular}
    \caption{Bulk and boundary gaps and brane tensions on the $c=1$ orbifold branch}
    \label{tab:orbifoldBranes}
\end{table} 

\begin{figure}[h]
    \centering
    \includegraphics[width=.6\textwidth]{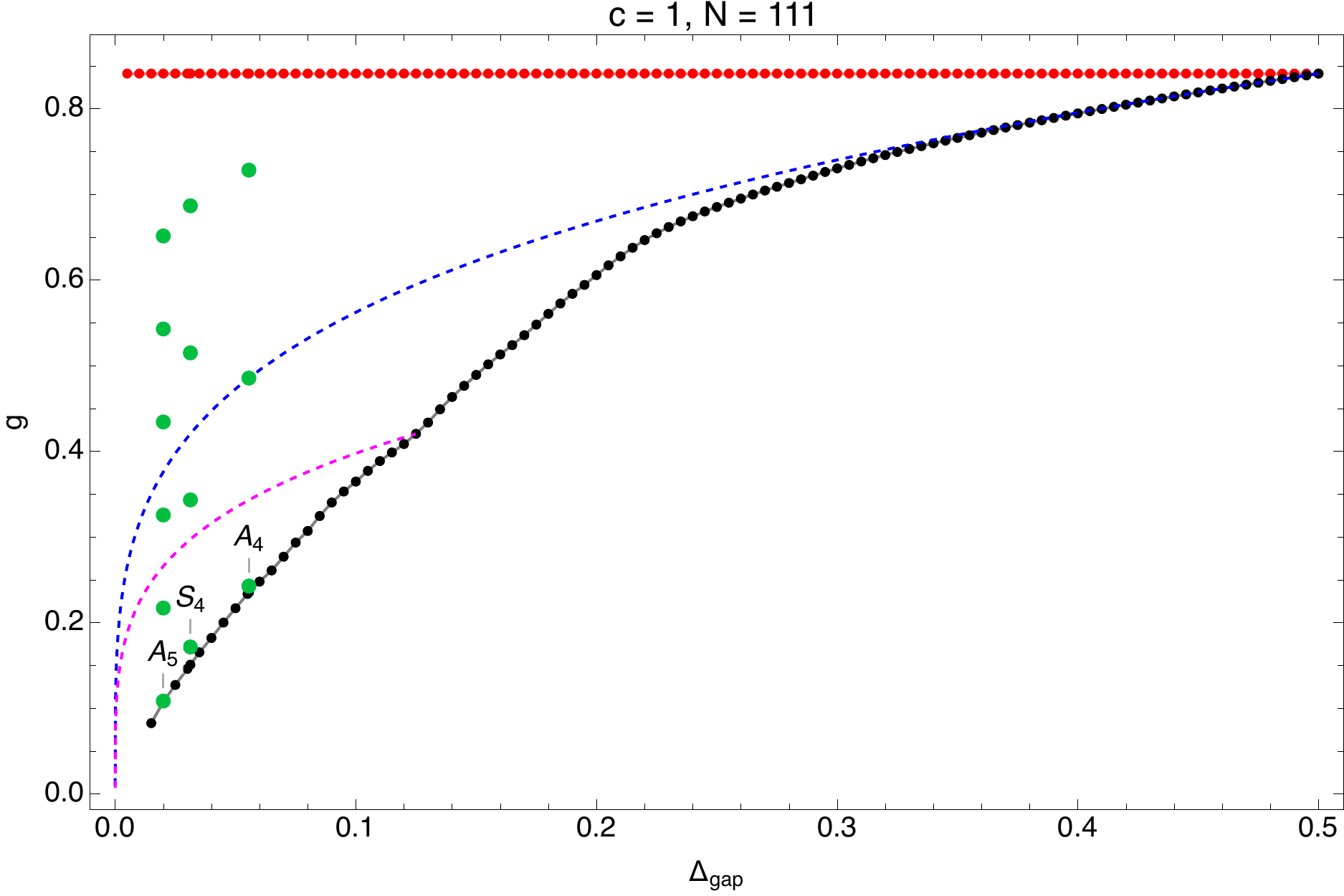}
    \caption{Upper (red) and lower (black) bounds on the $g$-function for stable boundary conditions in $c=1$ CFTs as a function of the gap in the bulk scalar sector. The dashed blue and magenta lines denote the curves $\Delta_{\rm gap}^{1/4}$ and $2^{-1/2}\Delta_{\rm gap}^{1/4}$, corresponding to the $g$-functions of the Dirichlet boundary conditions on the circle and orbifold branches, respectively. The latter truncates at $\Delta_{\rm gap} = {1\over 8}$ due to the moduli-independent gap in the twisted sector. The green points denote the $g$-function of the stable rational branes in the exceptional orbifold theories.}
    \label{fig:c1}
\end{figure}

Finally we come to the exceptional orbifolds that live at three isolated points T,O and I on the $c=1$ moduli space \cite{Ginsparg:1987eb}. They are described by the orbifold of the $SU(2)_1$ CFT by the non-anomalous non-abelian global symmetries $G=A_4,S_4,A_5 \subset SO(3)$ respectively, which are related to the exceptional simple Lie algebras $E_6,E_7,E_8$  by the Mckay correspondence. These CFTs are rational with maximal chiral algebras given by W-algebras of the type $\cW(2,9,16)$, $\cW(2,16)$ and $\cW(2,36)$ respectively \cite{Dijkgraaf:1989hb}. Their torus partition functions are defined by the diagonal modular invariant with respect to these chiral algebras. Consequently they contain elementary rational branes that are one-to-one correspondence with the chiral primaries: 21 for $G=A_4$, 28 for $G=S_4$ and 37 for $G=A_5$. Among them, the stable  rational branes are listed in Tables~\ref{tab:A4branes},~\ref{tab:S4branes} and \ref{tab:A5branes}, which are in one-to-one correspondence with irreducible representations of $\tilde G$ which is the binary lift of $G$ to $SU(2)$ (equivalently the stable rational branes are labelled by the irreducible linear and projective representations of $G$). The values of their brane tensions are listed there and also incorporated in the bootstrap bound plot in Figure~\ref{fig:c1}. 

Besides the branes that preserve the diagonal  W-algebras, the orbifold CFTs all contain regular branes \eqref{rbrane} defined by
\ie
|[h]_G\ra\equiv {1\over \sqrt{G}} \sum_{g\in G}|g h g^{-1}\ra 
\label{orbregularbrane}
\fe
which have tension $g=2^{-{1\over 4}}\sqrt{|G|}$ and live on a continous moduli space parametrized by $G$-orbits $[h]_G$ in $SU(2)$. Such regular branes are always unstable. They are elementary for generic $h\in SU(2)$ but split into elementary fractional branes at fixed orbits where $[h]_G$ has a nontrivial stablizer $G_h\subset G$. In particular, for $h\in \tilde G$, the regular branes (now labelled by conjugacy classes of $\tilde G$) break down to elementary rational branes of the exceptional orbifold CFT, which are in general unstable. For the special cases $h=1$ and $h=-1$, the resulting fractional rational branes are stable and labelled respectively by irreducible linear and projective representations of $G$, as in Tables~\ref{tab:A4branes}-\ref{tab:A5branes}. The fractional branes may admit marginal deformations and come with conformal families, in which case the brane tension stays the same while the boundary spectrum can develop a tachyon depending on the marginal parameter. In this case, we expect the fractional branes to be stable for a closed subset of the boundary conformal manifold. The above exhausts all stable branes in the exceptional orbifold CFTs that are accessible from fractionalizations and marginal deformations of the regular branes \eqref{orbregularbrane}. Because brane tensions are constant under boundary marginal deformations, the tension of all such stable branes coincides with those listed in Tables~\ref{tab:A4branes}-\ref{tab:A5branes}.

\begin{table}[!htb]
    \centering		\renewcommand{\arraystretch}{1.6}
    \begin{tabular}{|c|c|c||c|}
    \hline
       &   $|1_0\ra,|1_1\ra,|1_2  \ra$ & $|3\ra$ & $|2_0\ra,|2_1\ra,|2_2\ra$   \\\hline
    $\Delta_{\rm gap}$  &  \multicolumn{3}{c|}{$1\over 18$} \\\hline
      $h_{\rm gap}$   &  $9$ &   $1,1$
      &
      $1$  \\\hline
        $g$     & ${1\over \sqrt{12\sqrt{2}}} $  & ${3\over \sqrt{12\sqrt{2}}}$  & ${2\over \sqrt{12\sqrt{2}}}$ \\\hline
    \end{tabular}
    \caption{Stable rational branes in the $SU(2)_1/A_4$ orbifold CFT. They correspond to irreducible linear (left) and projective (right) representations of $A_4$. The label $|r_i\ra$ keeps track of the multiple inequivalent $r$-dimensional representations of $G$. For each stable brane,  we list the lowest scaling dimensions of the boundary operators, keeping track of their multiplicities.}
    \label{tab:A4branes}
\end{table}

\begin{table}[!htb]
    \centering		\renewcommand{\arraystretch}{1.6}
    \begin{tabular}{|c|c|c|c||c|c|}
    \hline
       &   $|1_0\ra,|1_1\ra$ & $|2\ra$ & $|3_0\ra,|3_1\ra$ & $|2_0\ra,|2_1\ra$ & $|4\ra $   \\\hline
    $\Delta_{\rm gap}$  &  \multicolumn{5}{c|}{$1\over 32$} \\\hline
      $h_{\rm gap}$   &  $16$ &   $9$
      &
      $1$ &
      $1$ &
      $1,1$  \\\hline
        $g$     & ${1\over \sqrt{24\sqrt{2}}} $  & ${2\over \sqrt{24\sqrt{2}}}$  & ${3\over \sqrt{24\sqrt{2}}}$ & ${2\over \sqrt{24\sqrt{2}}}$& ${4\over \sqrt{24\sqrt{2}}}$\\\hline
    \end{tabular}
    \caption{Stable rational branes in the $SU(2)_1/S_4$ orbifold CFT. The notations    are parallel to that in Table~\ref{tab:A4branes}.}
    \label{tab:S4branes}
\end{table}

\begin{table}[!htb]
    \centering		\renewcommand{\arraystretch}{1.6}
    \begin{tabular}{|c|c|c|c|c|c||c|c|c|c|}
    \hline
       &   $|1\ra$   & $|3\ra$ & $|3'\ra$ & $|4\ra$ & $|5\ra$ & $|2 \ra$ & $|2'\ra$ & $|4\ra $  & $|6\ra $   \\\hline
    $\Delta_{\rm gap}$  &  \multicolumn{9}{c|}{$1\over 50$} \\\hline
      $h_{\rm gap}$   &  $36$ &   $1$
      &
      $4$ &
      $1$ &
      $1$ & $1$ & $9$ & $1$ & $1,1$  \\\hline
        $g$     & ${1\over \sqrt{60\sqrt{2}}} $  & ${3\over \sqrt{60\sqrt{2}}} $  &
        ${3\over \sqrt{60\sqrt{2}}} $  &
        ${4\over \sqrt{60\sqrt{2}}} $  &
        ${5\over \sqrt{60\sqrt{2}}} $  &
        ${2\over \sqrt{60\sqrt{2}}} $  &
        ${2\over \sqrt{60\sqrt{2}}} $  &
        ${4\over \sqrt{60\sqrt{2}}} $  &
        ${6\over \sqrt{60\sqrt{2}}} $ \\\hline
    \end{tabular}
    \caption{Stable rational branes in the $SU(2)_1/A_5$ orbifold CFT. The notations are parallel to that in Table~\ref{tab:A4branes}.}
    \label{tab:A5branes}
\end{table} 

From Table~\ref{tab:circleBranes} and Table~\ref{tab:orbifoldBranes}, we see the bulk scalar gap $\Delta_{\rm gap}$ is sensitive to the target space modulus $R$ for the $c=1$ boson and its $\mZ_2^C$ orbifold. 
We can thus study the bounds on the $g$ function as a function of the target space modulus by treating the bulk scalar gap as a proxy for the latter. The gap $\Delta_{\rm gap}$ is maximized at the self-dual radius, $R=\sqrt{2}$, where there is an enhancement of the chiral algebra and the saturating theory is the $SU(2)_1$ WZW model. That this bound is optimal in the space of all unitary, compact $c=1$ CFTs was numerically observed in \cite{Collier:2016cls} and proven in \cite{Afkhami-Jeddi:2020ezh} by construction of an analytic functional whose action on the Virasoro characters is non-negative for values of the dimension above the gap and that has zeros on the spectrum of local operators in the saturating theory.  

We are now in a position to compute bounds on the $g$-function for stable boundary conditions in $c=1$ CFTs using the algorithm described in Section~\ref{subsec:linearProgramming}. An additional subtlety particular to $c=1$ is that in the regions where the functional must act positively, we must additionally impose positivity of the functional when acting on the degenerate characters at special values of the weights where the representation of the Virasoro algebra becomes degenerate. In particular, at weights given by $h= {n^2\over 4}$ for $n\in\mathbb{Z}_{\geq 0}$, the $c=1$ Virasoro algebra has null states at level $n+1$. The corresponding degenerate characters are given by
\begin{equation}
    \widehat\chi_{\rm deg}^{(n)}(t) = t^{1\over 4}q^{n^2\over 4}(1-q^{n+1}).
\end{equation}
All bounds computed on the $g$ function for $c=1$ have this additional positivity property incorporated.

In Figure~\ref{fig:c1} we present numerical upper and lower bounds on $g$ for stable boundary conditions as a function of the bulk gap. The resulting bounds exhibit some interesting features. We see that the upper bound is insensitive to the gap in the spectrum of bulk scalar primaries (we will return to this shortly). For sufficiently large bulk gap, the lower bound is well-approximated by the $g$-function of the stable Dirichlet boundary condition on the $S^1$ branch \cite{Elitzur:1998va} (c.f. Table \ref{tab:circleBranes})
\begin{equation}
    g_{\rm Dirichlet} = (2R)^{-\half} = \Delta_{\rm gap}^{1\over 4}.
\end{equation}

The bounds are also consistent with the stable boundary conditions on the orbifold branch, including the exceptional orbifolds, as described in Tables \ref{tab:circleBranes}-\ref{tab:A5branes}. In particular, at the maximal value of the gap in the $\mathbb{Z}_2$ orbifold branch, $\Delta_{\rm gap} = {1\over 8}$, the lower bound appears to be saturated by the stable Dirichlet boundary condition:
\begin{equation}
    \Delta_{\rm gap} = {1\over 8}: ~ g \geq 2^{-5/4}
\end{equation}
to within at least 40 digits of precision. On the other hand, in the case of the exceptional orbifolds, the lower bounds are not quite saturated by the corresponding minimal-tension rational branes. However, we find that if we feed the detailed information of the low-lying bulk spectrum into the numerics as described around equation (\ref{eq:incorporateLowLying}), then the lower bounds are indeed realized by these minimal tension rational branes. For instance, if we incorporate the bulk scalars with dimension $\Delta \leq 2$, then the lower bounds obtained at $N=127$ derivatives improve to the following
\begin{equation}
\begin{aligned}
     A_5:~ g & \geq 0.1085592604054384293742693646631371565\\
     S_4:~ g & \geq 0.1716472619922598129858564379334175563\\
     A_4:~ g & \geq 0.2427458858536617101418603137455420728,\\
\end{aligned}
\end{equation}
which should be compared to the numerical values of ${1\over \sqrt{60\sqrt{2}}}$, ${1\over \sqrt{24\sqrt{2}}}$ and ${1\over \sqrt{12\sqrt{2}}}$, respectively.

We also notice that at the maximal value of the bulk scalar gap $\Delta_{\rm gap} = \half$, corresponding to the free boson theory at the self-dual radius $R=\sqrt{2}$, the upper and lower bounds precisely coincide at
\begin{equation}
    g = 2^{-{1\over 4}},
\end{equation}
so that the $g$-function is uniquely specified for stable boundary conditions. The saturating $g$-function is realized by the cylinder partition function associated with the identity Cardy brane, which is the identity character of the $SU(2)_1$ current algebra in the open string channel
\begin{equation}
    Z^{SU(2)_1}_{\id\id}(t^{-1}) = {\Theta_{0,1}(i t)\over\eta(i t)} = {\theta_3(2it)\over \eta(it)}\,,
\end{equation}
where
\begin{equation}
    \Theta_{m,k}(i t) \equiv \sum_{n\in\mathbb{Z}}q^{k(n+{m\over 2k})^2}\,.
\end{equation}

We also notice in Figure~\ref{fig:c1} that the upper bound is completely flat, apparently independent of the assumed gap in the spectrum of bulk scalars. The reason is ultimately that the bounds must also apply to general linear combinations of solutions to the cylinder crossing equation. For example, consider the following fictitious cylinder partition function
\begin{equation}
    Z_{\rm fict}(t) = \sum_i a_i Z_{\alpha_i\alpha_i}(t)\,,
\end{equation}
where $\alpha_i$ label different stable boundary conditions (not necessarily associated with the same bulk CFT, however the bulk central charge is assumed to be the same) and the $a_i$ are positive coefficients such that $\sum_i a_i = 1$. Suppose the boundary condition $\alpha_1$ leads to the largest value of $g=g_1$. Then if we take $a_1 \gg a_{i>1}$, the $g$-function associated to $Z_{\rm fict}$ will be well approximated by $g_1$, while the gap in the bulk scalar sector will be given by the minimum of the bulk gaps among the cylinder partition functions associated to the boundary conditions $\alpha_i$. So in situations where the bulk gap can be made arbitrarily small while admitting stable conformal boundaries, the upper bound on $g$ must be insensitive to the value of the bulk gap. We emphasize that this argument is quite general and does not only apply to $c=1$, indeed we will observe this insensitivity of the upper bound to the assumed gap in the bulk spectrum in the other examples that follow. Note that integrality of the degeneracies in the boundary spectrum could in principle be used to rule out these fictitious solutions to the crossing equations, since these linear combinations do not in general preserve integrality of degeneracies of operators in the boundary channel.

In practice, we also observe  experimentally (in the present $c=1$ case and in all examples that follow) that the lower bound on the $g$-function is completely insensitive to the assumption of the gap in the boundary spectrum.
As far as we can tell, this phenomenon cannot similarly be simply explained by the fact that the bounds should also apply to linear combinations of cylinder partition functions. In cases where there is a boundary conformal manifold, the boundary gap can be tuned (in general within a finite range) without affecting the $g$-function.\footnote{This happens for example for the Dirichlet and Neumann boundaries for the $\mZ_2^C$ orbifold at $c=1$ (see Table~\ref{tab:orbifoldBranes}).}

\subsection{$c=2$ and the $SU(3)_1$ CFT }
For our next class of examples we study bounds on the $g$-function for $c=2$ CFTs. Already at $c=2$ there is no known classification of boundary states. A family of $c=2$ CFTs is provided by sigma models with $T^2$ target space, the theory of two free bosons based on an even self-dual lattice $\Lambda\in\mathbb{R}^{2,2}$. This family of CFTs, whose operator content organizes into representations of a $U(1)^2\times U(1)^2$ chiral algebra, is parameterized by a four-dimensional moduli space, labelled by two elements $\sigma,\rho$ of $\mathbb{H}/PSL(2,\mathbb{Z})$, which are respectively the complex structure and the complexified Kahler structure of the target $T^2$. These parameters can be written in terms of the metric $G$ and $B$-field flux of the target $T^2$ as
\begin{equation}
\begin{aligned}
     \rho &= B + i \sqrt{\det G}\\
     \sigma &= {G_{12}\over G_{11}} + i{\sqrt{\det G}\over G_{11}}.
\end{aligned}    
\end{equation}
If we work in a $T$-duality frame where $\sigma,\rho$ lie in the standard fundamental domain of $\mathbb{H}/PSL(2,\mathbb{Z})$, then the gap in the spectrum of primary operators is given in terms of the target space moduli as
\begin{equation}
    \Delta_{\rm gap} = {1\over 2\im\rho\im\sigma}.
\end{equation}
Like the $c=1$ case, this is maximized at a point of enhanced symmetry, namely the $\mathbb{Z}_3$ point where $\sigma = \rho = e^{\pi i\over 3}$. Here, $\Delta_{\rm gap} = {2\over 3}$, and the saturating theory is the $SU(3)_1$ WZW model. This turns out to be the maximal value of the gap in the space of all unitary, compact $c=2$ CFTs \cite{Collier:2016cls}.

\begin{figure}[h]
    \centering
    \includegraphics[width=.6\textwidth]{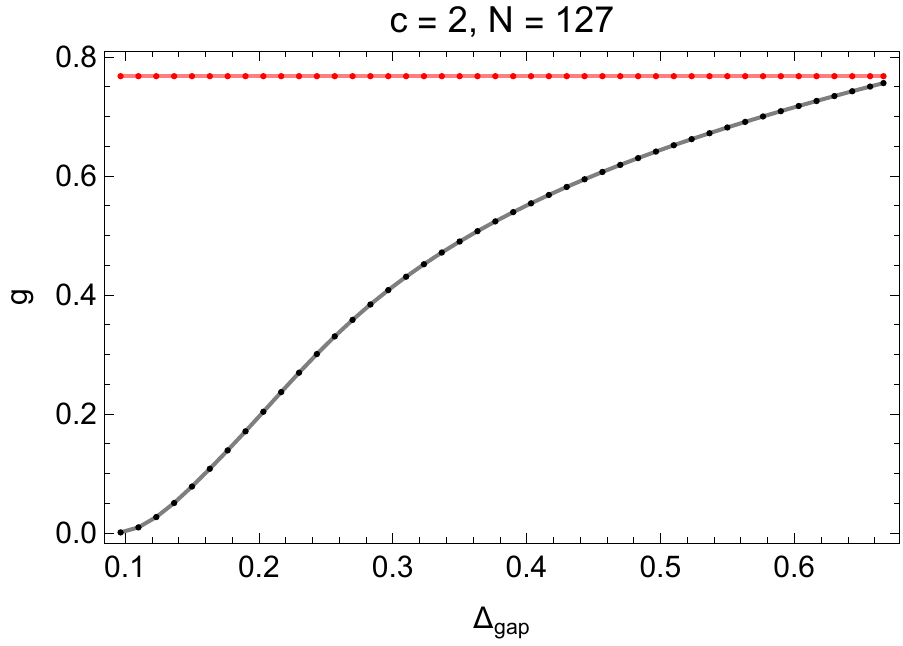}
    \caption{Upper (red) and lower (black) bounds on the $g$-function for stable boundary conditions in $c=2$ CFTs as a function of the gap in the bulk scalar sector.}
    \label{fig:c2}
\end{figure}

In Figure~\ref{fig:c2} we plot bounds on the $g$-function associated with stable boundary conditions for $c=2$ CFTs as a function of the gap in the spectrum of bulk primaries. Unlike the case of $c=1$, the upper and lower bounds are not coincident at the maximal value of the gap. In order to explore whether the $g$-function is uniquely specified for stable boundary conditions in the theory with the maximal gap, we incorporate the information of a number of low-lying bulk scalar primaries into the crossing equation as described in Section~\ref{subsec:linearProgramming}. The bulk scalars come in two towers of states spaced by even integers
\begin{equation}
    \Delta \in 
    \{2\mathbb{Z}_{\geq 0}\} \cup \{2/3 + 2\mathbb{Z}_{\geq 0}\}.
\end{equation}

\begin{table}[h]
    \centering
    \begin{tabular}{c|c|c}
        \# of bulk scalars incorporated & lower bound & upper bound\\
        \hline\hline
        0 & 0.7560986262 & 0.7677246687 \\
        \hline 
        2 & 0.7596267328 & 0.7598741664 \\
        \hline
        10 & 0.7598352810 & 0.7598356865 \\
        \hline
        50 & 0.7598356856 & 0.7598356857 \\
        \hline
    \end{tabular}
    \caption{Upper and lower bounds on the $g$ function for stable boundary conditions in the $SU(3)_1$ WZW model at $N=127$ derivative order with an increasing number of low-lying bulk scalars incorporated. The bounds appear to be convergent upon $3^{-{1\over 4}} = 0.75983568565\ldots$}
    \label{tab:c2}
\end{table}

In Table \ref{tab:c2} we present the resulting upper and lower bounds on $g$. As the number of low-lying bulk scalars fed into the crossing equation is increased, the bounds appear convergent upon
\ie 
g=3^{-{1\over 4}}
\fe
to within a high degree of numerical precision. This is precisely the boundary entropy associated with the identity rational brane $|B_\id\ra$ of the $SU(3)_1$ WZW model, with the saturating cylinder partition function given by the identity character of the $SU(3)_1$ current algebra
\ie
Z^{SU(3)_1}_{\id\id}(t^{-1})={\Theta_{0,3}(it)\Theta_{0,1}(it)+\Theta_{3,3}(it)\Theta_{1,1}(it)\over \eta(it)^2}
\fe
The same properties hold for the boundary conditions obtained from fusion with the ${SU(3)\times SU(3)\over \mZ_3}\rtimes \mZ_2$ invertible global symmetries on $|B_\id\ra$. They include in particular the other two   rational branes $|B_\varphi\ra $ and $|B_{\bar \varphi}\ra$ corresponding to the two $h={2\over 3}$ chiral primaries $\varphi,\bar\varphi$ which are generated from the $\mZ_3$ center symmetry acting on $|B_\id\ra$.

\begin{figure}[h]
    \centering
    \begin{subfigure}[t]{0.49\textwidth}
    \includegraphics[width=.95\textwidth]{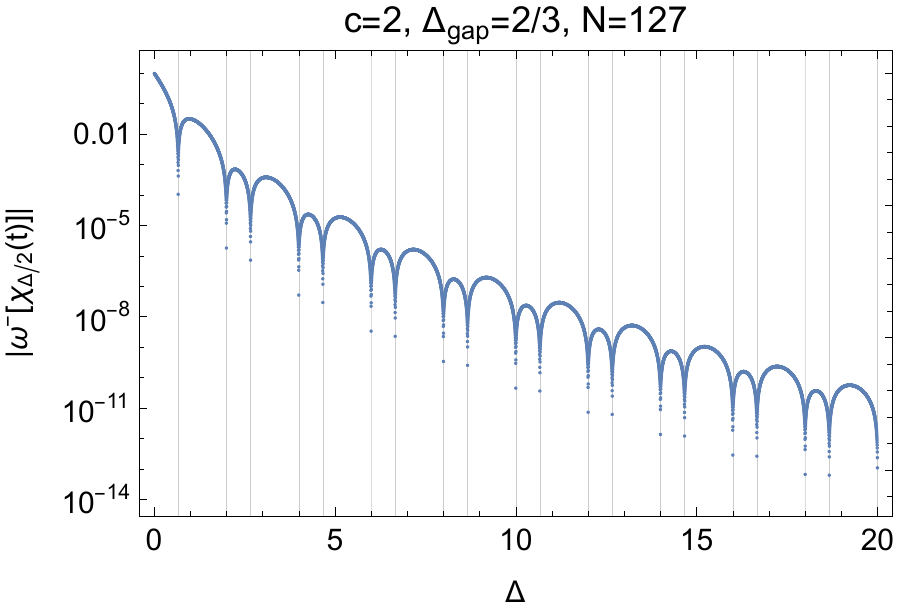}
    \caption{Bulk spectrum}
    \end{subfigure}
    \begin{subfigure}[t]{0.49\textwidth}
    \includegraphics[width=.95\textwidth]{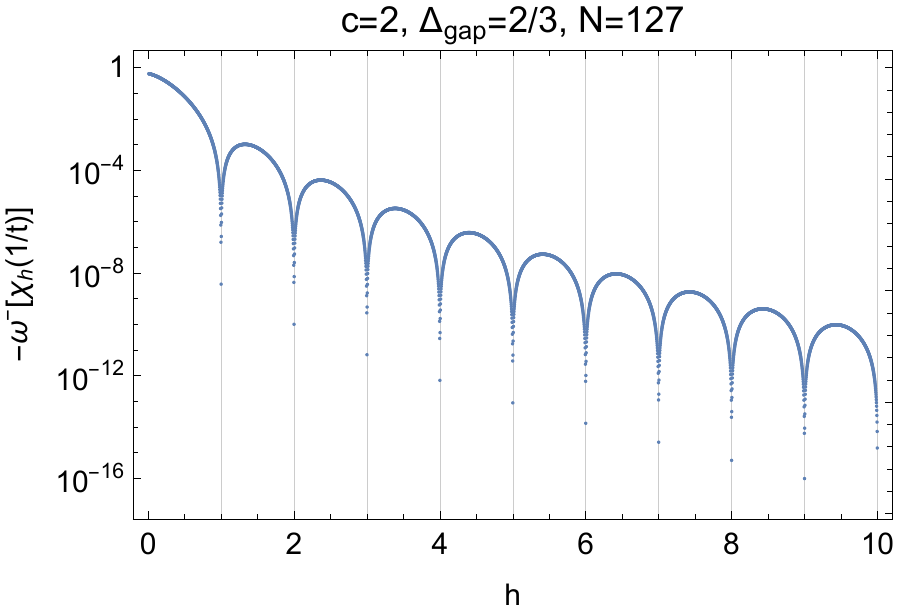}
    \caption{Boundary spectrum}
    \end{subfigure}
    \caption{Action of the functional $\omega^-$ that computes the lower bounds on $g$ for the $SU(3)_1$ WZW model on the bulk and boundary characters. We plot the absolute value of the functional acting on bulk characters because when we feed in the information of the low-lying scalars into the numerics, the resulting functional has single rather than double zeros at these dimensions.}
    \label{fig:c2Functional}
\end{figure}

Indeed, we see that the spectrum of this cylinder partition function is realized by the action of the corresponding numerical extremal functional acting on the bulk and boundary characters, as plotted in Figure~\ref{fig:c2Functional}. In particular we notice that the boundary spectrum is supported on positive integer weights, consistent with the fact that the saturating cylinder partition function is the $SU(3)_1$ vacuum character in the open string channel. 

A novel feature of this functional compared to extremal functionals that have been studied previously in modular bootstrap is that the bulk spectrum is supported on multiple towers of states with dimensions separated by even integers. It would be very interesting to construct this functional analytically using techniques of \cite{Hartman:2019pcd}.

\subsection{$(G_2)_1$ CFT}
The next example found to saturate modular bootstrap bounds on the gap that we will consider is the $(G_2)_1$ WZW model, with central charge $c = {14\over 5}$. The gap in the spectrum of primary operators in this theory is
\begin{equation}
    \Delta_{\rm gap} = {4\over 5},
\end{equation}
which was found to be optimal in the space of unitary, compact $c={14\over 5}$ CFTs in \cite{Collier:2016cls}. 

As in the case of $c=2$, it appears from the numerical bounds on $g$ that the information of the bulk gap alone is insufficient to uniquely specify the boundary entropy for stable boundary conditions in the $(G_2)_1$ WZW model. However, as shown in Table \ref{tab:g2}, when we incorporate the low-lying scalars with dimensions
\begin{equation}
\Delta \in \{2\mathbb{Z}_{\geq 0}\} \cup \{4/5 + 2\mathbb{Z}_{\geq 0}\}    
\end{equation}
into the cylinder crossing equation, the resulting bounds on $g$ are increasingly convergent upon 
\begin{equation}
    g = \left(\half - {\sqrt{5}\over 10}\right)^{1\over 4}\,.
    \label{G2num}
\end{equation}

\begin{table}[h]
    \centering
    \begin{tabular}{c|c|c}
         \# of bulk scalars incorporated & lower bound & upper bound\\
        \hline\hline
        0 & 0.7152115481 & 0.7409519929 \\
        \hline 
        2 & 0.7241441983 & 0.7252206408 \\
        \hline
        10 & 0.7250670675 & 0.7250731872 \\
        \hline
        50 & 0.7250731746 & 0.7250731771 \\
        \hline
    \end{tabular}
    \caption{Upper and lower bounds on the $g$ function for stable boundary conditions in the $(G_2)_1$ WZW model at $N=127$ derivative order with an increasing number of low-lying bulk scalars incorporated. The bounds appear to be convergent upon $\left(\half - {\sqrt{5}\over 10}\right)^{1\over 4} = 0.72507317708\ldots$}
    \label{tab:g2}
\end{table}

The $(G_2)_1$ WZW model has two rational branes $|B_\id\ra$ and $|B_\varphi\ra$ corresponding to the identity and the $h={2\over 5}$ chiral primary $\varphi$.\footnote{The $(G_2)_1$ CFT has a fusion category symmetry described by the Fibonacci category with a single simple TDL $W$ satisfying the fusion rule $W^2=1+W$.The quantum dimension of $W$ is $\la W\ra={\sqrt{5}+1\over 2}$. The two rational branes are related by fusion as $|B_\varphi \ra=W|B_\id\ra$. }
Their $g$-functions are given by
\ie
g_{\id}=\left(\half - {\sqrt{5}\over 10}\right)^{1\over 4}\,,\quad 
g_\varphi=\left (1 + {2\over \sqrt{5}}\right)^{1\over 4}\,.
\fe
We see the former precisely coincide with our bootstrap bound \eqref{G2num}. Indeed the second boundary state $|B_\varphi\ra$ is not stable, as it contains a $h={2\over 5}$ relevant operator in the open-string spectrum. 

\begin{figure}[h]
    \centering
    \begin{subfigure}[t]{.49\textwidth}
    \includegraphics[width=.95\textwidth]{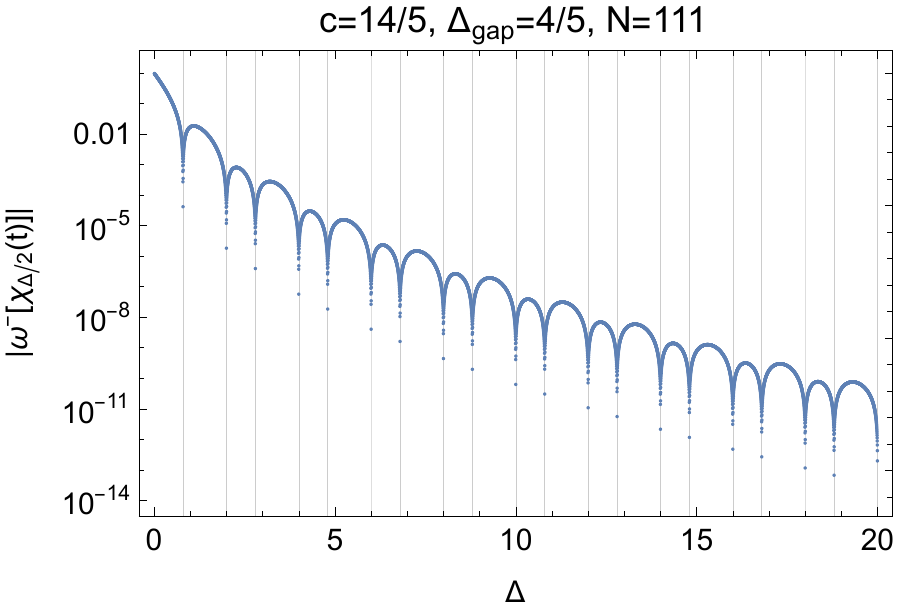}
    \caption{Bulk spectrum}
    \end{subfigure}
    \begin{subfigure}[t]{.49\textwidth}
    \includegraphics[width=.95\textwidth]{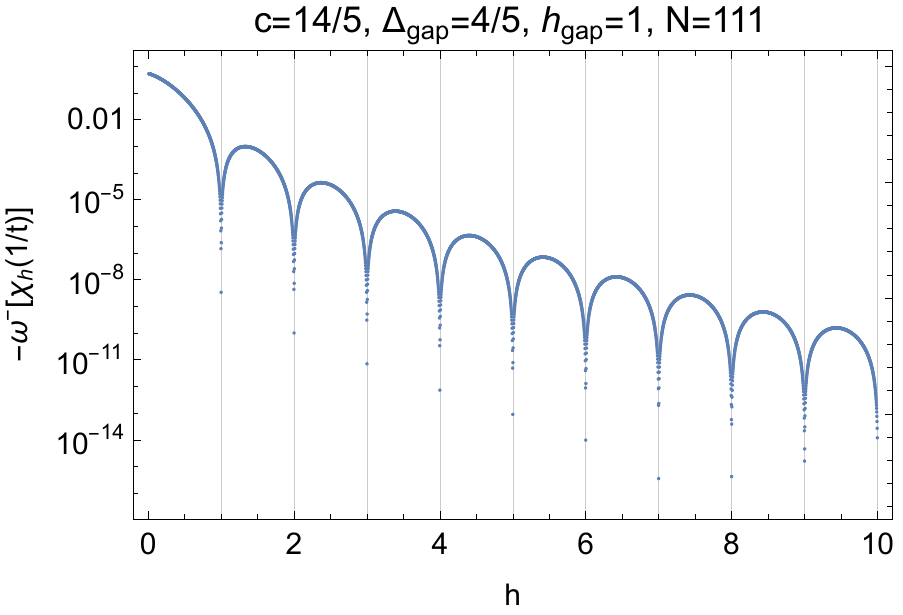}
    \caption{Boundary spectrum}
    \end{subfigure}
    \caption{Action of the functional $\omega^-$ that computes the lower bounds on $g$ for the $(G_2)_1$ WZW model on the bulk and boundary characters.}
    \label{fig:g2Functional}
\end{figure}

We can further compare the entire spectrum obtained from the optimal functional with the cylinder parition function for the stable brane $|B_\id\ra$.
The explicit form of the latter can be obtained using the following isomorphism \cite{Sugiyama:2001qh}
\ie
(G_2)_1\cong {SU(3)_1\times \text{three-state Potts}\over \mZ_3}\,.
\fe
Then the saturating cylinder partition function is given by the identity character of the $(G_2)_1$ chiral algebra, which can be written as
\ie
Z^{(G_2)_1}_{\id\id}(t^{-1})
=\chi_0^{\rm Potts}(t)\chi_{\bf 0}^{SU(3)_1}(t)+ \chi_{2/3}^{\rm Potts}(t)\chi_{\bf 3}^{SU(3)_1}(t)+ \chi_{2/3}^{\rm Potts}(t)\chi_{\bf \bar 3}^{SU(3)_1}(t)\,,
\fe
where $\chi_h^{\rm Potts}$ are W-algebra characters for primaries of weight $h$ in the three-state Potts CFT,
\ie
\chi_0^{\rm Potts}(t)=&
{q^{1\over 20}\over \eta(\tau)^2}\sum_{r,s=0}^\infty 
(-1)^{r+s} q^{{r(r+1)\over2}+{s(s+1)\over2}+4rs}
\left(
1-q^{4(1+r+s)}
\right)\,,
\\
\chi_{2/3}^{\rm Potts}(t)=&
{q^{43\over 60}\over \eta(\tau)^2}\sum_{r,s=0}^\infty 
(-1)^{r+s} q^{{r(r+1)\over2}+{s(s+1)\over2}+4rs}
\left(
q^{2r+s}-q^{1+2r+3s)}
\right)\,.
\fe
and $\chi_{\bf r}^{SU(3)_1}$ are $SU(3)_1$ Kac-Moody characters,
\ie
\chi_{\bf 0}^{SU(3)_1}(t)=&{\Theta_{0,3}(it)\Theta_{0,1}(it)+\Theta_{3,3}(it)\Theta_{1,1}(it)\over \eta(it)^2}\,,
\\
\chi_{\bf 3}^{SU(3)_1}(t)=\chi_{\bf \bar 3}^{SU(3)_1}(t)=&{\Theta_{2,3}(it)\Theta_{0,1}(it)+\Theta_{1,3}(it)\Theta_{1,1}(it)\over \eta(it)^2}\,.
\fe
Indeed, the spectrum of this saturating cylinder partition function is explicitly realized by the action of the numerical extremal functional on the bulk and boundary characters, as plotted in Figure~\ref{fig:g2Functional}. 

\subsection{$Spin(8)_1$ CFT}
In \cite{Collier:2016cls}, a kink was found in the curve describing the upper bound on the gap in the spectrum of Virasoro primaries at $c=4$, where the optimal bound on the gap was found to be $\Delta_{\rm gap} = 1$. This bound is saturated by the $Spin(8)_1$ WZW model.\footnote{This is also sometimes written as the $SO(8)_1$ WZW model in the literature. In general the $Spin(n)_k$ and $SO(n)_k$ WZW models are related by a $\mZ_2$ orbifold
\ie
SO(n)_k={Spin(n)_k/\mZ_2}\,,
\fe
where the $\mZ_2\subset Z(Spin(n))$ is the kernel of the projection from the double cover $Spin(n)$ to $SO(n)$. For $k=1$, the WZW model is self-dual under this $\mZ_2$ gauging,
\ie
SO(n)_1 \cong Spin(n)_1.
\fe
} Once again, we find that the information of the bulk gap is insufficient to pin down the $g$-function associated to stable boundary conditions in this CFT. By incorporating the information of low-lying bulk scalars with
\begin{equation}
    \Delta\in\mathbb{Z}_{\geq 0}
\end{equation}
into the cylinder crossing equation, the bounds on $g$ appear to be increasingly convergent upon
\ie 
g={1\over \sqrt{2}}.
\fe
See Table \ref{tab:c4} for the numerical bounds.

\begin{table}[h]
    \centering
    \begin{tabular}{c|c|c}
        \# of bulk scalars incorporated & lower bound & upper bound\\
        \hline\hline
        0 & 0.6801484630 & 0.7368753088 \\
        \hline
        2 & 0.7021195531 & 0.7077424981 \\
        \hline
        10 & 0.7069544116 & 0.7071069574 \\
        \hline
        50 & 0.7071058400 & 0.7071067812 \\
        \hline
        100 & 0.7071058572 & 0.7071067812 \\
        \hline
    \end{tabular}
    \caption{Upper and lower bounds on the $g$ function for stable boundary conditions in the $Spin(8)_1$ WZW model at $N=127$ derivative order with an increasing number of low-lying bulk scalars incorporated. The bounds appear to be convergent upon ${1\over \sqrt{2}} = 0.7071067811865\ldots$.}
    \label{tab:c4}
\end{table}

Yet again, we find that the saturating value for the $g$-function of stable boundary conditions is saturated by the identity rational brane $|B_\id\ra$ of the $Spin(8)_1$ WZW model. The corresponding cylinder partition function is given by the vacuum character of the $Spin(8)_1$ chiral algebra
\ie
Z^{Spin(8)_1}_{\id\id}(t^{-1})={\theta_3(it)^4+\theta_4(it)^4\over 2\eta(it)^4}.
\fe
The $Spin(8)_1$ CFT has three other rational branes $|\cB_{\varphi_i}\ra$ with $i=1,2,3$ associated with the three chiral primaries $\varphi_i$ of weight $h={1\over 2}$ that are permuted under the $S_3$ triality symmetry. In addition, they are also related by fusing the three $\mZ_2$ generators of the $\mZ_2\times \mZ_2$ center symmetry with $|B_\id\ra$. 
Consequently, all rational branes of the $Spin(8)_1$ CFT have the same $g$-function and cylinder partition function.

The $Spin(8)_1$ is self-dual under gauging the $\mZ_2\times \mZ_2$ center symmetry, thus it also has duality defects $\cN_i$ for each $\mZ_2$ subgroup generated by $\eta_i$ satisfying
\ie
\cN^2_i=1+\eta_i\,,~~
\cN_i \eta_i=\eta_i\cN_i =\cN_i\,,~~
\eta_i^2=1\,,~~ \eta_1\eta_2=\eta_3\,.
\fe
They give rise to conformal boundaries $|\cN_i\ra \equiv \cN_i |B_\id\ra $ by fusion. As reviewed in Section~\ref{sec:bdyfromTDL}, these boundary states has $g_{\cN_i}=\la \cN_i\ra g_\id=1$ and cylinder partition function (as in \eqref{PFfromTDL}),
\ie
Z^{Spin(8)_1}_{\cN_i\cN_i}(t^{-1})=Z^{Spin(8)_1}_{\id \id}(t^{-1})+Z^{Spin(8)_1}_{\id \varphi_i}(t^{-1})\,.
\fe
The second term on the RHS contains operators of dimension $h={1\over 2}$ which follows from the $Spin(8)_1$ fusion rule\footnote{They are the lowest weight boundary changing operators between boundaries $|B_\id\ra$ and $|B_{\varphi_i}\ra$.} and thus $|\cN_i\ra$ are elementary but unstable, in agreement with our bootstrap bound.

\subsection{$c=8$ and the $(E_8)_1$ CFT}\label{subsec:c8}
\begin{figure}
    \centering
    \includegraphics[width=.6\textwidth]{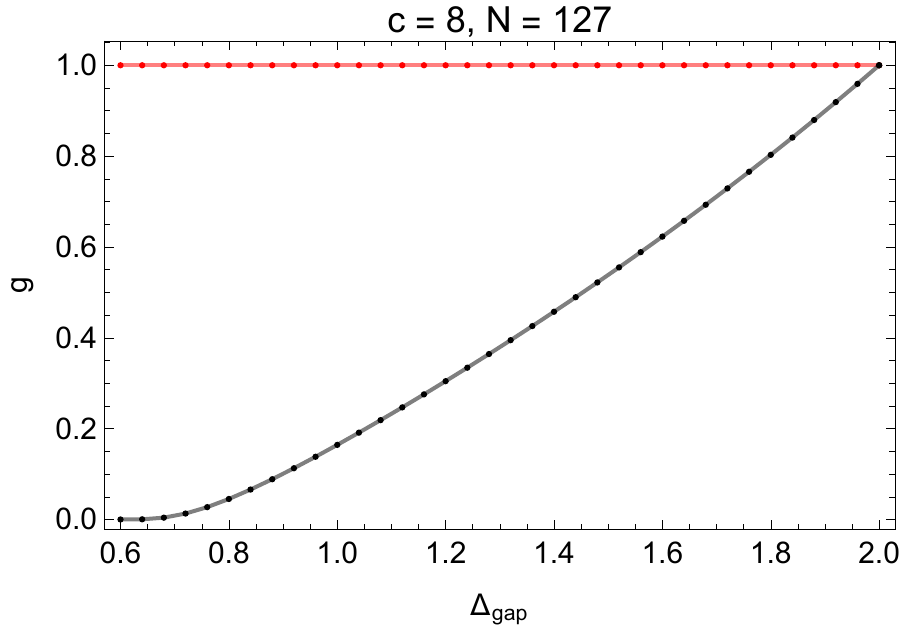}
    \caption{Upper (red) and lower (black) bounds on the $g$-function for stable boundary conditions in $c=8$ CFTs.}
    \label{fig:c8}
\end{figure}

Another kink was identified on the modular bootstrap bounding curve of \cite{Collier:2016cls}, this time for the bound on the spectrum of scalar Virasoro primaries (which is weaker than the overall gap bound for $c > 4$), at $c=8$, with $\Delta_{\rm gap}^{\ell = 0} = 2$. This bound is saturated by the $(E_8)_1$ WZW model. In Figure~\ref{fig:c8} we plot the bounds on the $g$-function for stable boundary conditions as a function of the bulk scalar gap and again find that they are nearly coincident at the maximal value of the bulk gap. Indeed, at $\Delta_{\rm gap} = 2$, we find the following upper and lower bounds on $g$
\begin{equation}
    N=127:~0.9999998668 < g < 1.0000001332.
\end{equation}
Thus it is tempting to conjecture that at this value of the bulk gap, $g$ is uniquely specified for stable conformal boundaries to be
\begin{equation}
    g = 1
\end{equation}
for stable boundary conditions in the $(E_8)_1$ WZW model. Indeed, the lower bound is insensitive to the value of the gap in the spectrum of boundary primaries, so this suggests that \emph{all} boundary conditions (not necessarily stable) for the $(E_8)_1$ WZW model satisfy $g \geq 1$. In the next section we will prove that this is the case by constructing analytically the extremal functional that proves optimality of this bound.

Once again, this value of $g$ is realized by the identity Cardy brane in the $(E_8)_1$ WZW model, with the cylinder partition function given by the identity character of the $(E_8)_1$ chiral algebra
\ie
Z^{(E_8)_1}_{\id\id}(t^{-1})={\theta_2(it)^8+\theta_3(it)^8+\theta_4(it)^8\over 2\eta(it)^8}.
\fe

There are some special features of this solution to the cylinder crossing equation that will allow us to prove analytically that it saturates this bound in Section~\ref{sec:c8Saturation} using the analytic functionals discussed in Section~\ref{sec:analyticFunctionals}. In particular the spectrum of bulk scalar primaries has support only on a single tower of  scaling dimensions spaced by even integers. Moreover, the spectrum of conformal weights of bulk primaries is simply the bulk spectrum of dimensions rescaled by a factor of two 
\begin{equation}
    \Delta \in 2 \mathbb{Z}_{\geq 0}, ~ h\in \mathbb{Z}_{\geq 0}
\end{equation}
so that the cylinder partition function is invariant under the modular $S$ transformation.
In particular, the bulk gap is twice the boundary gap implied by the stability assumption. In fact, this is an example of the more general phenomenon that in scenarios where the boundary gap assumed to compute the upper bound and the bulk gap assumed to compute the lower bound on $g$ are precisely correlated $h_{\rm gap} = {\Delta_{\rm gap}\over 2}$, then the resulting upper and lower bounds are inversely related.\footnote{To see this, note that the upper bound on $g$ is computed by a functional $\omega^+$ such that
\begin{equation}\label{eq:gUpperBound}
    g^2 \leq g_+ = \min\left( {\omega^+[\widehat\chi_{\rm vac}(t^{-1})]\over\omega^+[\widehat\chi_{\rm vac}(t)]}\right), \quad \omega^+[\widehat\chi_{\rm vac}(t)] > 0, \quad \omega^+[\widehat\chi_{h}(t^{-1})] \geq 0, \quad \omega^+[\widehat\chi_{\Delta\over 2}(t)]\geq 0,
\end{equation}
for $h\geq h_{\rm gap}$, $\Delta \geq 0$. Similarly, the lower bound is computed by $\omega^-$ such that
\begin{equation}\label{eq:gLowerBound}
    g^2 \geq g_- = \max\left( {-\omega^-[\widehat\chi_{\rm vac}(t^{-1})]\over -\omega^-[\widehat\chi_{\rm vac}(t)]}\right), \quad -\omega^+[\widehat\chi_{\rm vac}(t)] > 0, \quad \omega^-[\widehat\chi_{h}(t^{-1})] \geq 0, \quad \omega^-[\widehat\chi_{\Delta\over 2}(t)]\geq 0,
\end{equation}
for $\Delta \geq \Delta_{\rm gap}$, $h\geq 0$. In the case that the assumed gaps in the bulk and boundary sectors are identical $h_{\rm gap} = {\Delta_{\rm gap}\over 2}$, then the roles of the bulk and boundary spectra in defining the functional $\omega^-$ in (\ref{eq:gLowerBound}) are precisely exchanged compared to the case of $\omega^+$ in (\ref{eq:gUpperBound}), and so the resulting upper and lower bounds are inversely related
\begin{equation}
    g_- = (g_+)^{-1}.
\end{equation}
Clearly, if these assumptions are to be mutually compatible then it must be that $g_+ \geq 1$.\label{longFootnote}
} Here we see they coincide at $g=1$.

\subsection{$(F_4)_1$, $(E_6)_1$ and $(E_7)_1$ CFTs}
For the sake of completeness, here we study the upper and lower bounds on the $g$-function for the remainder of the RCFTs with WZW models at level one with global symmetry given by the remainder of the Deligne-Cvitanovi\'c exceptional series. In \cite{Bae:2017kcl} it was shown that these theories also saturate modular bootstrap bounds on the gap in the spectrum of twists, so one may wonder whether the $g$-function is also pinned down for stable boundary conditions in these theories as well. Here we will present preliminary numerical evidence that in these theories the boundary entropy for stable boundary conditions is once again given by that of the identity Cardy brane of the WZW model. We should caution however that although the upper bound converges convincingly to these special values of $g$, the convergence of the lower bound as the derivative order of the numerical functional is increased is much slower than in the previously considered cases and so we cannot convincingly conclude that $g$ is indeed pinned down for stable boundary conditions in these theories.

Moreover, we will find that the saturating value of the $g$-function is given by one of the saturating values already discovered in our study of boundary conditions in the RCFTs previously considered in this section. The reason is due to a factorization property of the $(E_8)_1$ vertex operator algebra (VOA) into products of these other VOAs in the Deligne-Cvitanovi\'c exceptional series pairwise. A consequence of this factorization property is that the modular $S$ matrices are identical (recall that the RCFTs considered in this section for which $g$ is pinned down by the annulus crossing equation are all two-character diagonal modular invariants), so that in particular the boundary entropies associated with the Cardy branes are the same.

In Table \ref{tab:f4e6e7} we present numerical upper and lower bounds on the $g$-function for stable boundary conditions in the $(F_4)_1$, $(E_6)_1$ and $(E_7)_1$ WZW models obtained from functionals with an increasing number of derivatives acting on the cylinder crossing equation. The upper bounds are convincingly convergent upon the unique $g$-functions for stable boundary conditions in the $(G_2)_1$, $SU(3)_1$ and $SU(2)_1$ WZW models (which, recall, is that of the identity Cardy brane), respectively, due to the previously mentioned coincidence of their modular $S$ matrices. The lower bounds are potentially also convergent upon these values, but the convergence of the lower bounds as the derivative order of the numerical functional is increased is slower than in the previously considered cases. We present the bounds as a function of derivative order of the functional in order to illustrate this slow convergence.

\begin{table}[h]
    \centering
    \begin{subtable}[t]{0.45\textwidth}
    \begin{tabular}{c|c|c}
        $N$ & lower bound & upper bound\\
        \hline\hline
        31 & 0.7166237724 & 0.7251179795\\
        \hline
        63 & 0.7236008653 & 0.7250733783\\
        \hline
        95 & 0.7246845577 & 0.7250731800\\
        \hline
        127& 0.7249485217 & 0.7250731772\\
        \hline
    \end{tabular}
    \caption{$(F_4)_1$}
    \end{subtable}
    \begin{subtable}[t]{0.45\textwidth}
    \begin{tabular}{c|c|c}
        $N$ & lower bound & upper bound  \\
        \hline\hline
        31 & 0.7354975594 & 0.7599988478 \\
        \hline
        63 & 0.7516884345 & 0.7598370623 \\
        \hline
        95 & 0.7561788534 & 0.7598357206 \\
        \hline
        127& 0.7579342364 & 0.7598356872\\
        \hline
    \end{tabular}
    \caption{$(E_6)_1$}
    \end{subtable}\\
    \par\bigskip
    \begin{subtable}[t]{0.45\textwidth}
    \begin{tabular}{c|c|c}
        $N$ & lower bound & upper bound  \\
        \hline\hline
        31 & 0.7645274707 & 0.8416243547\\
        \hline
        63 & 0.7977452405 & 0.8409098737\\
        \hline
        95 & 0.8105293308 & 0.8408969900\\
        \hline
        127& 0.8174066935 & 0.8408964559\\
        \hline
    \end{tabular}
    \caption{$(E_7)_1$}
    \end{subtable}
    \caption{Upper and lower bounds on the $g$-function for stable boundary conditions in the $(F_4)_1$ (top left), $(E_6)_1$ (top right), and $(E_7)_1$ (bottom) WZW models obtained at various derivative orders $N$, with the information of the first 100 bulk scalars incorporated into the crossing equation. The bounds are potentially convergent upon $(\half -{\sqrt{5}\over 10})^{1/4}$, $3^{-1/4}$ and $2^{-1/4}$, which are also the unique $g$-functions for stable boundaries in the $(G_2)_1$, $SU(3)_1$ and $SU(2)_1$ WZW models, respectively.}
    \label{tab:f4e6e7}
\end{table}

\subsection{$|\text{Monster}|^2$ CFT}
Here we will explore bounds on conformal boundary conditions in the holomorphically factorized $|\text{Monster}|^2$ CFT whose Hilbert space is given by that of the tensor product of two chiral Monster CFTs. The chiral Monster CFT is the $\mathbb{Z}_2$ orbifold of the $c=24$ chiral CFT associated to the Leech lattice $\Lambda_L$, with parition function
\ie
Z_{\rm Monster}(\tau)=J(\tau)\equiv j(\tau)-744\,.
\fe
Relatedly, the $|\text{Monster}|^2$ CFT is the $\mZ_2\times \mZ_2$ orbifold of the sigma model on the Leech torus $\mR^{24}/\Lambda_L$ with a specific $B$-field \cite{Dixon:1988qd}, such that the partition function factorizes,
\ie
Z_{|\rm Monster|^2}(\tau,\bar \tau )=|J(\tau)|^2\,.
\fe
Although neither the Monster nor $|\rm Monster|^2$ to the best of our knowledge saturate any bootstrap bounds on gaps in the operator spectrum (for CFTs that are not necessarily holomorphic or holomorphically factorized), we will see that, similarly to the case of the $(E_8)_1$ WZW model, it will serve as an interesting laboratory for the development of analytic functionals that will allow us to prove rigorous and optimal bounds on the admissable boundary conditions. Since conformal boundaries in the $|\text{Monster}|^2$ CFT are in one-to-one correspondence with conformal defects (interfaces) in the chiral Monster CFT by the folding trick, the bounds we derive will apply automatically to both cases.

The $|\text{Monster}|^2$ CFT has a global symmetry generated by two copies of the Monster group ${\bf M}$ acting separately on the holomorphic and anti-holomorphic sides. Since the CFT has a single conformal block (i.e. trivial MTC) with respect to the maximal chiral algebra, namely the Monster VOA $V^\natural$, there is a single rational brane which is nothing but the identity brane $|\id\ra$ that preserves the diagonal $V^\natural$ has annulus partition function
\ie
Z_{\id \id}^{|\rm Monster|^2}(t)=J(it)\,,
\fe
and brane tension $g=1$.
 The Monster symmetries $(g_L,g_R)$ with $g_L,g_R\in {\bf M}$ rotates the identity brane by \cite{Craps:2002rw}
\ie
(g_L,g_R) |\id  \ra =   |g_L g_R^{-1} \ra\,,
\fe
to produce a class of branes $|g\ra$ that are in one-to-one correspondence with group elements  $g\in {\bf M}$.\footnote{The $|g\ra$ branes correspond to the $g$ symmetry defects in the chiral Monster CFT after unfolding. In particular, the identity brane correspnds to the transparent (trivial) defect.} The original rational brane $|\id  \ra$ corresponds to the trivial group element and is invariant under the diagonal ${\bf M}$ generated by $(g,g)$. Because they are related by an invertible symmetry to the identity brane $|\id  \ra$, they are guaranteed to have the same boundary operator spectrum and $g=1$.\footnote{The $|g\ra$ boundaries arise from the fusion product of the $g$ symmetry defect with the identity brane $|\id \ra$.
On the cylinder with identical boundaries given by $|g\ra$, the $g$ defects can be pulled off from the boundaries and annihilate each other, producing the same partition function with $|\id \ra$ as boundaries. This would be different when one considers a non-invertible topological defect $\cL$ which can fuse with $|\id \ra$ to produce a new boundary state with a different open string spectrum.} Moreover the cylinder partition functions between different $|g\ra$ boundaries are give by \cite{Craps:2002rw},
\ie
Z^{|\rm Monster|^2}_{g_1 g_2}(t)=T_{g_1^{-1}g_2}(it)
\label{cylMT}
\fe
where $T_g(\tau)$ denotes the Mckay-Thompson series which has an integral Fourier expansion
\ie
T_g(\tau)={1\over q}+\sum_{n=1}^\infty a_n q^n\,,\quad a_n\in \mZ 
\,,
\label{MTs}
\fe
and concides with $J(\tau)$ for $g=1$. They plays a important role in the Monster moonshine. 

Athough there is no complete classification of conformal boundaries in the $|{\rm Monster}|^2$ CFT, it was proven in \cite{Craps:2002rw} that the $|g\ra$ branes form a complete set of elementary branes that preserve the ${\bf M}$-invariant subalgebra $\cW \subset V^\natural$.\footnote{The first nontrivial generator of $\cW$ comes at spin 12.} Furthermore, thanks to the orbifold description $(\mR^{24}/\Lambda_L)/(\mZ_2\times \mZ_2)$, one can identify D-branes from the $(\mR^{24}/\Lambda_L)$ CFT that survive the orbifold. As usual for the orbifold constructions, such branes fall into two classes, the regular (bulk) branes that have continous moduli and the fractional branes stuck at the $2^{24}$ fixed points. Here the geometric elementary D-branes (D$p$-branes for $p=0,1,\dots,24$) before orbifolding all have tension $g=1$ because the Leech lattice is unimodular and elementary D-branes correspond to primitive sublattices (thus $\det G=1$ in \eqref{sigbranegTd}).\footnote{In partcular, these D-branes transform into one another under the global symmetry of the Leech torus CFT given by the Conway group Co$_0$.} Consequently the regular branes in the $\mZ_2\times \mZ_2$ orbifold theories have $g=2$ whereas the fractional branes have $g=1$.\footnote{Because one of the $\mZ_2$ in the orbifold acts as a T-duality on all 24 directions of the Leech torus, the branes that survive the orbifold always involve a combinations of D$p$-D$(24-p)$ branes.} All these ``geometric'' D-branes are stable. The regular branes become reducible at isolated points on their moduli space and decompose into the fractional branes. The latter produce a strict subset of the branes $|g\ra$ \cite{Craps:2002rw}. 

Another way to identify conformal boundaries for the $|\rm Monster|^2$ CFT is to consider topological defects in the chiral Monster CFT. The symmetry defects associate with the Monster group $\bf M$ in the chiral theory reproduce the $|g\ra$ branes after folding. However there is a much richer zoo of topological defects that are just starting to be explored (see \cite{Lin:2019hks} for an example). A useful way to identify topological defects in a CFT is to study how it responses to discrete gauging (see for example \cite{Thorngren:2021yso}). Here the Monster symmetry is known to have various non-anomalous subgroups that can be gauged \cite{Johnson-Freyd:2017ble}. For simplicity, we will focus on the cyclic subgroups generated by elements $g\in {\bf M}$.\footnote{There are 194  conjugacy classes of group elements in $\bf M$ and 172  conjugacy classes of cyclic subgroups. Out of the 172 classes, 121 of them have Fricke generators and 51 do not. The 82 conjugacy classes of non-anomalous cyclic groups with Fricke generators correspond to those with $h=1$ in Table 2 of \cite{Conway:1979qga}. The 38 conjugacy classes of non-anomalous cyclic groups without Fricke generators can be found in Appendix 8 of \cite{2018}.}

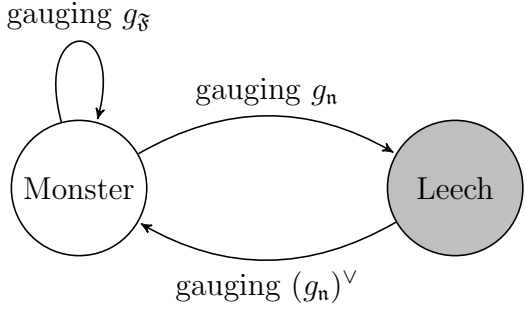
\begin{figure}[!htb]
    \centering
 \begin{tikzpicture}[->,>=stealth',shorten >=1pt,auto,node distance=5cm,
                semithick,scale=1,transform shape]
\tikzstyle{every state}=[text=black,minimum size=18mm,inner sep=0pt]

\node[state]    (A)                    {Monster};
\node[state,fill=gray!50]    (B) [right of=A]       {Leech};

\path (A) edge [loop above] node {gauging $g_\mathfrak{F}$} (A)
          edge [bend left]  node {gauging $g_\mathfrak{n}$} (B)
      (B)
          edge [bend left]  node {gauging $(g_\mathfrak{n})^\vee$} (A);
\end{tikzpicture}
    \caption{Gauging non-anomalous cyclic subgroups of $\bf M$ in the chiral Monster CFT. Depending on whether the subgroup contains a Fricke element $g_\mathfrak{F}$ or not (generated by a non-Fricke element $g_\mathfrak{n}$), the resulting CFT is either the chiral Monster itself or the chiral Leech CFT. In the latter case, gauging the dual symmetry generated by $(g_\mathfrak{n})^\vee$ recovers the chiral Monster CFT \cite{Tuite:1990us,Tuite:1993hy,Paquette:2017xui,2018}. }
    \label{fig:Fricke}
\end{figure}

A distinguished class of non-anomalous cyclic subgroups of $\bf M$ are those generated by the so-called Fricke elements of $\bf M$ whose orders are given by the prime factors of $|\bf M|$ below \cite{fricke1922elliptischen,ogg1975automorphismes},
\ie
\mathfrak{F}=\{2,3,5,7,11,13,17,19,23,29,31,41,47,59,71\}.
\label{fricke}
\fe
We denote these cyclic subgroups as $\mZ_{pA}$ and the Fricke generator by $g_\mathfrak{F}$.\footnote{They are labelled as  conjugacy classes $p+$ in Table 2 of \cite{Conway:1979qga}.} An element $g\in {\bf M}$ is Fricke if the corresponding Mckay-Thompson series $T_g(\tau)$ is invariant under the Fricke involution (an element of $PSL(2,\mR)$),
\ie
T_{g}(-{1/(h o(g) \tau) })=T_{g}(\tau)\,,
\fe
where $o(g)$ is the order of $g$ and $h|24$ is a positive integer. 
Furthermore, $T_{g}(\tau)$ is singular at $\tau\to 0$ ($q\to 1$ in \eqref{MTs}) if and only if $g$ is Fricke. The non-anomalous case corresponds to $h=1$, which is the case for the Fricke elements $g_{\mf{F}}$ in the list \eqref{fricke}.

The chiral Monster CFT responds differently under gauging a non-anomalous cyclic subgroup of $|\bf M|$ depending on whether it is generated by a Fricke element or not, as summarized in Figure~\ref{fig:Fricke}. In particular 
it is  self-dual under gauging the discrete $\mZ_{pA}$ subgroups generated by a Fricke element $g_\mf{F}$ \cite{Tuite:1990us,Tuite:1993hy,Paquette:2017xui,2018}. By general arguments in \cite{Thorngren:2021yso}, this implies the existence of   $\mZ_{pA}$ duality defects $\cN_p$ with quantum dimension $\sqrt{p}$, which correspond to new conformal boundaries with $g=\sqrt{p}$ in the $|\rm Monster|^2$ CFT, which we label as $|\cN_p\ra$ (see Section~\ref{sec:bdyfromTDL}). The annulus partion function of these ``non-geometric'' branes are given by
\ie
Z_{\cN_p\cN_p}( t)=\sum_{g\in \mZ_{pA}} Z_{g\id}(t) =\sum_{g\in \mZ_{pA}} T_{g}(it)
\fe
which follows from the fusion rule of the duality defect $\cN_p$ (see \eqref{PFfromTDL}) and in the last step we have used the relation \eqref{cylMT}. The boundary operator spectrum follows from the modular S-transform. From the properties of the Fricke elements and their Mckay-Thompson series reviewed above, we see the open string channel with boundary $|\cN_p\ra$ at low temperature is dominated, in addition to $T_1(\tau)=J(\tau)$, by the S-transformed Mckay-Thompson series associated to the Fricke elements in $\mZ_{pA}$,
\ie
T_{g_\mf{F}}(-1/\tau)={1\over q^{1/p}}+\cO(q)\,.
\fe
Therefore we conclude immediately that while $|\cN_p\ra$ defines elementary boundary states in the $|\rm Monster|^2$ CFT, they have a boundary gap 
\ie
h^{\rm gap}_{\cN_p \cN_p}={p-1\over p}\,,
\fe
and thus are always unstable, in agreement with our bootstrap bounds below.

In Table \ref{tab:c24} we present numerical upper and lower bounds on the $g$ function for stable boundary conditions in the $|\text{Monster}|^2$ CFT, with only the information of the bulk gap $\Delta_{\rm gap} = 4$ fed into the numerics. In fact, the lower bound is independent of the assumption on the boundary gap, and so applies for all conformal boundaries of the $|\text{Monster}|^2$ CFT, stable or not. 

The lower bounds in Table \ref{tab:c24} appear to be convergent upon 1. Indeed, using more detailed information about the bulk spectrum in order to obtain more convincing numerical evidence, \cite{Friedan:2013bha} conjectured that all conformal boundaries in the $|\text{Monster}|^2$ CFT obey 
\begin{equation}\label{eq:g1Conjecture}
    g \geq 1.
\end{equation}
While we could input more information about the bulk spectrum to acquire more convincing evidence for this conjecture, the novel result here is the upper bound on $g$ for stable branes (which would also be slightly improved by feeding in more information about the bulk spectrum to the numerics). 

In Section~\ref{sec:c24Saturation}, we will in fact prove the conjecture (\ref{eq:g1Conjecture}) by analytically constructing the extremal functional.

We note in passing that if we were to make a stronger assumption on the boundary gap than the minimal one required for stability of the conformal boundary condition, then the upper bound on $g$ would improve. For example, if we assume $h_{\rm gap} = 2$, then we find numerical upper bounds on $g$ convincingly convergent upon 1. Indeed, in Section~\ref{sec:c24Saturation}, we will analytically construct a functional that proves that $g \leq 1$ whenever $h_{\rm gap} \geq 2$ for $c=24$, regardless of the bulk spectrum.

\begin{table}[h]
    \centering
    \begin{tabular}{c|c|c}
        $N$ & lower bound & upper bound  \\
        \hline\hline
        63 & 0.9938113725 & 3904117.1366\\
        \hline
        95 & 0.9990226855 & 3352366.1444\\
        \hline
        127 & 0.999800478 & 3297998.9342\\
        \hline
    \end{tabular}
    \caption{Upper and lower bounds on the $g$-function for stable boundary conditions in $c=24$ CFTs with $\Delta_{\rm gap} = 4$ at various derivative orders $N$. The lower bounds are insensitive to the assumption of any gap in the boundary spectrum and so rigorously bound the tension of branes in the $|\text{Monster}|^2$ CFT from below, regardless of stability.}
    \label{tab:c24}
\end{table}

\section{Analytic Bounds and Optimal Solutions}\label{sec:analyticFunctionals}
Let us now explain how to prove the exact saturation of the upper and lower bounds on $g$ for central charges $c=8$ and $c=24$. Our main tools will be the analytic functionals reviewed in Section \ref{subsec:analyticFunctional}, which can be applied to the problem at hand thanks to the observations of Section \ref{subsec:ZfromG}.

\subsection{Saturation at $c=8$}\label{sec:c8Saturation}
Let us start by stating the results. Suppose $c=8$ and $\hgap \geq 1$. Then there is a rigorous upper bound $g\leq 1$ independent of $\Dgap$. Similarly, suppose $c=8$ and $\Dgap \geq 2$. Then there is a rigorous lower bound $g\geq 1$ independent of $\hgap$. In both cases, there is a unique partition function which saturates the bound. This is the cylinder partition function associated with the identity Cardy brane in the $(E_8)_1$ WZW model
\begin{equation}
    Z(t) = \frac{E_4(it)}{\eta(it)^{8}}\,.
    \label{eq:ZUnique8}
\end{equation}
Here $E_4(\tau)$ is the holomorphic Eisenstein series. The bulk scalar spectrum consists of the identity and positive even integer dimensions $\Delta =2,4,\ldots$. The boundary spectrum has identity and primaries of positive integer dimensions $h=1,2,\ldots$.

Given the discussion of Sections \ref{subsec:analyticFunctional} and \ref{subsec:ZfromG}, it is straightforward to construct the analytic functionals which prove these inequalities. To prove the upper bound on $g$ under the boundary gap assumption, we will use the functional
\begin{equation}
    \omega^{\text{upper}}_{c=8}[\cF(t)] = -\beta^{+}_{\frac{1}{2}}\left[\frac{\widehat{\cF}(z)+\widehat{\cF}(1-z)}{2}\right]+\beta^{-}_{\frac{1}{2}}\left[\frac{\widehat{\cF}(z)-\widehat{\cF}(1-z)}{2}\right]\,.
    \label{eq:omegaUpper}
\end{equation}
Here $\beta^{+}_{\Df}$ and $\beta^{-}_{\Df}$ are the analytic functionals for the 1D correlator bootstrap, discussed in Section \ref{subsec:analyticFunctional}. Here, we use them with $\Df = c/16=1/2$. We used the definition
\begin{equation}
    \widehat{\cF}(z) = \left[2^8 z (1-z)\right]^{-\frac{c}{24}}\cF(t(z))\,,
    \label{eq:FHat}
\end{equation}
where $t(z)$ is given in \eqref{eq:zToTau}. The prefactor arises from the conversion between a partition function and a four-point function of twist operators, see \eqref{eq:GfromZ}.

Recall that $\beta^{+}_{\Df}$ and $\beta^{-}_{\Df}$ exhibit a characteristic pattern of double zeros when acting on 1D conformal blocks. Recall also from \eqref{eq:expansion1} and \eqref{eq:expansion2} that both the bulk and boundary characters can be expanded in towers of 1D conformal blocks with even integer shifts. It follows that $\omega^{\text{upper}}_{c=8}$ inherits the structure of zeros when acting on characters:
\begin{enumerate}
    \item $\omega^{\text{upper}}_{c=8}[\chi_{\frac{\Delta}{2}}(t)]$ has double zeros at $\Delta=2n+2$ for integer $n\geq 0$.
    \item $\omega^{\text{upper}}_{c=8}[\chi_{h}(t^{-1})]$ has a simple zero with a negative slope at $h=1$ and double zeros at   $h=n+1$ for integer $n\geq 1$.
\end{enumerate}

\begin{figure}[h]
    \centering
    \includegraphics[width=.7\textwidth]{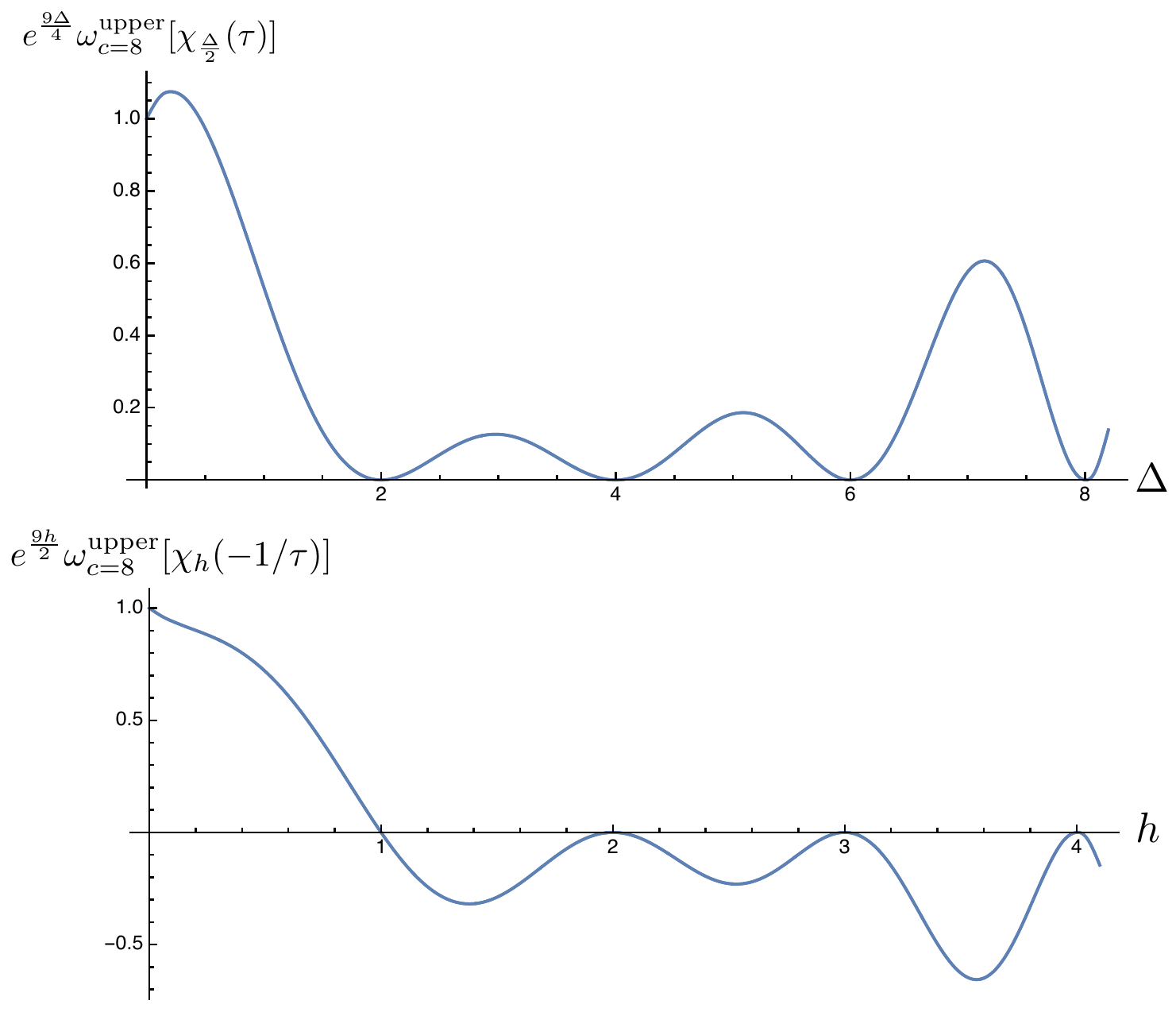}
    \caption{The action of the functional $\omega^{\text{upper}}_{c=8}$, defined in \eqref{eq:omegaUpper}, on the bulk (upper plot) and boundary (lower plot) characters. For clarity, we plot $e^{9\Delta/4}\omega^{\text{upper}}_{c=8}[\chi_{\frac{\Delta}{2}}(\tau)]$ and $e^{9h/2}\omega^{\text{upper}}_{c=8}[\chi_{h}(-1/\tau)]$. The numerical functional at $N=111$ is indistinguishable from the analytic curve by the naked eye.}
    \label{fig:Functional-upper-c8}
\end{figure}

In particular, the first derivative of $\omega^{\text{upper}}_{c=8}[\chi_{\frac{\Delta}{2}}(t)]$ at $\Delta = 2$ cancels between the first and second term in \eqref{eq:omegaUpper}. Furthermore, thanks to the positivity of coefficients in \eqref{eq:expansion1} and \eqref{eq:expansion2}, the action of $\omega^{\text{upper}}_{c=8}$ on characters also inherits positivity from the action on 1D conformal blocks:
\begin{enumerate}
    \item $\omega^{\text{upper}}_{c=8}[\chi_{\frac{\Delta}{2}}(t)]\geq 0$ for all $\Delta>0$.
    \item $\omega^{\text{upper}}_{c=8}[\chi_{h}(t^{-1})]\leq 0$ for all $h\geq 1$.
\end{enumerate}
Finally, it is not hard to check that 
\begin{equation}
    \omega^{\text{upper}}_{c=8}[\chi_{\text{vac}}(t)] = \omega^{\text{upper}}_{c=8}[\chi_{\text{vac}}(t^{-1})] > 0\,.
\end{equation}
Figure~\ref{fig:Functional-upper-c8} shows the action of $\omega^{\text{upper}}_{c=8}$ on bulk and boundary characters. When we apply $\omega^{\text{upper}}_{c=8}$ to the modular crossing equation, the positivity properties just stated imply $g\leq 1$ under the assumption $\hgap\geq 1$.

Let us now prove uniqueness of the saturating solution. Let $Z(t)$ be a solution with $c=8$ satisfying $g=1$ and $\hgap\geq 1$. All nonzero scaling dimensions in both channels must sit at zeros of the functional $\omega^{\text{upper}}_{c=8}$. Therefore, both $Z(t)$ and $Z(t^{-1})$ admit an expansion in $q^{n-\frac{1}{3}}$ with $n\in\mathbb{Z}_{\geq 0}$. Consider $Z_{\pm}(t) = [Z(t) \pm Z(t^{-1})]/2$. It is easy to see that $\eta(it)^{8} Z_{+}(t)$ is a modular form for $SL(2,\mathbb{Z})$ of weight $4$. Since the space of these modular forms is one-dimensional, we have $\eta(it)^{8} Z_{+}(t) = E_4(it)$.

On the other hand, $\eta(it)^{8}Z_{-}(t)$ transforms like a modular form of weight $4$, except that it picks a minus sign under the action of $S\in SL(2,\mathbb{Z})$. However, we have $(S T)^3 = I$. The only way this identity is consistent with the transformations of $\eta(it)^{8}Z_{-}(t)$ is if in fact $Z_{-}(t) = 0$.\footnote{This is just the statement that $SL(2,\mathbb{Z})$ has no nontrivial one-dimensional representation where $T=1$ and $S=-1$.} It follows that \eqref{eq:ZUnique8} is the unique saturating partition function.

The argument proving $g\geq 1$ under the assumption $\Dgap\geq 2$ is obtained by switching the roles of the bulk and boundary channels. We use the functional
\begin{equation}
\begin{aligned}
     \omega^{\text{lower}}_{c=8}[\cF(t)] &=
    \beta^{+}_{\frac{1}{2}}\left[\frac{\widehat{\cF}(z)+\widehat{\cF}(1-z)}{2}\right]+\beta^{-}_{\frac{1}{2}}\left[\frac{\widehat{\cF}(z)-\widehat{\cF}(1-z)}{2}\right] =\\
    &=-\omega^{\text{upper}}_{c=8}[\cF(t^{-1})]\,.
\end{aligned}
    \label{eq:omegaLower}
\end{equation}
It follows straigthforwardly from the discussion of $\omega^{\text{upper}}_{c=8}$ that $\omega^{\text{lower}}_{c=8}$ satisfies the required positivity properties to prove $g\geq 1$ if $\Dgap\geq 2$. The argument showing uniqueness of the saturating solution is also identical.

\subsection{Saturation at $c=24$}\label{sec:c24Saturation}
The situation for $c=24$ is very similar to $c=8$. Indeed, suppose $c=24$ and $\hgap \geq 2$. Then we claim that we must have $g\leq 1$ independently of $\Dgap$. Similarly, suppose $c=24$ and $\Dgap \geq 4$. Then there is a lower bound $g\geq 1$ independent of $\hgap$. In both cases, there is a unique partition function which saturates the bound. This is the cylinder partition function for the identity brane in the $|\mathrm{Monster}|^2$ CFT
\begin{equation}
    Z(t) = j(it) - 744\,,
    \label{eq:monsterZ}
\end{equation}
where
$j(\tau)$ is the modular $j$-invariant
\begin{equation}
    j(\tau) = \frac{1728 E_4^3(\tau)}{E_4^3(\tau)-E_6^2(\tau)} = \frac{1}{q} + 744 + 196884q+ O(q^2)\,.
\end{equation}

\begin{figure}[h]
    \centering
    \includegraphics[width=.7\textwidth]{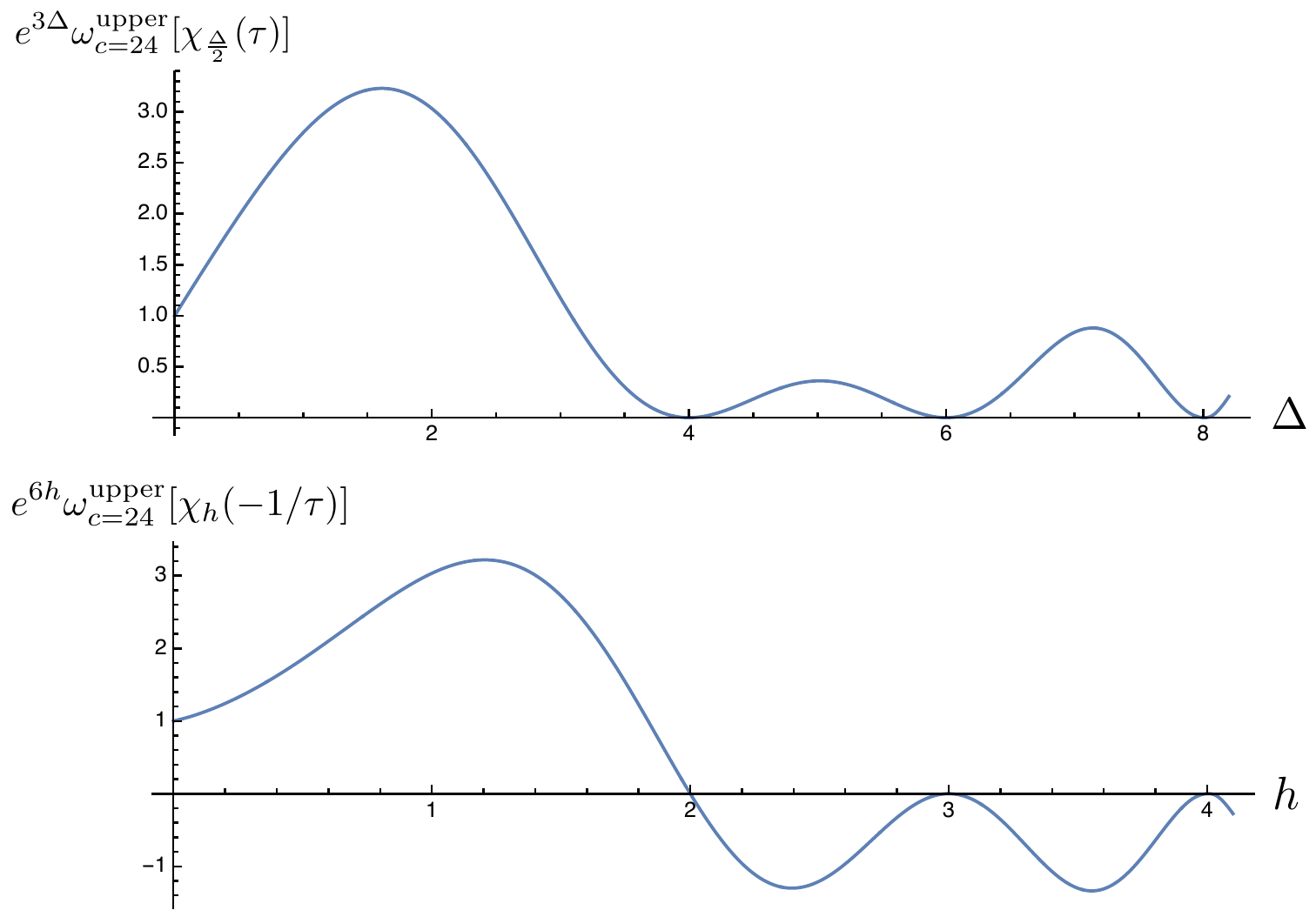}
    \caption{The action of the functional $\omega^{\text{upper}}_{c=24}$, defined in \eqref{eq:omegaUpper24}, on the bulk (upper plot) and boundary (lower plot) characters. For clarity, we plot $e^{3\Delta}\omega^{\text{upper}}_{c=24}[\chi_{\frac{\Delta}{2}}(\tau)]$ and $e^{6h}\omega^{\text{upper}}_{c=24}[\chi_{h}(-1/\tau)]$. The numerical functional at $N=111$ agrees with the analytic curve to an excellent accuracy.}
    \label{fig:Functional-upper-c24}
\end{figure}

To prove these results, we use analogous functionals to those of the previous section, but this time with $\Df = c/16 = 3/2$
\begin{equation}
    \begin{aligned}
         \omega^{\text{upper}}_{c=24}[\cF(t)] &= -\beta^{+}_{\frac{3}{2}}\left[\frac{\widehat{\cF}(z)+\widehat{\cF}(1-z)}{2}\right]+\beta^{-}_{\frac{3}{2}}\left[\frac{\widehat{\cF}(z)-\widehat{\cF}(1-z)}{2}\right]\\
         \omega^{\text{lower}}_{c=24}[\cF(t)] &= -\omega^{\text{upper}}_{c=24}[\cF(t^{-1})]\,.
    \end{aligned}
    \label{eq:omegaUpper24}
\end{equation}
The structure of zeros of $\omega^{\text{upper}}_{c=24}$ is
\begin{enumerate}
    \item $\omega^{\text{upper}}_{c=24}[\chi_{\frac{\Delta}{2}}(t)]$ has double zeros at $\Delta=2n+4$ for integer $n\geq 0$.
    \item $\omega^{\text{upper}}_{c=24}[\chi_{h}(t^{-1})]$ has a simple zero with a negative slope at $h=2$ and double zeros at   $h=n+2$ for integer $n\geq 1$.
\end{enumerate}
Similarly, the positivity properties become
\begin{enumerate}
    \item $\omega^{\text{upper}}_{c=24}[\chi_{\frac{\Delta}{2}}(t)]\geq 0$ for all $\Delta>0$.
    \item $\omega^{\text{upper}}_{c=24}[\chi_{h}(t^{-1})]\leq 0$ for all $h\geq 2$.
\end{enumerate}
We also find $\omega^{\text{upper}}_{c=24}[\chi_{\text{vac}}(t)] = \omega^{\text{upper}}_{c=24}[\chi_{\text{vac}}(t^{-1})] > 0$. These properties together guarantee that the upper bound holds. The argument for the lower bound is again identical. Figure~\ref{fig:Functional-upper-c24} shows the action of $\omega^{\text{upper}}_{c=24}$ on bulk and boundary characters.

To prove uniqueness of the $Z(t)$ with $c=24$, $g=1$ and $\hgap\geq 2$, note that the pattern of zeros of $\omega^{\text{upper}}_{c=24}$ implies that both $Z(t)$ and $Z(t^{-1})$ admit an expansion of the form
\begin{equation}
    \frac{1}{q}+\sum\limits_{n=1}^{\infty} a_n q^n\,.
\end{equation}
This means that $Z_{+}(t) = [Z(t)+Z(t^{-1})]/2$ is invariant under $SL(2,\mathbb{Z})$ and also admits such an expansion. Such a function is unique and equal to $Z_{+}(t) = j(it) - 744$. The argument that $Z_{-}(t) = [Z(t)-Z(t^{-1})]/2$ must vanish is the same as for $c=8$, and so uniqueness of \eqref{eq:monsterZ} is established.

Finally, let us remark on the relationship of the results of this section to the sphere packing problem. Reference \cite{Hartman:2019pcd} explained that the functionals $\beta^{+}_{\Df}$ and $\beta^{-}_{\Df}$ give rise respectively to the Fourier-even and Fourier-odd part of the extremal functions used in the solution of the sphere packing problem in dimension 8 \cite{sphere8} and 24 \cite{sphere24}. In the same work, $\beta^{-}_{\Df}$ was applied to the modular bootstrap equation for the torus partition function. However, the physical role of $\beta^{+}_{\Df}$ remained unclear. Here, we explained that when we consider the annulus partition function, both $\beta^{+}_{\Df}$ and $\beta^{-}_{\Df}$ are needed and appear on equal footing, since the bulk and boundary spectrum are a priori independent. This is also the case for sphere packing. In fact, the combination \eqref{eq:omegaUpper} is precisely the same as the one used in the works on the sphere packing problem.

\section{Universal Bounds on Brane Tension}\label{sec:universalBounds}

Throughout this paper we have focused on deriving bounds on the $g$-function for stable boundary conditions in specific CFTs, or within certain conformal manifolds of CFTs. However, with appropriate assumptions on the boundary spectrum, one can derive a nontrivial upper bound on the $g$-function without any assumptions on the bulk spectrum.\footnote{Indeed, as remarked at the end of Section~\ref{subsec:c1}, this bound is insensitive to any assumption on the gap in the bulk spectrum.} Similarly, with suitable assumptions on the spectrum of bulk scalars, there exists a nontrivial lower bound on the $g$-function that does not require any assumptions on the boundary spectrum beyond unitarity. In this section we will explore such bounds on $g$ as a function of the central charge with both numerical and analytic techniques, with various assumptions on the bulk or boundary spectra.

\subsection{Universal Upper Bounds on Stable Branes}\label{subsec:universalUpper}
For $c<25$, there is a nontrivial upper bound on the $g$-function associated to stable boundary conditions, namely those for which the gap in the spectrum of boundary operators is greater than one. The reason that this bound only exists for $c<25$ is that, as commented in Section~\ref{subsec:linearProgramming}, one needs to assume $h_{\rm gap} > {c-1\over 24}$ in order for an upper bound on $g$ to exist in the first place.

\begin{figure}[h]
    \centering
    \includegraphics{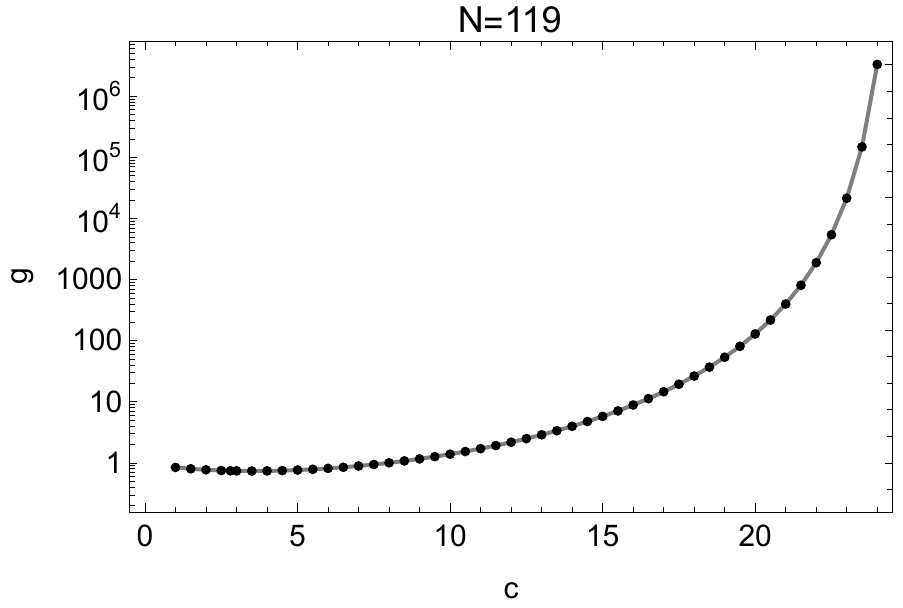}
    \caption{The upper bound on the $g$-function for stable boundary conditions as a function of central charge for $c \leq 25$. The bound diverges as $c$ approaches 25 as that is the point at which the $h_{\rm gap} \geq {c-1\over 24}$ threshold needed for the existence of the upper bound starts to exceed the stability threshold $h_{\rm gap} = 1$.}
    \label{fig:universalUpper}
\end{figure}

In Figure~\ref{fig:universalUpper} we plot the upper bound on the $g$-function for stable boundary conditions for $1<c<25$ obtained at a fixed value of the derivative order. As expected, the bounds become significantly weaker as the central charge approaches 25. In the case that the bulk CFT is part of a worldsheet string theory, this nontrivial upper bound implies that the spectrum of D-branes in the theory cannot have arbitrarily large tension.

\subsection{Universal Lower Bounds on Branes with a Bulk Gap}\label{subsec:universalLower}
\begin{figure}
    \centering
    \includegraphics{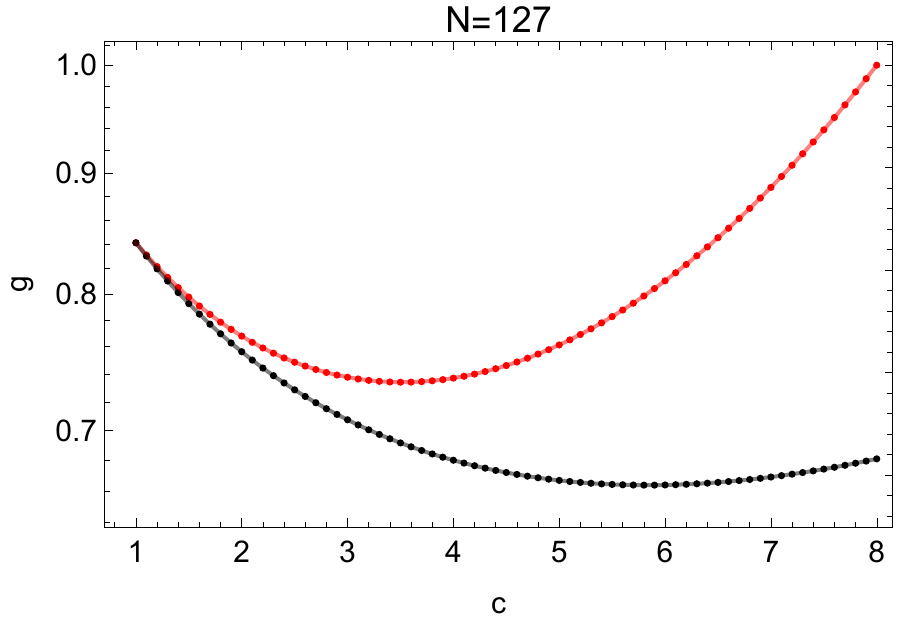}
    \caption{Bounds on the $g$ function for $1\leq c \leq 8$. The lower bounds (black) are computed assuming a gap in the spectrum of bulk scalars of $\Delta_{\rm gap} = {c\over 6} + {1\over 3}$, which is close to the maximal gap for the range of central charges displayed. They are superposed with the upper bounds (red) computed assuming stability of the corresponding boundary condition.}
    \label{fig:universalBoth}
\end{figure}

As was already established by \cite{Friedan:2013bha}, provided the gap in the spectrum of bulk scalar operators exceeds $c-1\over 12$, one can derive a nontrivial lower bound on the $g$-function without needing to impose any assumptions on the boundary spectrum. As a proof of principle, in Figure~\ref{fig:universalBoth} we plot such lower bounds assuming a gap in the bulk spectrum $\Delta_{\rm gap} = {c\over 6} + {1\over 3}$ for $1\leq c \leq 8$ (for these values of the central charge, the assumed value of $\Delta_{\rm gap}$ is close to the maximal one allowed by modular invariance). We see that for CFTs with gaps greater than or equal to this value, the $g$-function of admissable stable boundary conditions is constrained to lie in a narrow window.

Let us provide some intuition for the necessity of a bulk scalar gap for a lower bound of the brane tension for CFTs with a sigma model description. As reviewed in Section~\ref{sec:sigma}, the D$0$-brane has a $g$-function \eqref{sigbraneg} that inversely proportional to the volume $V$ of the target space manifold. Consequently, in cases where the volume $V$ is a conformal moduli, at large $V$, the D$0$-brane tension $g_0\sim V^{-{1\over 2}}$ and the bulk scalar gap $\Delta_{\rm gap}\sim V^{-{2\over d}}$. Therefore the brane tension can be lowered arbitarily close to $0$ at the expense of vanshing $\Delta_{\rm gap}$. Curiously the sigma model picture also suggests a diverging boundary gap $h_{\rm gap}$ for the D0 brane in the large volume limit. This is because the open string winding modes are associated with geodesics on the target space which become increasingly long in this limit
\cite{Gao:2013mn}. On the other hand, for D$p$-branes wrapping smooth submanifolds of  dimension $1\leq p< {d\over 2}$ in the target space, their tension behaves as $V^{2p-d\over 2d}$ in the large volume limit, and the boundary gap closes $h_{\rm gap}\sim V^{-{2\over d}}$ due to open string momentum modes along the brane.

\subsection{Entropy Bound for General $c$ from Analytic Functionals}\label{subsec:analyticGeneralc}
In this subsection, we will extend the analytic functional method away from $c=8$ and $c=24$. As a result, we will be able to derive upper and lower bounds on $g$ for general $c>1$. However, the resulting bounds will only be optimal for $c=8$ and $c=24$. 

\begin{figure}[h]
    \centering
    \includegraphics[width=.6\textwidth]{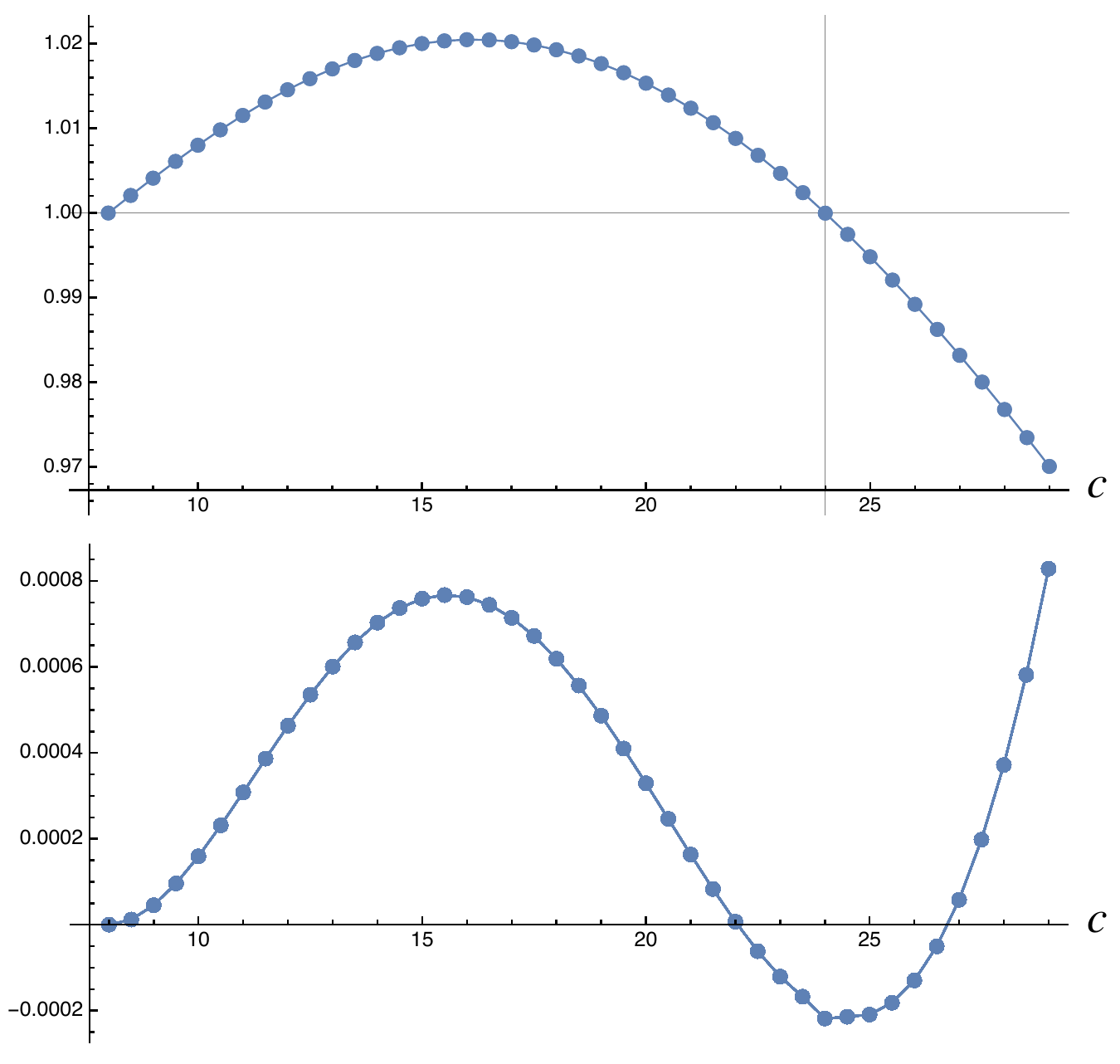}
    \caption{Upper plot shows the upper bound on the boundary entropy $g$, given by \eqref{eq:gAnalyticUniversalUpper}, coming from the analytic functionals under the assumption $h_{\text{\rm gap}}\geq (c+8)/16$. Lower plot shows the difference $g_{\text{ana.}}-g_{\text{num.}}$ between the analytic upper bound \eqref{eq:gAnalyticUniversalUpper} and the corresponding bound derived using the standard numerical bootstrap under the assumption $h_{\text{\rm gap}}\geq (c+8)/16$ at $N=127$.}
    \label{fig:Universal-bounds}
\end{figure}

Recall from Section \ref{subsec:analyticFunctional} the functionals $\beta^{+}_{\Df}$ and $\beta^{-}_{\Df}$. These functionals exist for any $\Df>0$, but so far we have only used them for $\Df=1/2$ and $\Df=3/2$. It is natural to consider the extension of \eqref{eq:omegaUpper} and \eqref{eq:omegaUpper24} to general $c>1$
\begin{equation}
    \begin{aligned}
         \omega^{\text{upper}}_{c}[\cF(t)] &= -\beta^{+}_{\frac{c}{16}}\left[\frac{\widehat{\cF}(z)+\widehat{\cF}(1-z)}{2}\right]+\beta^{-}_{\frac{c}{16}}\left[\frac{\widehat{\cF}(z)-\widehat{\cF}(1-z)}{2}\right]\\
         \omega^{\text{lower}}_{c}[\cF(t)] &= -\omega^{\text{upper}}_{c}[\cF(t^{-1})]\,,
    \end{aligned}
    \label{eq:omegaUpperGeneral}
\end{equation}
where $\widehat{F}(z)$ is defined in \eqref{eq:FHat}. $\omega^{\text{upper}}_{c}$ has the following structure of zeros when acting on the bulk and boundary characters
\begin{enumerate}
    \item $\omega^{\text{upper}}_{c}[\chi_{\frac{\Delta}{2}}(t)]$ has double zeros at $\Delta=2n+(c+8)/8$ for integer $n\geq 0$.
    \item $\omega^{\text{upper}}_{c}[\chi_{h}(t^{-1})]$ has a simple zero with a negative slope at $h=(c+8)/16$ and double zeros at $h=n+(c+8)/16$ for integer $n\geq 1$.
\end{enumerate}
It also has the following positivity properties
\begin{enumerate}
    \item $\omega^{\text{upper}}_{c}[\chi_{\frac{\Delta}{2}}(t)]\geq 0$ for all $\Delta>0$.
    \item $\omega^{\text{upper}}_{c}[\chi_{h}(t^{-1})]\leq 0$ for all $h\geq (c+8)/16$.
\end{enumerate}
It follows from applying $\omega^{\text{upper}}_{c}$ to the modular crossing equation that under the assumption $\hgap\geq(c+8)/16$, $g$ satisfies the upper bound
\begin{equation}
    g^2 \leq \frac{\omega^{\text{upper}}_{c}[\chi_{\text{vac}}(t^{-1})]}{\omega^{\text{upper}}_{c}[\chi_{\text{vac}}(t)]}\,.
    \label{eq:gAnalyticUniversalUpper}
\end{equation}
Bound on $g$ in \eqref{eq:gAnalyticUniversalUpper} is plotted in the upper portion of Figure~\ref{fig:Universal-bounds}. The lower portion of the figure shows the difference $g_{\text{ana.}}-g_{\text{num.}}$ between \eqref{eq:gAnalyticUniversalUpper} and the corresponding numerical bound obtained at $N=127$ under the assumption $\hgap\geq(c+8)/16$. We see that the numerical bound is stronger except for a small region around $c=24$. The numerical bound must become at least as strong as \eqref{eq:gAnalyticUniversalUpper} in the limit of $N\rightarrow\infty$ for any $c$. Since \eqref{eq:gAnalyticUniversalUpper} is optimal at $c=8,24$, we expect that $g_{\text{ana.}}-g_{\text{num.}}$ exhibits double zeros at $c=8,24$ in the limit $N\rightarrow\infty$.

We can go through the analogous argument using $\omega^{\text{lower}}_{c}$. This time, we find that assuming $\Dgap\geq(c+8)/8$, there is a lower bound
\begin{equation}
    g^2 \geq \frac{-\omega^{\text{lower}}_{c}[\chi_{\text{vac}}(t^{-1})]}{-\omega^{\text{lower}}_{c}[\chi_{\text{vac}}(t)]} = 
    \frac{\omega^{\text{upper}}_{c}[\chi_{\text{vac}}(t)]}{\omega^{\text{upper}}_{c}[\chi_{\text{vac}}(t^{-1})]}\,.
    \label{eq:UBGeneral}
\end{equation}
Note that the lower bound is the reciprocal of the upper bound. It is instructive to rewrite the upper bound as follows
\begin{equation}
    \frac{\omega^{\text{upper}}_{c}[\chi_{\text{vac}}(t^{-1})]}{\omega^{\text{upper}}_{c}[\chi_{\text{vac}}(t)]} =
    \frac{-\beta^{+}_{\frac{c}{16}}(\text{vac})-\beta^{-}_{\frac{c}{16}}(\text{vac})}{-\beta^{+}_{\frac{c}{16}}(\text{vac})+\beta^{-}_{\frac{c}{16}}(\text{vac})}\,.
    \label{eq:LBGeneral}
\end{equation}
It turns out that for all $c>1$, we have $\beta^{+}_{\frac{c}{16}}(\text{vac})<0$. On the other hand, the sign of $\beta^{-}_{\frac{c}{16}}(\text{vac})$ is negative for $8<c<24$ and positive otherwise. It follows that the upper bound \eqref{eq:UBGeneral} is above 1 for $8<c<24$ and below 1 otherwise. This means that the upper and lower bounds \eqref{eq:UBGeneral} and \eqref{eq:LBGeneral} are incompatible for $c\in(1,8)\cup(24,\infty)$. This is not a contradiction but simply the statement that when $c$ is in this range, there is no consistent partition function with both $\hgap\geq(c+8)/16$ and $\Dgap\geq(c+8)/8$. The same result follows more directly by considering $[Z(t)+Z(t^{-1})]/2$ and applying the upper bound on the gap for the spinless modular bootstrap proved in \cite{Hartman:2019pcd}.

It is interesting to ask what is the limiting behavior of the bound \eqref{eq:gAnalyticUniversalUpper} as $c\rightarrow\infty$. To find it, we need to evaluate $\beta^{+}_{\frac{c}{16}}(\text{vac})$ and $\beta^{-}_{\frac{c}{16}}(\text{vac})$ at large $c$. This computation can be done using the saddle-point approximation of the contour integrals which define $\beta^{\pm}$, as explained in detail in Section 7.2 of reference \cite{Hartman:2019pcd}. In terms of the 1D correlator cross-ratio $z$, the saddle point for both $\beta^{+}$ and $\beta^{-}$ is located at the value $z_0\in(0,1)$ where the following function is maximized
\begin{equation}
    \log[z(1-z)]-2\pi t(z)\,,
\end{equation}
with $t(z)$ given by \eqref{eq:zToTau}. While there is no analytic formula for $z_0$, numerically we have $z_0 \approx 0.887578\ldots$. The detailed saddle-point calculation then reveals that
\begin{equation}
    \lim_{c\rightarrow\infty}\frac{-\beta^{+}_{\frac{c}{16}}(\text{vac})-\beta^{-}_{\frac{c}{16}}(\text{vac})}{-\beta^{+}_{\frac{c}{16}}(\text{vac})+\beta^{-}_{\frac{c}{16}}(\text{vac})} = \frac{1-z_0}{z_0}\,.
\end{equation}
It follows that under the assumption $h_{\text{\rm gap}}\geq (c+8)/16$, the (non-optimal) analytic upper bound \eqref{eq:gAnalyticUniversalUpper} asymptotes to the constant
\begin{equation}
    g \leq \sqrt{\frac{1-z_0}{z_0}} \approx 0.355895\ldots
\end{equation}
as $c\rightarrow\infty$.

\subsection{Large $c$ and the End-of-the-world Branes in Pure Gravity}

Here we consider annulus bootstrap bounds on the boundary entropy $\log g$ coming from the Cardy condition at large central charge $c\gg 1$ and look for implications for the holographic dual of the BCFT. 

A class of well-studied BCFTs is described by the semi-classical Einstein gravity on the bulk manifold $\cM$ 
that terminates on an end-of-the-world (ETW) brane \cite{Karch:2000gx,Takayanagi:2011zk,Fujita:2011fp,Nozaki:2012qd}. In particular, the BCFT vacuum corresponds to a portion of the AdS$_3$ solution bounded by the ETW brane  over a two-manifold $\cQ$ that ends on the CFT boundary in the asymptotic region.  In general, the bulk theory may involve matter fields both in the bulk and on the ETW brane worldvolume. However it is tempting to postulate a quantum theory of ``pure'' AdS$_3$ gravity with ETW branes such that the only degrees of freedom come from the metric. In this case, the coupled bulk theory is described by the gravity action \cite{Takayanagi:2011zk}
\ie
S={1\over 16\pi G_N} \int_\cM d^3 x \sqrt{-g} \left(R+{2\over \ell^2}\right)+{1\over 8\pi G_N}  
\int_{\pa \cM} d^2 y\sqrt{-h} K 
-{1\over 8\pi G_N}  \int_{\cQ} d^2 y\sqrt{-h} T\,,
\label{pGaction}
\fe
where $G_N$ is the gravitational constant and $\ell$ is the curvature length of the bulk geometry. Under the AdS/CFT dictionary, the central charge of the dual CFT is
\ie
c={3\ell \over 2G_N}\gg 1\,.
\fe
In the last two terms of \eqref{pGaction}, 
we have included the Gibbons-Hawking term on both the ETW brane and the asymptotic boundary, and a constant energy density $T$ on the ETW brane which is also known as the tension of the ETW brane. 

Let us parametrize the AdS$_3$ metric in the following form, 
\ie
ds^2=d\rho^2+\cosh^2{\rho\over \ell} ds^2_{\rm AdS_2} \,,\quad ds^2_{\rm AdS_2}=\ell ^2{dz^2-dt^2\over z^2}
\label{AdS3}
\fe
with $\rho,t \in \mR$ and $z\in \mR_+$. The vacuum solution to \eqref{pGaction} is a restriction of \eqref{AdS3} to the sub-region $\rho \leq \rho_*$
determined by the ETW brane tension as \cite{Takayanagi:2011zk}
\ie
T={1\over \ell } \tanh {\rho_*\over \ell}\,.
\fe

In this case, the ETW brane worldvolume manifold $\cQ$ is the AdS$_2$ slice at $\rho=\rho_*$, which ends on the asymptotic boundary as ${z/\cosh {\rho\over \ell }}\to 0$. Consequently, the geometry \eqref{AdS3} is dual to the vacuum of the BCFT on the half space parametrized by $(t,z> 0)$ at the asymptotic boundary. To determine the boundary entropy, it is convenient to perform a conformal transformation in the Euclidean geometry such that the asymptotic boundary is a disk. The $g$-function is determined by the disk partition function of the BCFT, which 
follows from the regularized bulk on-shell action  \eqref{pGaction} after Wick rotation. The resulting boundary entropy is \cite{Takayanagi:2011zk}
\ie
\log g_{\rm ETW}={\rho_*\over 4G}={\ell \over 4G} {\rm arctanh}(T \ell )={c\over 6}{\rm arctanh}(T \ell )\,.
\fe
We see the boundary entropy grows monotonically in $T$ which satisifies $|T\ell|\leq 1$. Intuitively this is because for larger tension $T$, the ETW brane encloses more of the bulk geometry.  In the following,
we would like to diagnose whether such ETW branes are fully consistent with pure gravity beyond the semi-classical limit. Our main tool here is the annulus bootstrap which will be exploited to derive constraints on ETW branes with certain assumptions on the  spectrum in pure gravity to be elaborated below.

In the strictest setting, pure gravity on AdS$_3$ has a spectrum that consists of the $\Delta=0$ vacuum state and a continnum of heavy states with $\Delta\gtrsim {c\over 12}$  to account for the (spinning) BTZ black holes, and they correspond to Virasoro primaries in the dual CFT. In the Euclidean theory, they are described by smooth saddle points of the gravitational path integral.
The Virasoro descendants come from boundary gravitons that capture the metric fluctuations around these classical configurations. This notion of pure gravity generalizes naturally in the presence of an ETW brane with  constant tension $T$ and no additional worldvolume matter. Since the bulk geometry is always locally AdS$_3$, the ETW brane will simply excise part of the bulk outside the oriented two manifold $\cQ$ that is anchored on the asymptotic boundary. The precise location (embedding) of $\cQ$ is determined by the equation of motion from   \eqref{pGaction}, which leads to the following junction condition on the induced metric $h_{ab}$ and extrinsic curvature $K_{ab}$ on $\cQ$,
\ie
K_{ab}-K h_{ab}=-Th_{ab} 
~\Leftrightarrow~ K_{ab}=T h_{ab}\,.
\label{junction}
\fe
Similar to the local operator spectrum of the dual CFT, we assume the boundary operator spectrum for the ETW brane to have a unique vaccuum at $h_{\rm ETW}=0$ and a continuum at $h_{\rm ETW}\gtrsim {c\over 24}$. The former is simply described by the excised AdS$_3$ solution \eqref{AdS3} in the bulk. The latter are responsible for excised geometries that involve BTZ black holes. 

To see this explicitly, it is convenient to work with the Euclidean signature, with an asymptotic boundary $S^1_\tau \times I_x$ where $\tau$ is the periodic Euclidean time and $I_x$ an interval along the spatial direction on the boundary. The cylinder partition function with the same boundary conditions correspond to the bulk Euclidean path integral with an ETW brane anchored at the two ends of $S^1_\tau \times I_x$. The solutions for such ETW branes satisfying \eqref{junction} can be found in \cite{Takayanagi:2011zk,Fujita:2011fp} whose explicit form we won't need here but we will note the following qualitative feature. Let's denote the usual radial coordinate for asymptotic AdS$_3$ geometries by $z>0$ such that the asymptotic boundary is at $z\to 0$.
For empty AdS$_3$, the ETW brane is connected and reaches some maximum value $z=z_{\rm max}(T)$ determined by the tension before turning back to join the other end of the boundary cylinder.
However in the BTZ geometry with a horizon at $z=z_H$, the ETW brane contain two disconnected components that end on two points in the $x$ direction at the horizon. Under the AdS/CFT dictionary, the scaling dimension of the dual boundary operator is related to the ADM mass $M_{\rm ADM}$,
\ie
M_{\rm ADM} ={\pi \over L} \left( L_0-{c\over 24} \right)
\fe
where $L$ is the length of the interval $I_x$.\footnote{This is a consequence of the relation between the Brown-York tensor \cite{Brown:1992br} and the stress energy tensor of the dual CFT \cite{Balasubramanian:1999re,Myers:1999psa,deHaro:2000vlm}.}
Here $M_{\rm ADM}$ is the conserved charge in the $\tau$ direction for the geometry exercised by the ETW brane. For BTZ black holes, explicit computation following \cite{Balasubramanian:1999re} shows $M_{\rm ADM}\geq 0$, and thus the gap $h_{\rm ETW}\geq {c\over 24}$ in the boundary operator spectrum beyond the identity.

After these preparations, we are ready to test the existence of ETW branes using annulus bootstrap with the following assumptions on the BCFT data,
\begin{enumerate}
    \item Bulk scalar gap $\Delta_{\rm gap}={c\over 12}$\label{item:bulkgap},
      \item Boundary gap $h_{\rm gap}^{\rm ETW}={c\over 24}$\label{item:boundarygap},
      \item Boundary entropy $\log g_{\rm ETW}={c\over 6} {\rm arctanh} (T\ell)$.
\end{enumerate}
We emphasize that conditions \ref{item:bulkgap} and \ref{item:boundarygap} reflect a particularly crude assumption about the spectrum of the dual of pure gravity in the semiclassical limit, and do not incorporate one-loop corrections to the black hole threshold suggested by the explicit sum over saddles in the gravitational path integral \cite{Maloney:2007ud}. Indeed, as we have seen, if $\Delta_{\rm gap}<{c-1\over 12}$ or $h_{\rm gap} < {c-1\over 24}$, then the lower and upper bounds (respectively) on $g$ as computed by the functional method will cease to exist (see  Section~\ref{subsec:linearProgramming}). What we find in this section is that if one assumes gaps that are even an order-one amount above these thresholds, then there are nontrivial upper and lower bounds on the tension of ETW branes in pure gravity. These upper and lower bounds can be evaded by the inclusion of additional operators at or below the thresholds $h = {c-1\over 24}$ and $\Delta = {c-1\over 12}$, respectively.

\begin{figure}[h]
    \centering
    \includegraphics{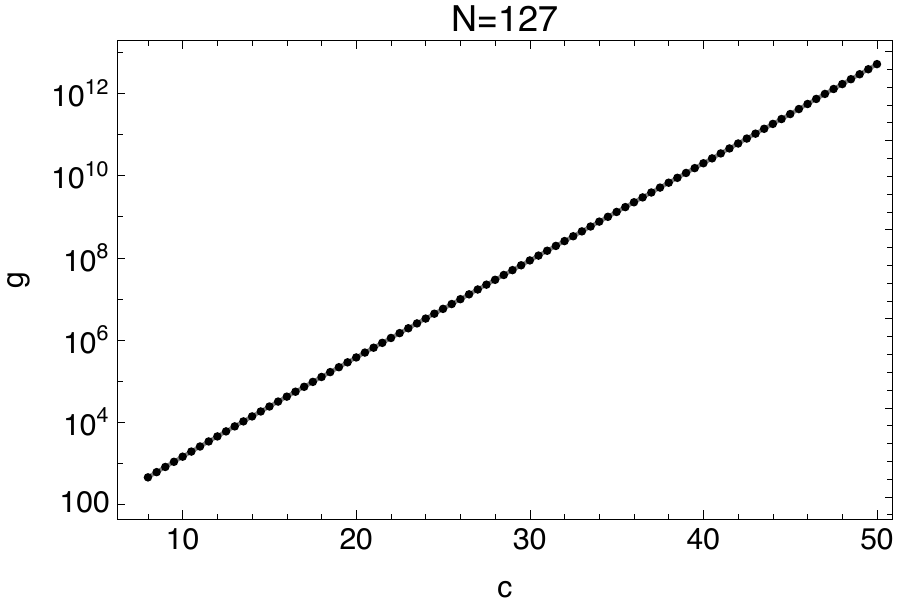}
    \caption{Upper bounds on the $g$-function associated with end-of-the-world branes in pure gravity, with $h_{\rm gap} \geq {c\over 24}.$ The lower bounds on $g$ with $\Delta_{\rm gap} \geq {c\over 12}$ are precisely the reciprocal of these lower bounds.}
    \label{fig:ETW}
\end{figure}

The numeric bounds from the annulus bootstrap are displayed in Figure~\ref{fig:ETW}. We find that at large $c$, condition 1 implies a lower bound on the boundary entropy that is linear in $c$, 
\ie
\log g_{\rm ETW} \gsim -{c\over 6}\times 3.29... \,,
\fe
whereas condition 2 implies an upper bound
\ie
\log g_{\rm ETW} \lsim {c\over 6}\times 3.29 \,.
\fe
Therefore we find that the semiclassical ETW branes are consistent with a putative spectrum of ``pure'' gravity (as defined by conditions \ref{item:bulkgap} and \ref{item:boundarygap} above) at the quantum level only if $|{\rm arctanh}(T\ell)|\lsim 3.29$.\footnote{We emphasize that these are only necessary conditions for the ETW branes to make sense in the pure gravity. Further constraints can come from considering the gravitational one-loop partition function with an ETW brane as studied in \cite{Suzuki:2021pyw}.} 

Alternatively, if one insists on the existence of ETW branes for the entire range of tensions $-1\leq T\ell\leq 1$, our bootstrap bounds imply that there must be non-identity bulk scalar and boundary operators below the corresponding gaps at $c\over 12$ and $c\over 24$. This is reminiscient of recent results obtained by studying the lightcone modular bootstrap \cite{Benjamin:2019stq} which predict that the strict pure gravity with the partition function given in \cite{Maloney:2007ud} is not consistent at the quantum level due to the tension between modular invariance and unitarity. It has been suggested that this non-unitarity can be cured by including states with twist  $\min (h,\bar h)\leq {c-1\over 32}$, which have natural intepretations in the gravitational path integral as a sequence of conical defects in AdS$_3$ \cite{Benjamin:2020mfz}, however we should note that there is also a very intriguing proposal involving off-shell configurations that may not lead to additional operators with twists significantly below the naive black hole threshold in the semiclassical limit \cite{Maxfield:2020ale}.\footnote{However, we should say that the scalar sector of the bulk spectrum, which is the only sector relevant for the annulus bootstrap, is uncontrolled in the proposal of \cite{Maxfield:2020ale}.} 

Our results should be viewed as complementary to recent work studying the relationship between bootstrap bounds on holographic BCFT and gravitational physics of bulk ETW branes. In \cite{Reeves:2021sab} it was shown that the existence of a bulk causal structure associated with a gravitating ETW brane places highly non-generic constraints on the spectrum of boundary operators in holographic BCFT. In \cite{Belin:2021nck} it was found that typical boundary states preserving the full chiral algebra in symmetric orbifolds of irrational seed CFTs correspond to ETW branes with super-Planckian tension, indicating the absence of a nice semiclassical bulk dual. 

We emphasize that here we have not utilized modular invariance of the bulk operator spectrum and consequently
the bounds here are much weaker.  
One expects to get stronger bounds by combining annulus bootstrap wih the modular invariance of torus partition function. We leave this to future investigation.

\section{Outlook}
In this work, we have studied bounds on the boundary entropy which are a consequence of unitarity and consistency of the annulus partition function. After making various assumptions about the bulk and boundary gaps, we saw that the numerical bounds are often nearly saturated by interesting physical theories. We were able to prove exact saturation analytically at $c=8$ and $c=24$ by the $(E_8)_1$ and $|\rm Monster|^2$ CFTs respectively.

Our work gives rise to many concrete technical questions regarding 2d BCFTs. An obvious open problem is to prove the exact saturation of the bootstrap bounds beyond the solved cases of $c=8$ and $c=24$. These cases are special because both the bulk and boundary spectrum consists of a single family of scaling dimensions with equal spacing. Constructing extremal functionals for more general spectra will require new ideas. Such ideas would also be relevant in the context of sphere packing, for example for proving universal optimality of the hexagonal lattice using linear programming bounds.

It would also be interesting to investigate the relationship between the boundary gap and the boundary entropy in more detail. Our results suggest that for a fixed $c$, the boundary gap can be made arbitrarily large at the expense of decreasing $g$ (see Section~\ref{sigma} for an intuitive picture in the context of sigma models). It is an open question whether the boundary gap can be made arbitrarily large subject to unitarity and consistency of the annulus partition function.

Our analysis can be naturally extended in several ways. Firstly, it would be interesting to combine the annulus bootstrap with additional constraints, such as the consistency of bulk two-point functions in the presence of a conformal boundary \cite{Cardy:1991tv}. The latter has been studied separately in the context of CFTs in general $d$ in a number of interesting works, see e.g. \cite{Liendo:2012hy,Gliozzi:2015qsa,Behan:2020nsf}. Another constraint which was not used in our analysis is the integrality of the coefficients appearing in the boundary channel. In that regard, the annulus bootstrap is a natural next step for extending the findings of \cite{Kaidi:2020ecu} for the torus partition function.

It is instructive and straightforward to augment our set-up by imposing the presence of additional symmetries, such as discrete and continuous (generalized) global symmetries and supersymmetry. The resulting constraints can be expected to teach us new lessons about a variety of fascinating questions, such as the classification of branes in the $|\rm Monster|^2$ CFT (and correspondingly interfaces in the chiral Monster CFT), or in the K3 SCFT and more general supersymmetric non-linear sigma models.

We hope to report on some of these directions in the near future.

\section*{Acknowledgements}

We thank Alexandre Belin, Clay Cordova, Michael Douglas, Matthias Gaberdiel, Hao Geng, Jeffrey Harvey, Daniel Jafferis, Zohar Komargodski, Edward Mazenc, Marco Meineri and Cumrun Vafa for useful discussions and correspondences. We also thank Matthias Gaberdiel and Ying-Hsuan Lin for helpful comments on a draft. SC is especially grateful to Xi Yin for suggesting the problem and for collaboration at an early stage of this project. We are grateful to Enrico Brehm for pointing out some typos in an earlier version of the paper. 
The work of YW is  supported in part by the Center for Mathematical Sciences and Applications and the Center for the Fundamental Laws of Nature at Harvard University.
In the case of SC and DM, this work was performed in part at the Aspen Center for Physics, which is supported by National Science Foundation grant PHY-1607611. This work was partially supported by a grant from the Simons Foundation. DM gratefully acknowledges funding provided by Edward and Kiyomi Baird as well as the grant DE-SC0009988 from the U.S. Department of Energy. Some of the numerical computations in this work were performed using the Odyssey cluster supported by the FAS Division of Science research computing group at Harvard University.

\newpage
\bibliographystyle{JHEP}
\bibliography{bref}

\end{document}